\documentclass[11pt]{article}
\usepackage{graphicx,rotating}
\usepackage{comment}
\usepackage{physics}
\usepackage{ragged2e}
\usepackage{cite}
\usepackage{amsmath, amsfonts,amssymb}
\usepackage[export]{adjustbox}

\def\Title#1{\begin{center} {\Large #1 } \end{center}}
\def\Author#1{\begin{center}{ \sc #1} \end{center}}
\def\Address#1{\begin{center}{ \it #1} \end{center}}

\newcommand\pubblock{\begin{flushright}   
ANL-186965, INT-PUB-24-029, MSUHEP-24-008, SMU-PHY-24-01
\\ \pubdate  \end{flushright}}
\newenvironment{Abstract}{\begin{quotation}  }{\end{quotation}}

\usepackage[a4paper,top=1.5cm,bottom=2cm,left=1.5cm,right=1.5cm,bindingoffset=0mm]{geometry}

\newcommand\pubdate{\today}

\usepackage{url} 
\usepackage{hyperref}
\usepackage{xcolor} 
\usepackage{multirow}
\definecolor{nicered}{rgb}{0.5,0.,0.}
\definecolor{nicegreen}{rgb}{0.,0.5,0.}
\definecolor{niceblue}{rgb}{0.,0.,0.5}
\definecolor{darkpink}{rgb}{0.8,0.47,0.47}
\hypersetup{colorlinks,citecolor=nicegreen,linkcolor=nicered,urlcolor=niceblue}
\usepackage{enumitem} 
\usepackage[normalem]{ulem}
\setlist{nolistsep} 
\usepackage{setspace}
\usepackage{bibspacing}

\newcommand{\MeV}{\textrm{MeV}}
\newcommand{\GeV}{\textrm{GeV}}
\newcommand{\TeV}{\textrm{TeV}}
\newcommand{\PeV}{\textrm{PeV}}
\newcommand{\rev}[1]{#1}

\usepackage{ulem}

\begin{document}
\begin{titlepage}

\pubblock
\vfill

\Title{New results in the CTEQ-TEA global analysis of parton distributions in the nucleon}
\vfill

\Author{
A. Ablat,$^1$ A. Courtoy,$^{2}$ S. Dulat,$^{1,6}$ M. Guzzi,$^3$ T.~J.~Hobbs,$^4$ T.-J. Hou,$^5$ J. Huston,$^6$ K.~Mohan,$^6$ H.-W. Lin,$^6$ P. Nadolsky,$^7$ I. Sitiwaldi,$^1$ K. Xie,$^6$ M. Yan,$^8$ C.-P. Yuan$^6$}
\Address{
$^1$Xinjiang University, Urumqi, Xinjiang 830046 China\\
$^2$IF UNAM, Apartado Postal 20-364, 01000 Ciudad de M\'exico, Mexico \\
$^3$Kennesaw State University, Kennesaw, GA 30144, USA \\
$^4$Argonne National Laboratory, Argonne, IL 60439, USA \\
$^5$University of South China, Hengyang, Hunan 421001, China\\
$^6$Michigan State University, East Lansing, MI 48824, USA \\
$^7$Southern Methodist University, Dallas, TX 75275-0181, USA\\ 
$^8$ Peking University, Beijing 100871, China
}

\vfill
\begin{Abstract}
This report summarizes the latest developments in the CTEQ-TEA global
analysis of parton distribution functions (PDFs) in the nucleon. The
focus is on recent NNLO fits to high-precision LHC data at 8 and 13
TeV, including Drell-Yan, jet, and top-quark pair production,  
pursued on the way toward the release of the new generation of
CTEQ-TEA general-purpose PDFs. The report also discusses advancements
in statistical and numerical methods for PDF determination and
uncertainty quantification, highlighting the importance of robust and
replicable uncertainties for high-stakes observables. Additionally, it
covers phenomenological studies related to PDF determination, such as
the interplay of experimental constraints, exploration of correlations
between high-$x$ nucleon sea and low-energy parity-violating
measurements, fitted charm in the nucleon, the photon PDF in the
neutron, and simultaneous SMEFT-PDF analyses.  
\end{Abstract} 

\end{titlepage}

\section{Introduction \label{sec:Intro}}
Parton distribution functions (PDFs) $f_a(x,Q)$ characterize the
internal structure of initial-state hadrons in high-energy collisions
in perturbative quantum chromodynamics (PQCD). State-of-the-art
predictions for hard-scattering cross sections across many processes
at the CERN Large Hadron Collider (LHC) and other experiments require
determinations of PDFs of commensurate precision. CTEQ-TEA -- the
Tung Et Al. group in the CTEQ collaboration -- pursues a comprehensive
research program on the determination of PDFs in the nucleon from a
global analysis of QCD data. In the last two years, the CTEQ-TEA group
focused on the development of a new generation of general-purpose PDFs
that will replace the previous comprehensive generation CT18 of
next-to-next-to-leading order (NNLO) PDFs from our group publicly released 
in 2019 \cite{Hou:2019efy}. Such PDFs are intended for use across a wide
range of applications including studies of Higgs and electroweak
precision observables as well as new physics searches.  In parallel
with this effort, the group carried out multi-prong physics studies
and development of methodology and computational tools to aid various
PDF applications.

Together with MSHT \cite{Bailey:2020ooq}, NNPDF \cite{NNPDF:2021njg},
and other groups \cite{H1:2021xxi,ATLAS:2021vod,Alekhin:2024bhs},
CTEQ-TEA obtains the NNLO PDFs by fitting a large collection of diverse
experimental data sets. 
\rev{Global PDF analyses at NLO are pursued as well \cite{Accardi:2023gyr,Sato:2019yez,Moffat:2021dji}.}
Such analysis is a multi-step process, which
includes implementation of precise theoretical cross sections
(currently predicted at NNLO for many processes and increasingly at
N$^3$LO in DIS
~\cite{Vermaseren:2005qc,Moch:2004xu,Moch:2008fj,Davies:2016ruz,Blumlein:2022gpp} 
(including for massive quarks \cite[ and references
  therein]{Ablinger:2024qxg}), Drell-Yan
process~\cite{Baglio:2022wzu,Duhr:2020sdp,Duhr:2021vwj,Chen:2021vtu,Chen:2022lwc},
Higgs
production~\cite{Anastasiou:2015vya,Mistlberger:2018etf,Dulat:2018bfe,Dreyer:2016oyx,Cieri:2018oms},
and jet production in DIS~\cite{Gehrmann:2018odt,Currie:2018fgr},
although a complete N$^3$LO fit is still in the future);
identification of sensitive, mutually consistent new experimental data
sets using preliminary fits and fast techniques, such as
\texttt{ePump}~\cite{Schmidt:2018hvu,Hou:2019gfw} and
sensitivities~\cite{Wang:2018heo,Hobbs:2019gob} developed by the
CTEQ-TEA group; inclusion of lattice QCD constraints on the PDFs
\cite{Hou:2022onq}; development and implementation of statistical
methods for representative estimation of uncertainties on
PDFs~\cite{Courtoy:2022ocu}, including various sources of aleatory and
epistemic uncertainties; production of the final error PDF sets that
capture the cumulative uncertainty from all such sources.

This article reviews various results that were obtained along the
described avenues on the path toward the next release of the precision
CTEQ-TEA PDF ensemble. We showcase connections and synergies of
extensive ongoing efforts reported in our recent articles on
standalone developments. The upcoming global analysis will integrate
these coordinated developments into the release of new CTEQ-TEA PDFs.

In Sec.~\ref{sec:highdensity}, we present new, highly dense grids for
general-purpose PDFs of the CT18 NNLO family customized for precision
studies.  Section~\ref{sec:LHCdata} reviews the recently completed
studies of new data sets from various energies of the Large Hadron
Collider considered for the inclusion in the new CTEQ-TEA global
fit. Section~\ref{sec:AdvancedStatistics} touches upon the development
of advanced statistical techniques for quantification of PDF
uncertainties.  Following Sec.~\ref{sec:smallx} with a discussion of
recent developments in low-$x$ PDF phenomenology, we turn to several
PDF studies for which high $x$ is more relevant: a preliminary
analysis of low-energy electron-scattering measurements to the
light-quark sea at high $x$ (Sec.~\ref{sec:PVsea}); the recent CT18FC
study of the nonperturbative charm (Sec.~\ref{sec:FC}); a
just-released QCD+QED analysis of the neutron's electromagnetic
structure (Sec.~\ref{sec:neutron}); and a joint fit of PDFs and
coefficients of the effective field theory extension of the Standard
Model (SMEFT) in Sec.~\ref{sec:SMEFT}.  In Sec.~\ref{sec:Conclusions},
we provide a general outlook on these recent CTEQ-TEA activities and
ongoing efforts.


\section{Tabulated CT18 NNLO parton distributions with enhanced precision \label{sec:highdensity}}
\begin{table}[p]
{ \centering
\begin{tabular}{ cc }     
\begin{tabular}{|c|cc||c|cc|}
\hline intervals in x & CT18 & CT18up & intervals in x & CT18 & CT18up
\\ \hline $[ 10^{-10}, 10^{-9}]$ & 1 & 1 & $[0.1, 0.2]$ & 7 & 18 \\ $[
  10^{-9}, 10^{-8}]$ & 11 & 11 & $[0.2, 0.3]$ & 6 & 16 \\ $[ 10^{-8},
  10^{-7}]$ & 12 & 12 & $[0.3, 0.4]$ & 5 & 12 \\ $[ 10^{-7}, 10^{-6}]$
& 11 & 11 & $[0.4, 0.5]$ & 3 & 13 \\ $[ 10^{-6}, 10^{-5}]$ & 12 & 12 &
$[0.5, 0.6]$ & 6 & 15 \\ $[ 10^{-5}, 10^{-4}]$ & 11 & 15 & $[0.6,
  0.7]$ & 6 & 12 \\ $[ 10^{-4}, 10^{-3}]$ & 12 & 23 & $[0.7, 0.8]$ & 8
& 11 \\ $[ 10^{-3}, 10^{-2}]$ & 11 & 23 & $[0.8, 0.9]$ & 14 & 17 \\ $[
  10^{-2}, 10^{-1}]$ & 12 & 40 & $[0.9, 1.0]$ & 15 & 38 \\ \hline
\multicolumn{4}{|r|}{Total} & 161 & 300 \\ \hline
\end{tabular}

\centering
\begin{tabular}{|lr|cc|}
\hline \multicolumn{2}{|c|}{intervals in Q} & CT18 & CT18up\\ \hline
$[ Q_{0}$,& $m_{c}]$ & 2 & 4 \\ $[ m_{c}$,& $m_{b}]$ & 8 & 11 \\ $[
  m_{b}$,& $m_{t}]$ & 14 & 18 \\ $[m_{t}$, & $Q_{\max}]$ & 13 & 16
\\ \hline \multicolumn{2}{|r|}{Total} & 37 & 49 \\ \hline
\end{tabular}
\end{tabular}\par}
\caption{Numbers of nodes within the shown intervals in $x$ (left) and
  $Q$ (right) in the CT18 and CT18up NNLO LHAPDF grids.}
\label{table:grids}
\end{table}

\begin{figure}[p]
    \centering
    \includegraphics[width=0.49\textwidth]{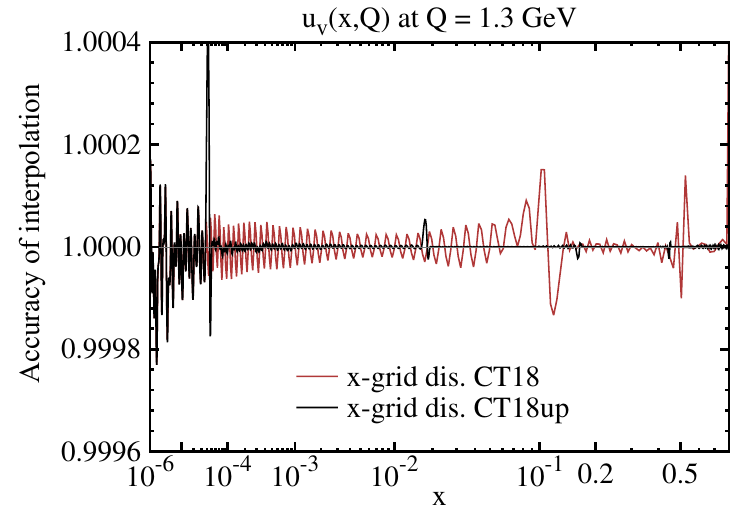}
    \includegraphics[width=0.49\textwidth]{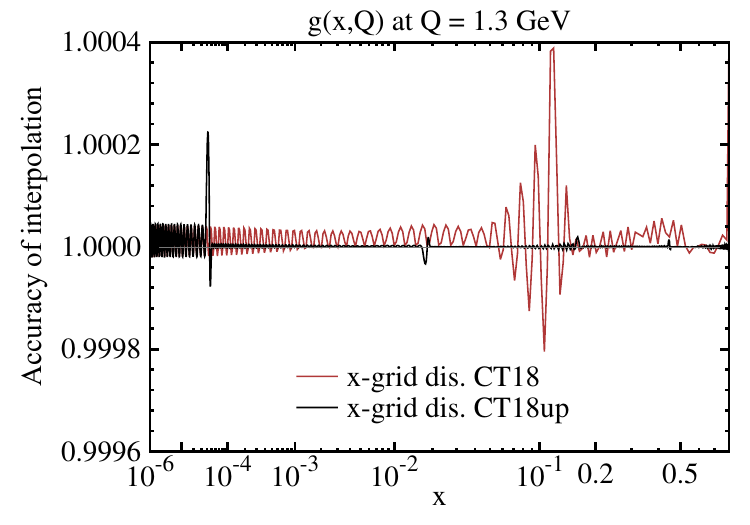}
    \caption{Ratio of PDFs obtained from the LHAPDF grids to those
      returned directly from the CTEQ fitting code, for up valence
      (left) and gluon (right) PDFs at the initial evolution scale
      $Q=1.3$ GeV. The interpolation accuracy for the 2019 CT18 grids
      is shown in red; for the high-density grids (``CT18up''), in
      black.}
    \label{fig:highdensity_x}
\end{figure}

\begin{figure}[p]
    \centering
    \includegraphics[width=0.49\textwidth]{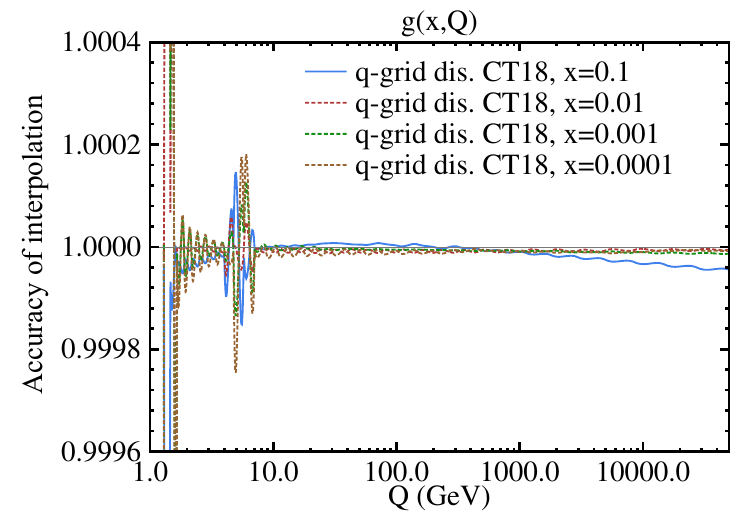}
    \includegraphics[width=0.49\textwidth]{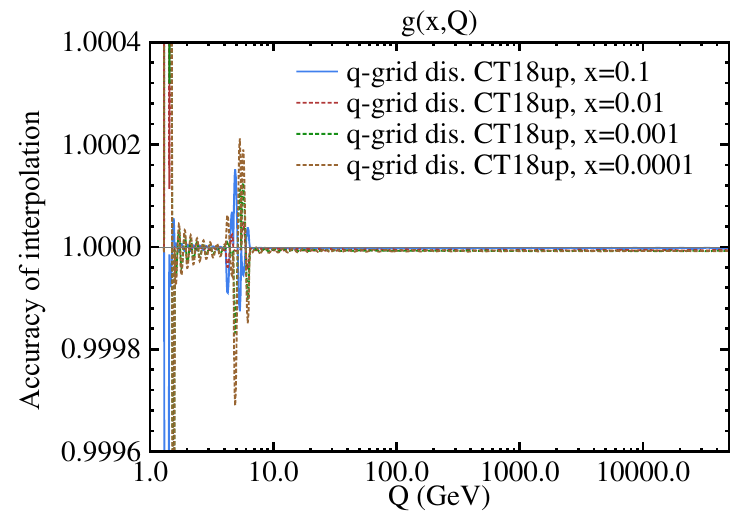}
    \caption{Same as Fig.~\ref{fig:highdensity_x}, for the CT18 gluon
      PDF as a function of $Q$ at $x=0.1, 0.01, 0.001, 0.0001$ (in,
      respectively, blue, red, green and mustard) for the CT18 (left)
      and CT18up (right) grids. }
    \label{fig:highdensity_Q}
\end{figure}

Since the release of CT18 NNLO and NLO PDFs in 2019
\cite{Hou:2019efy}, they have been used in many computations of
growing kinematic reach and precision. The PDF parametrizations are
provided to a broad range of users in the form of interpolated tables
as a function of the partonic momentum fraction $x$ and QCD
factorization scale $Q$ above 1.3 GeV. When generating these tables,
one balances between the precision of interpolation (in all cases
sufficient for general-purpose applications), size of the files, and
speed of the computation. 

As the LHC enters the decade of high-luminosity runs, raising
requirements for precision over the whole domain of momentum
fractions, the CTEQ-TEA group released higher-density versions of
tabulated grids, named ``CT18up'', for all NNLO PDFs of the CT18
family. Namely, the new grids for CT18, CT18Z, CT18A, and CT18X NNLO
error PDF sets with up to 5 active quark flavors at the QCD coupling
strength $\alpha_s(M_Z)=0.118$, $10^{-9}\leq x \leq 1$, and $1.3 \leq
Q \leq 10^{5}$ GeV are published online as described in the Data Availability  
Statement. The original grids of the CT18 family published in 2019,
which are both smaller and moderately faster to run, also remain available
and can be used when very high precision in interpolation is not necessary. 
We also provide the grids for the
counterpart best-fit NNLO PDFs for alternative $\alpha_s(M_Z)$ values
in the ranges 0.110-0.124 and 0.116-0.120, as well as the central PDFs
with up to 3, 4, 6 active quark flavors and for CT18 LO PDFs
\cite{Yan:2022pzl}.

The new grids contain 1.9 times more interpolation nodes than the 2019
CT18 ones. Table~\ref{table:grids} lists the node counts in the CT18
and CT18up NNLO grids in various of $x$ and $Q$ intervals.
Figure~\ref{fig:highdensity_x} illustrates improvements in the
accuracy of interpolation over $x$ on the example of the valence up
and gluon distributions. Here we plot the ratios of PDFs from the
LHAPDF grids and those obtained from the CTEQ fitting code directly at
500 specific values of $x$ using a sliding 4-point interpolation
routine distributed with .pds files at \cite{CTwebsite}. While the
2019 CT18 NNLO PDFs (in red) achieve interpolation accuracy below $5
\times 10^{-4}$ -- sufficient for the vast majority of applications --
the CT18up version (in black) pushes the deviations below $10^{-4}$
everywhere, and even lower in the $x$ intervals where the respective
PDFs are large. Similar improvements are seen for all PDF flavors.

Interpolation over $Q$ is also improved with a $30\%$ increase of the
number of $Q$ nodes -- see the right Table~\ref{table:grids} and
Fig.~\ref{fig:highdensity_Q}. While the CT18up PDFs \rev{continue to} use a single $Q$
grid over the whole interval $1.3 < Q < 100,000$ GeV, in CT18up the
interpolation is better across the whole $Q$ range due to the higher
density of nodes. A minor loss of accuracy due to switching from 4 to
5 active flavors is reduced to a smaller $Q$ interval of $\pm 1$ GeV
around the bottom-quark mass $m_b^{\rm pole}=4.75$ GeV. 
\rev{
The choice between using CT18 or CT18up PDFs should be made on the case-by-case basis,
with the CT18 grids providing acceptable predictions in most situations.
The CT18up
grids provide a better interpolation of the small PDF species, such as
the antiquark PDFs in the $x\to 1$ limit and the radiatively generated
small asymmetries between heavy-flavor quark and antiquark PDFs.
The denser CT18up grids, together with the original smaller CT18 grids,
allow the user to estimate the error due to PDF interpolation and to make interpolation more
uniform when using various interpolation algorithms available in the LHAPDF library \cite{LHAPDF6}
and other codes \cite{Nagar:2019gij,Diehl:2021gvs}.
}


\section{Inclusion of the new LHC data \label{sec:LHCdata}}
\begin{figure}[b]
    \centering
    \includegraphics[width=0.50\textwidth]{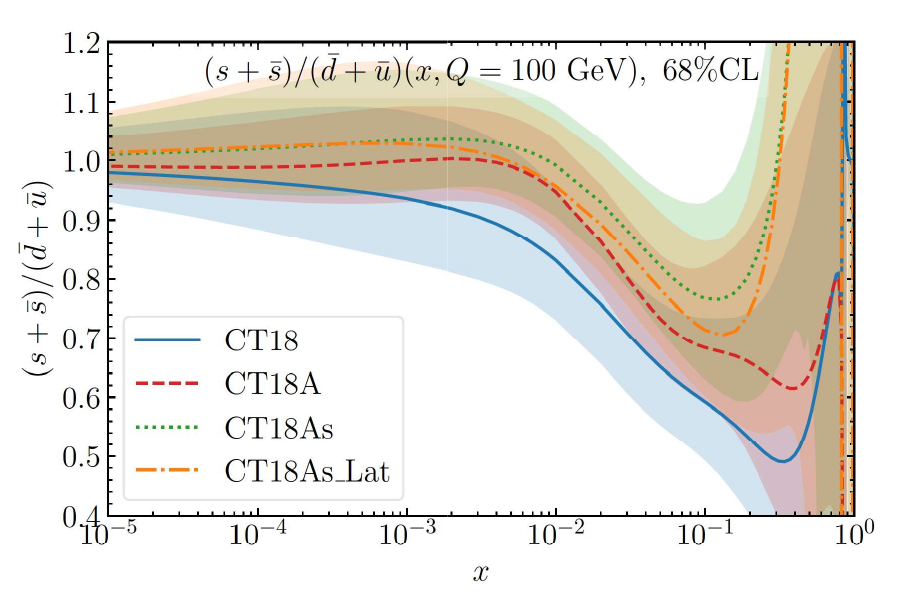}
    \includegraphics[width=0.48\textwidth]{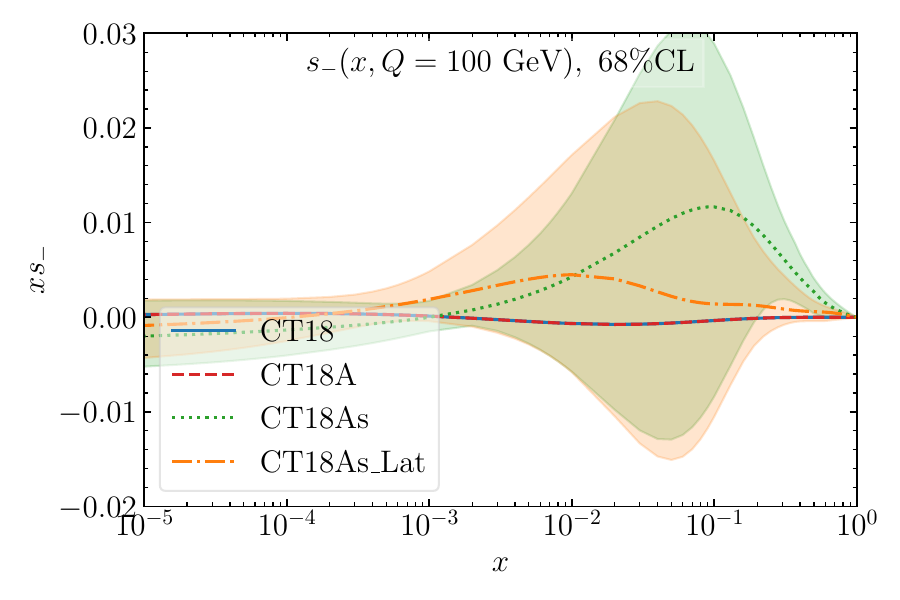}
    \caption{Comparison of the ratios
      $R_s=(s+\bar{s})/(\bar{d}+\bar{u})$ of strange and non-strange
      sea quark PDFs, and the strangeness asymmetry $s_-=s-\bar{s}$ at
      $Q=100~\GeV$ in the CT18~\cite{Hou:2019efy},
      CT18A~\cite{Hou:2019efy}, CT18As~\cite{Hou:2022onq} and
      CT18As\_Lat~\cite{Hou:2022onq} NNLO fits.}
    \label{fig:CT18AsLat}
\end{figure}

To reliably separate contributions of PDFs for distinct partonic
flavors $a$, the PDF fit must include experimental data sets of
diverse types and across a wide range of energies.  After the release
of CT18 PDFs~\cite{Hou:2019efy}, more and more precise data come out
from the measurements in the LHC Run II, including production of
vector bosons $W,Z$, top-quark pairs, inclusive jets and dijets, as
well as heavy flavors. As a part of preparation to the release of our
next PDFs, we invest into the selection of the most sensitive and
mutually consistent data from the recent measurements and into
exploration of their uncertainties.  In this section, we summarize the
key findings in our recent studies about the impact from post-CT18 LHC
precision data on production of lepton pairs~\cite{Sitiwaldi:2023jjp},
top-quark pairs\cite{Ablat:2023tiy}, and hadronic jets
\cite{Ablat:2024Jet}.


\subsection{The post-CT18 LHC Drell-Yan data and sea quark PDFs \label{sec:LHCDYdata}}

In the CT18 analysis~\cite{Hou:2019efy}, we found that the ATLAS 7 TeV
$W,Z$ precision data set (ATL7WZ)~\cite{ATLAS:2016nqi} has substantial
preference for a strangeness PDF $s(x,Q)$ of a larger magnitude at
$x\approx 0.02$, $Q\approx M_Z$, which is not fully consistent with
some other data sets, such as HERA I+II combined DIS cross
sections~\cite{H1:2015ubc} and NuTeV dimuon
measurement~\cite{Mason:2006qa,NuTeV:2007uwm}.  To quantify the effect
of this pull as well as of the alternative scale choices in
deep-inelastic scattering that are different from the ones in the CT18
analysis, we released a separate error PDF ensemble called CT18Z.  The
intermediate CT18A ensemble, also released within the same PDF family,
included the ATL7WZ data set while keeping the nominal DIS scale as in
CT18.

In those fits, the strangeness PDF was assumed to be equal to its
antiquark counterpart, $s(x,Q_0) = \bar{s}(x,Q_0)$ at the initial
scale of DGLAP evolution.  Since the experiments may impose different
pulls on $s$ and $\bar s$, in \cite{Hou:2022onq} we produced an
updated version of CT18A with $s(x,Q_0)\neq \bar s(x,Q_0)$ and more
flexible parametrization for the strangeness, as well as another
version in which the inputs from lattice QCD were included to
constrain the asymmetry $s_-\equiv s-\bar s$ at $Q_0$ and $0.3 \leq x
\leq 0.8$. These new fits were named CT18As and CT18As\_Lat,
respectively.  Figure~\ref{fig:CT18AsLat} illustrates the changes in
the ratio $R_s$ of strange and non-strange antiquark PDFs (left) and
strangeness asymmetry (right) observed in these fits. Inclusion of the
ATL7WZ data set in CT18A increases the $R_s$ ratio compared to CT18,
which does not include this data set.  The changes made in CT18As and
CT18As\_Lat lead to additional increases in $R_s$, while the large-$x$
lattice data disfavor a large $s_-$ value at $x>0.3$ that would
otherwise be preferred by the fitted combination of the experiments.
Moreover, the experiments are not fully consistent among themselves in
their preferences for the sea quark PDFs, as can be concluded based on
the methods discussed in Sec.~\ref{sec:L2}.

\begin{figure}[b]
    \centering
    \includegraphics[width=0.49\textwidth]{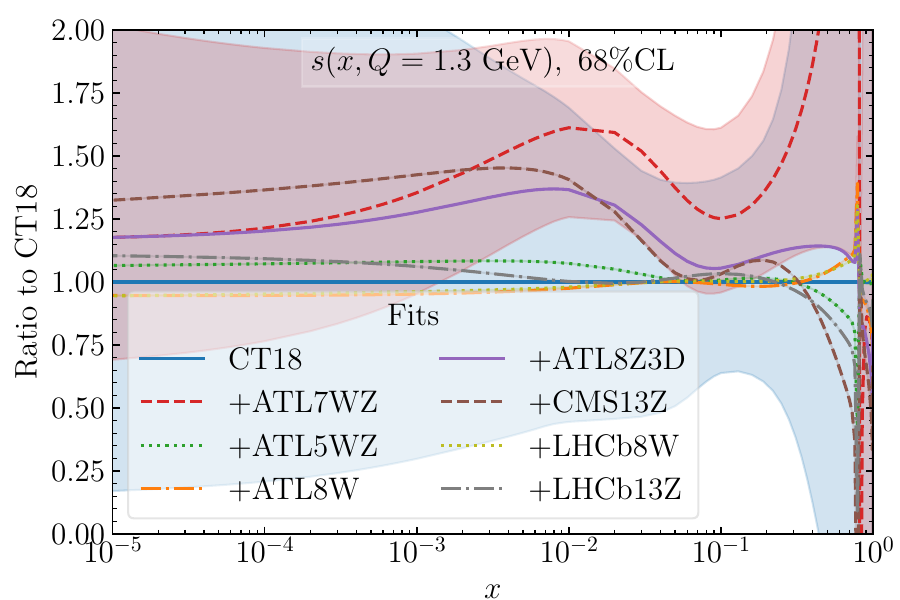}
    \includegraphics[width=0.49\textwidth]{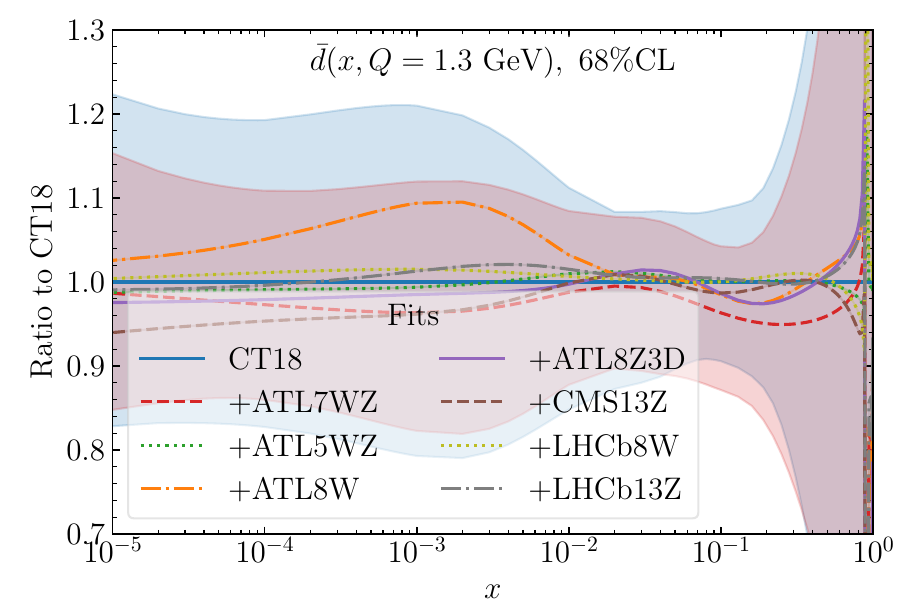}
    \caption{The impact of individual post-CT18 LHC Drell-Yan data
      sets on the CT18 $s$ and $\bar{d}$ PDFs at $Q=1.3~\GeV$. We add
      one data set at a time. The blue and red bands indicate the CT18
      and CT18+ATL7WZ=CT18A error bands at 68\% confidence level (CL),
      respectively. }
    \label{fig:CT18DYfit}
\end{figure}

\begin{figure}[tb]
    \centering
    \includegraphics[width=0.49\textwidth]{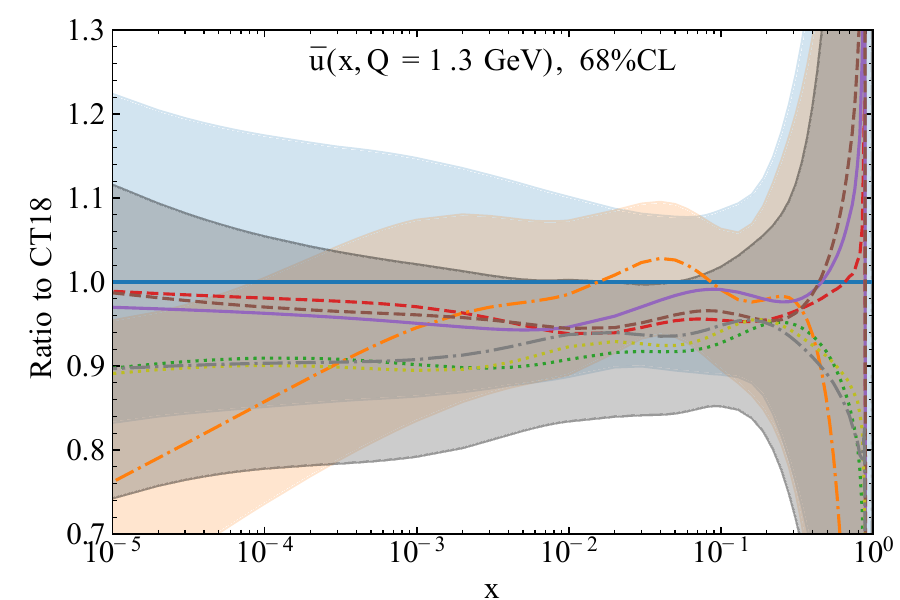}
    \includegraphics[width=0.49\textwidth]{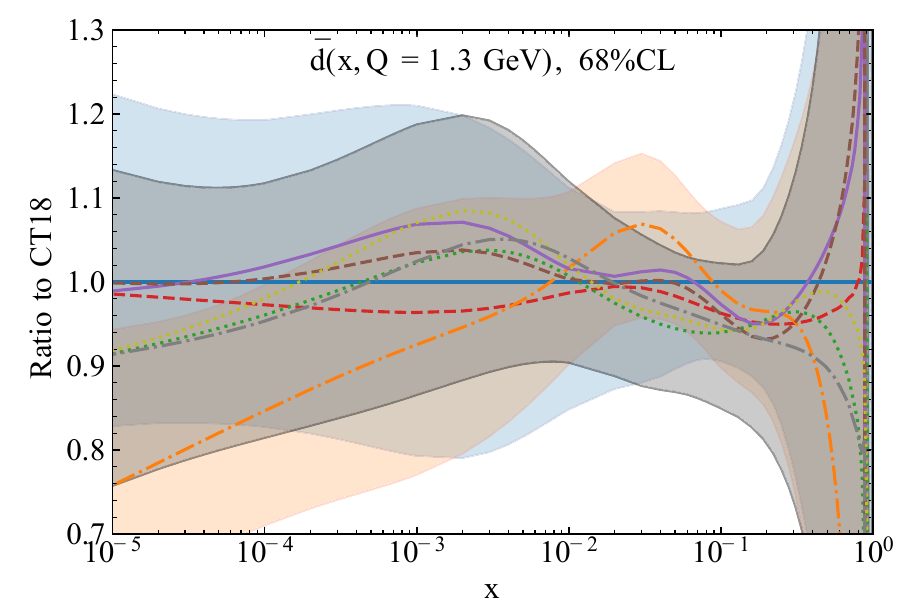}
    \\ \includegraphics[width=0.49\textwidth,
      valign=c]{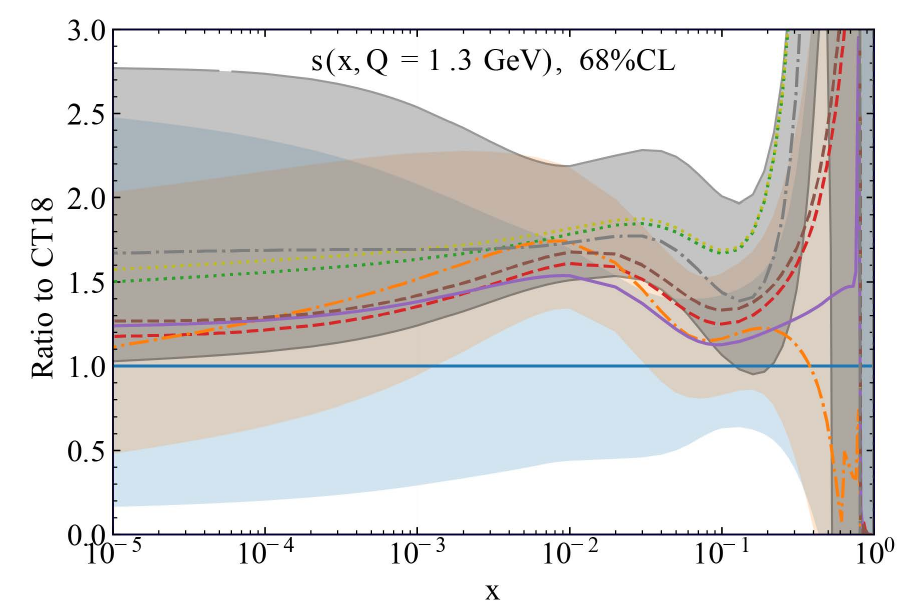}
    \includegraphics[width=0.49\textwidth,valign=c]{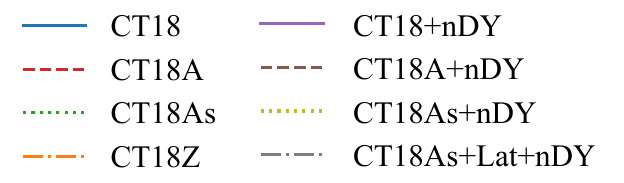}
    \caption{The impact of including post-CT18 Drell-Yan data sets on
      $\bar u$, $\bar d$, and $s$ PDFs at $Q=1.3$ GeV. We show the
      central PDFs without and with the new Drell-Yan (nDY) data sets
      in the frameworks of CT18~\cite{Hou:2019efy},
      CT18A~\cite{Hou:2019efy}, CT18As~\cite{Hou:2022onq}, and
      CT18As+Lat~\cite{Hou:2022onq} analyses, plotted as the ratios to
      the respective CT18 PDFs. We also show the 68\% CL PDF error
      bands for the CT18 and CT18As+nDY+Lat PDF ensembles, as well as
      for the CT18Z PDFs, which includes the $x$-dependent saturation
      scale as well as the ATLAS 7 TeV $W,Z$ data.}
    \label{fig:CT18DYfitAll}
\end{figure}

In Ref.~\cite{Sitiwaldi:2023jjp}, we addressed the following question:
do the even newer LHC measurements of vector boson production support
such behavior?  We examined the impact of several post-CT18 LHC
Drell-Yan data sets, including the ATLAS 5.02 TeV $W,Z$ boson
production (ATL5WZ)~\cite{ATLAS:2018pyl}, ATLAS 8 TeV $W$ boson
production (ATL8W)~\cite{ATLAS:2019fgb}, ATLAS 8 TeV triple
differential $Z$ production (ATL8Z3D)~\cite{ATLAS:2017rue}, CMS 13 TeV
$Z$ production (CMS13Z)~\cite{CMS:2019raw}, LHCb 8 TeV $W$ production
(LHCb8W)~\cite{LHCb:2016zpq}, and LHCb 13 TeV $Z$ production
(LHCb13Z)~\cite{LHCb:2021huf}.
\rev{
The theoretical predictions are obtained with the APPLgrid~\cite{Carli:2010rw} 
at NLO together with the NNLO/NLO $K$-factors calculated with \texttt{MCFM}~\cite{Campbell:2019dru} and \texttt{ResBos2}~\cite{Isaacson:2022rts}.
}
We concluded that these data sets
considerably affect the less constrained antiquark PDFs, in particular
those for the strange and down-type antiquarks plotted in
Fig.~\ref{fig:CT18DYfit} as functions of $x$ at the initial scale
$Q_0=1.3~\GeV$.  These PDFs and their uncertainties are depicted in
the form of ratios to the respective central PDFs from the CT18 NNLO
fit to examine how the PDFs change as we add the indicated data sets,
one at a time, to the CT18 NNLO global analysis.  We were most
interested if the new data sets provide a consistent pattern of pulls
on the strangeness PDF. In this exercise, we neglected the strangeness
asymmetry at the initial low scale, i.e., we assumed $s(x,Q_0)=\bar
s(x,Q_0)$ .

As can be seen from the left subfigure of Fig.~\ref{fig:CT18DYfit},
the pattern of preferences of the individual DY data sets remains
quite complex, with some new data sets preferring an enhanced
strangeness, and others opposing it. For instance, including either
the $Z$-boson triple differential distribution at the ATLAS 8 TeV
(ATL8Z3D) or $Z$ boson production at the CMS 13 TeV (CMS13Z) enhances
the strangeness in the direction consistent with the pull of
ATL7WZ. On the contrary, data sets on ATLAS 8 TeV $W$-boson production
(ATL8W) and LHCb $W$ production at 8 TeV and $Z$ production at 13 TeV
(LHCb8W and LHCb13Z) pull $s(x,Q)$ in the opposite direction, implying
the internal tension between the $W$ and $Z$ data sets. Analogous
opposing trends can be also seen in the pulls on $\bar d(x,Q)$ in the
right subfigure of Fig.~\ref{fig:CT18DYfit}, with the upward pull of
ATL8W going against the other experiments.

To estimate the impact on the flavor composition in the sea quark
sector, Fig.~\ref{fig:CT18DYfitAll} shows the candidate $\bar u$,
$\bar d$, and $s$ PDFs at $Q=1.3~\GeV$ before and after augmenting the
CT18 and CT18A~\cite{Hou:2019efy}, as well as CT18As and CT18As\_Lat
data sets \cite{Hou:2022onq} with {\it all} new Drell-Yan (``nDY'')
data sets referenced in Fig.~\ref{fig:CT18DYfit}. Recall that the
CT18As analysis descends from the CT18A one by allowing $s$ and $\bar
s$ PDFs to be non-equal, while CT18As\_Lat adds to CT18As the lattice
data to constrain the $s-\bar s$ difference at the initial scale $Q_0$
and $0.3 < x < 0.8$.  In addition to the central PDFs from 8 various
fits, Fig.~\ref{fig:CT18DYfitAll} depicts the error bands for CT18,
CT18Z, and CT18As+Lat+nDY fits, with the latter implementing all new
data sets and lattice data at once. An analogous comparison for PDFs
at $Q=100~\GeV$ can be found in Ref.~\cite{Sitiwaldi:2023jjp}.

The general trend for the final CT18As+Lat+nDY PDF set (gray
dot-dashed line) is that the $\bar u$ is suppressed across the whole
$x$ range, becoming largely consistent with CT18Z at small and large
$x$.  Some suppression also happens in the $u$ PDF at $x < 0.05$ (not
shown here).  The $\bar{d}$ PDF, as well as the $d$ one, get
moderately larger at $0.005 < x < 0.01$ and smaller elsewhere.  This
trend reflects the competition between the opposing pulls dominated by
the ATL7WZ and ATL8W data sets, already observed in the CT18A+nDY
central set (brown dashes) that lies between the CT18A (red dashes)
and CT18+nDY (violet solid) ones.

In contrast, the strangeness PDF accumulates the upward pulls from
ATL7WZ and other post-CT18 Drell-Yan data sets, especially ATL8Z3D and
CMS13Z. Consequently, it is larger in CT18As+Lat+nDY than {\it both}
CT18 and CT18Z strangeness PDFs.  In the CT18As fit with its more
flexible (anti)strangeness parametrizations, the strangeness can be
pulled even further in the positive direction. Meanwhile, the PDF
error bands shrink upon the inclusion of the post-CT18 precision
Drell-Yan data thanks to its significant constraining power.

\begin{figure}[b]
    \centering
    \includegraphics[width=0.49\textwidth]{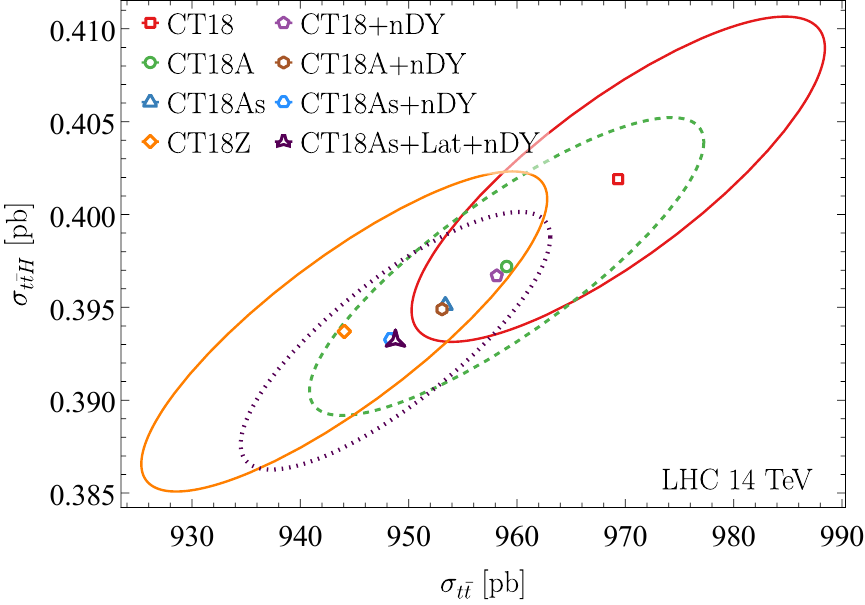}
    \includegraphics[width=0.49\textwidth]{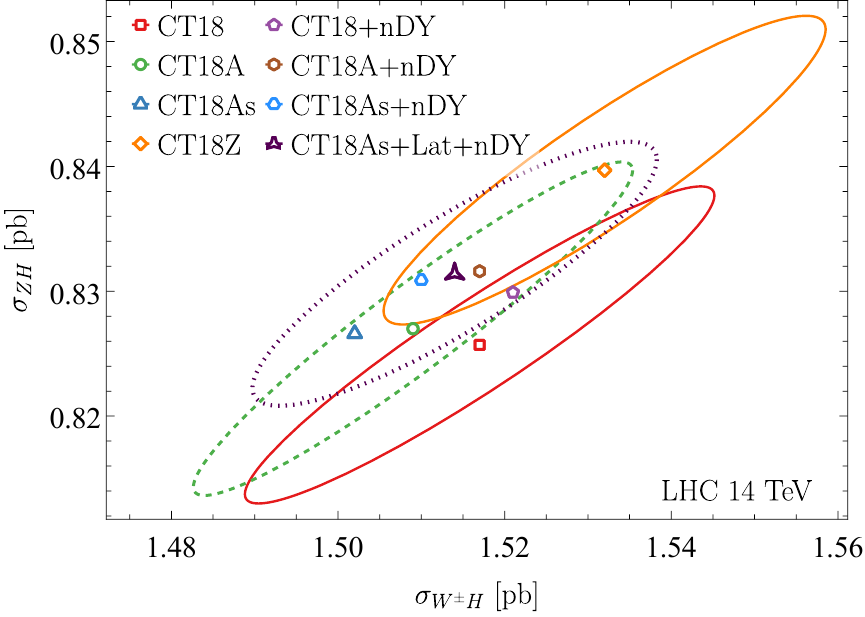}
    \caption{The correlation ellipses in the 68\% CL between the
      inclusive $t\bar{t}$ and $t\bar{t}H$, $W^\pm H$ and $ZH$
      production cross sections at the 14 TeV LHC for  $m_t^{\rm pole}=173.3$ GeV.}
    \label{fig:Corr}
\end{figure}

To illustrate some phenomenological implications, Fig.~\ref{fig:Corr}
presents the correlation ellipses for the inclusive $t\bar{t}$ and
$t\bar{t}H$, $W^\pm H$ and $ZH$ production cross sections at the LHC
14 TeV. Here the $t\bar{t}$ cross section is calculated with
\texttt{Top++}~\cite{Czakon:2011xx} at NNLO and with soft gluons
resummed up to the NNLL level, by setting the renormalization and
factorization scales equal to \rev{the top-quark pole mass $m_t=173.3$ GeV}. The
$t\bar{t}H,W^\pm H,ZH$ cross sections are calculated at NLO with
\texttt{MadGraph\_aMC@NLO}~\cite{Alwall:2014hca,Frederix:2018nkq}
interfaced with the \texttt{PineAPPL} library~\cite{Carrazza:2020gss},
with the scale set to the partonic collision energy
$\sqrt{\hat{s}}$. We show the 68\% CL ellipses for CT18, CT18Z, CT18A,
and CT18As+Lat+DY ensembles, as well as the central predictions for
these and four other ensembles.  The positive correlation between the
$t\bar{t}$ and $t\bar{t}H$ ($W^\pm H$ and $ZH$) cross sections mainly
reflects their shared $gg$ ($q\bar{q}$) parton luminosity in the
dominant production channel. For CT18As+Lat+DY, the ellipses in both
subfigures are reduced in size thanks to including the post-CT18
Drell-Yan data, reflecting the shrinkage of the underlying PDF error
bands. The respective central $t\bar{t}$ and $t\bar{t}H$ cross
sections are pulled toward CT18Z ones mainly due to the reduction of
gluon PDFs, as shown in Fig.~\ref{fig:CT18DYfitAll}. The
$(W^{\pm}H,ZH)$ correlation ellipse shares a similarity with the
$(W^\pm,Z)$ one presented in Ref.~\cite{Sitiwaldi:2023jjp}. For these
cross sections, the post-CT18 Drell-Yan data sets displace the
CT18As+Lat+DY correlation ellipse along the minor axis due to the
enhancement of the strangeness PDF. However, the displacement along
the major axis is milder for $\sigma_{W^\pm H}+\sigma_{ZH}$ than for
$\sigma_{W^\pm}+\sigma_{Z}$ shown in Ref.~\cite{Sitiwaldi:2023jjp},
mainly because $\sigma_{W^\pm H}+\sigma_{ZH}$ production is correlated
with the gluon PDF at larger typical $x$ values than in $W^\pm,Z$
production, where the gluon PDF is less influenced by the inclusion of
the new DY data.

\subsection{Top-quark pair production at the LHC 13 TeV \label{sec:LHC13ttbar}}
Another recent study \cite{Ablat:2023tiy} focused on eligible $t\bar
t$ production measurements at the LHC at 13 TeV, with the primary goal
to determine an optimal baseline selection of absolute $t\bar t$
differential cross sections that would contribute to the determination
of gluon PDF at large $x$, where it remains relatively weakly
constrained by the already included experiments.  In particular,
top-quark pair production at the LHC can probe the gluon PDF already
at $x\gtrsim 0.01$, but most helpfully at $x>0.1$.  Measurements of
$t\bar t$ differential cross sections at the LHC thus complement
production of hadronic jets in that they are also sensitive to the
gluon PDF in the intermediate to large-$x$ region, yet assess it using
a variety of separate experimental techniques \cite{Nadolsky:2008zw}.
Though the $t\bar t$ and inclusive high-$p_T$ jet production
measurements largely overlap in the $x-Q$ plane, their matrix elements
and phase-space suppression are different, and hence their dominant
constraints on the gluon arise at different values of $x$.  However,
PDF extraction may be challenged by the presence of tensions between
experiments (which pull the gluon in inconsistent directions) and by
the presence of strong correlations between the top-quark mass $m_t$,
the strong coupling $\alpha_s$, and the gluon PDF itself. The relevant
statistical and systematical uncertainties published by the
experimental collaborations can be expressed in terms of either the
covariance matrix or nuisance parameter representation. In general,
conversion from the covariance matrix to nuisance parameters is not
unique: complete information on the statistical, uncorrelated, and
correlated systematic uncertainties is critical for maximizing
constraints on the PDFs.  These features and other aspects of the new
$t\bar t $ measurements at 13 TeV have been examined in detail
in~\cite{Ablat:2023tiy}.

{\bf Eligible $t\bar t $ measurements at 13 TeV} include
high-precision absolute differential cross sections from the ATLAS and
CMS collaborations. In particular, we studied differential cross
section measurements in the dilepton~\cite{CMS:2018adi} and
lepton+jets~\cite{CMS:2021vhb} channels at CMS, as well as ATLAS 13
TeV measurements in the lepton+jets~\cite{ATLAS:2019hxz} and
all-hadronic~\cite{ATLAS:2020ccu} channels.  Among these data sets,
the CMS lepton+jets measurements~\cite{CMS:2021vhb} have a higher
integrated luminosity of 137 fb$^{-1}$~\cite{CMS:2021vhb} and better
control on the experimental uncertainties.

\rev{
Our theory predictions at NNLO in QCD are obtained with two methods. For the CMS 13 TeV dilepton channel and the ATLAS 13 TeV lepton+jets channel with the same histogram bins, NNLO versions of \texttt{fastNLO}
tables~\cite{Czakon:2017dip} based on the
\texttt{STRIPPER} subtraction~\cite{Czakon:2015owf,Czakon:2014oma} are available. 
For the other data sets, we generated the NLO \texttt{APPLgrid} tables~\cite{Carli:2010rw} with \texttt{MCFM}~\cite{Campbell:2015qma,Campbell:2012uf}, and the NNLO/NLO $K$-factors with \texttt{MATRIX}~\cite{Grazzini:2017mhc} using the $q_T$ subtraction approach~\cite{Catani:2019hip,Catani:2007vq}. We have examined the sensitivity to the top-quark mass by comparing the fits with $m_t^{\rm pole}=172.5$ GeV and 173.3 GeV, and found the PDF impact is minimal.   
Electroweak corrections at NLO, either taken from Ref.~\cite{Czakon:2017dip} or calculated with \texttt{MadGraph\_aMC@NLO}~\cite{Pagani:2016caq}/\texttt{MCFM}~\cite{Campbell:2016dks}, were also considered when available.
Their overall impact on PDF determination was negligible, given the
current size of experimental errors.  In addition, the dependence
on the QCD scale, quantifying a part of the uncertainty due to missing
higher orders, was explored by performing independent fits with
central-scale choices for the relevant top-quark predictions 
set to $H_T/4$, $H_T/2$, and $H_T$, respectively. As done in
the previous CT18 PDF analysis, the central set of the final post-CT18
PDFs adopted the QCD scale that leads to the best quality of the fit.
}

We started by adding the new LHC 13 TeV $t\bar t$ data set, one at a
time, to the CT18~\cite{Hou:2019efy} baseline data set and gauging
their impact first with the \texttt{ePump}
code~\cite{Schmidt:2018hvu,Hou:2019gfw} to preselect the most
promising data sets and then by performing full global fits with the
selected data sets and accounting for bin-by-bin statistical
correlations among the distributions (when those were available) as
well as by varying the central QCD scales.  We collected the most
sensitive distributions into two optimal combinations of measurements,
selected so as to maximize the extracted information about the
gluon PDF and minimize the tensions among the data sets in the extended baseline. The two optimal combinations were labeled ``CT18+nTT1'' and ``CT18+nTT2''\footnote{The new data sets in CT18+nTT1 and CT18+nTT2 are selected from the ATLAS and CMS measurements after a combined study of impact of each single kinematic distribution with the \texttt{ePump} code and through a individual global fits to identify distributions with maximal sensitivity and optimal $\chi^2/N_{pt}$.}.     
\rev{In each combination, the newly added differential distributions are selected to be mutually consistent, statistically independent, and agree with the NNLO theory, as explained in Ref.~\cite{Ablat:2023tiy}.}
The CT18+nTT1 combination includes
the ATL13had $y_{t\bar t}$, CMS13ll $y_{t\bar t}$, CMS13lj $m_{t\bar
  t}$, and ATL13lj $y_{t\bar t}$ distributions (resolved in terms of
the CMS bins), while the CT18+nTT2 includes the same distributions
from the ATL13had, CMS13ll, CMS13lj measurements, and the $y_{t\bar
  t}$ + $y^B_{t\bar t}$ + $m_{t\bar t}$ + $H^{t\bar t}_T$ combination
without statistical correlations from ATL13lj where $H_T^{t\bar{t}}$ is
the scalar sum of the transverse momenta of the hadronic and leptonic top quarks, and $y^B_{t\bar t}$ is the rapidity distribution for the boosted topology 
(see Ref.~\cite{Ablat:2023tiy} for more details). 

\begin{figure}[t]
\centering
\includegraphics[width=0.49\textwidth]{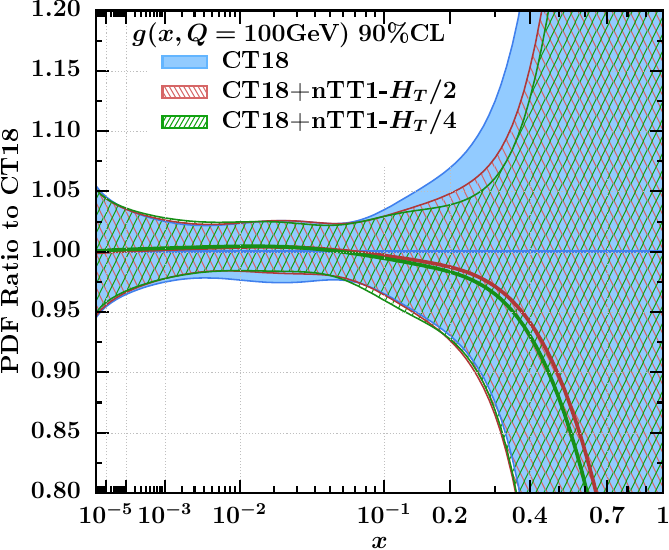}
\includegraphics[width=0.49\textwidth]{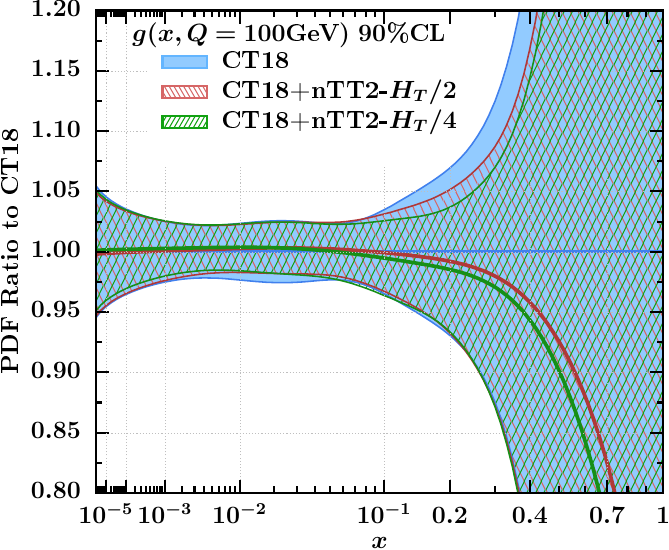}\\ (a) \hspace{3in}
(b)
\caption{Gluon PDFs and uncertainties in (a) the CT18+nTT1 and (b)
  CT18+nTT2 global QCD analyses at NNLO, plotted as ratios to CT18
  NNLO.  Hatched error bands represent the $H_T/2$ (red) and $H_T/4$
  (green) choices for the central scale in the 13 TeV $t \bar t$
  theory predictions. PDF uncertainties are evaluated at the $90\%$
  CL.}
\label{fig:nTT12}
\end{figure}

Figure~\ref{fig:nTT12}
shows the impact of these new $t\bar t$ data combinations on the
post-CT18 gluon central value and its uncertainty. We also illustrate
the changes in the gluon central value due to the different scale
choices in the 13 TeV $t\bar t$ theory predictions. The new data
prefer a softer gluon in the $x\geq 0.1$ region as compared to CT18,
driven mainly by the CMS13lj-$m_{t\bar t}$ data and achieving the
lowest $\chi^2$ when the central scale is set to $H_T/2$. The overall
quality-of-fit of the CT18+nTT1 and CT18+nTT2 fits is essentially the
same as that of CT18, with $\chi^2/N_{\rm pt} \approx 1.16$.
 
\subsection{Including inclusive jets and dijets\label{sec:LHCJets}}
Post-CT18 high-statistics measurements of hadronic jet observables at
the LHC \cite{ ATLAS:2013pbc,CMS:2015jdl, 
  ATLAS:2017ble,CMS:2016jip,CMS:2021yzl,ATLAS:2013jmu,CMS:2012ftr,CMS:2017jfq}
offer multiple avenues for constraining the proton structure.  The
global PDF analyses commonly fit inclusive differential cross sections
of single jets or jet pairs, which are well-understood theoretically
and notably are sensitive to the gluon distribution over a wide range
of $x$. The jet differential cross sections have been published as
functions of diverse kinematic variables. For single-inclusive jets,
these are the jet's transverse momentum $(p_T^j)$ and absolute
rapidity $(|y^j|)$. For a jet pair, these are commonly the dijet
invariant mass ($m_{12}$) together with the half-rapidity separation
$y^*\equiv |y_1 - y_2|/2$ \cite{ATLAS:2013jmu,ATLAS:2017ble} or the
largest absolute rapidity $y_{\max}={\rm sign}(| \max(y_1, y_2) | - |
\min (y_1, y_2|) \cdot \max (|y_1|, |y_2|) $ \cite{CMS:2012ftr} of the
harder and softer jets; or possibly the average jet's transverse
momentum $p_{T,{\rm avg}} \equiv ( p_{T,1} + p_{T,2} )/2$ together
with the half-rapidity separation $y^*$ or the dijet boost $y^b \equiv
|y_1 + y_2 |/2$) \cite{CMS:2017jfq}. Choosing between the specific
representations of jet data helps the fitter to optimize the
sensitivity to specific PDF combinations and to control the systematic
factors present both in the experimental and theoretical parts of the
analysis. Here we summarize a recent study \cite{Ablat:2024Jet} by the
CTEQ-TEA group aimed at selecting an optimal combination of (di)jet
production measurement and analysis settings for the next generation
of PDFs.

As in the $t\bar t$ production case, we first investigated the
approximate impact of individual (di)jet data sets by including them one at a time. We
excluded the CMS 7 TeV~\cite{CMS:2014nvq}, ATLAS 7
TeV~\cite{ATLAS:2014riz} and CMS 8 TeV~\cite{CMS:2016lna} inclusive
jet data from the CT18 baseline fit \cite{Hou:2019efy} to avoid
double-counting, thus obtaining a ``CT18mLHCJet'' data set. We then
refitted the CT18mLHCJet data together with one of the new single-jet
or dijet sets~\cite{ ATLAS:2013pbc,CMS:2015jdl, 
  ATLAS:2017ble,CMS:2016jip,CMS:2021yzl,ATLAS:2013jmu,CMS:2012ftr,CMS:2017jfq}
and examined the resulting changes in the PDFs. Exploring these new
data within the global fit requires fast theoretical computations at
NNLO QCD accuracy during the $\chi^2$ minimization procedure. 
\rev{
The theoretical predictions are obtained with the \texttt{APPLfast} \cite{Britzger:2022lbf,Currie:2016bfm} grids\footnote{\texttt{APPLfast} provides fast interpolation grids at NNLO QCD based on the \texttt{APPLgrid} \cite{Carli:2010rw} and \texttt{fastNLO} \cite{Kluge:2006xs,Wobisch:2011ij,Britzger:2012bs} methods.} from the \texttt{Ploughshare} repository~\cite{PloughshareProject},
which are calculated with the leading colour and leading $N_f$ approximation~\cite{Britzger:2022lbf}. The subleading colour contribution has been calculated recently~\cite{Czakon:2019tmo,Chen:2022tpk} but not in the form of fast interpolation grids needed for the global fit. The subleading colour contributions to single-inclusive jet and dijet double differential distributions were found to be small compared to the current experimental uncertainties~\cite{Chen:2022tpk}. Their impact was mild~\cite{Cridge:2023ozx} for the dijet triple differential distributions measured by CMS 8 TeV~\cite{CMS:2017jfq} but this data set was not selected for our final combined fit ``CT18+IncJet'' described below. 
}

\begin{figure}[t]
 \centering
 \includegraphics[width=0.49\textwidth]{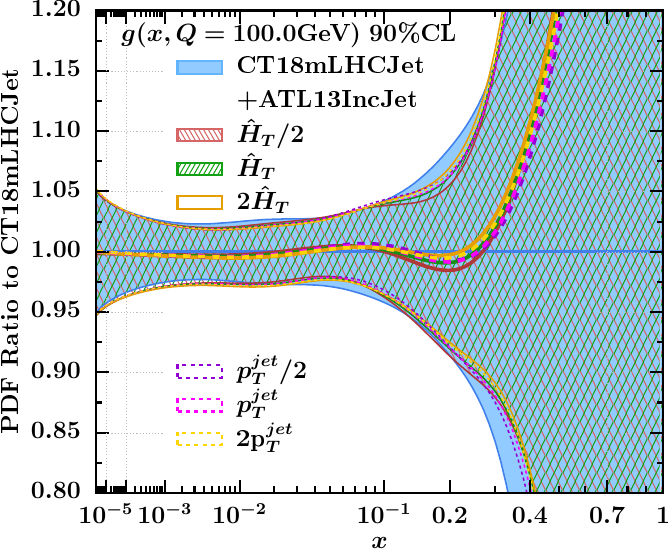}
 \includegraphics[width=0.49\textwidth]{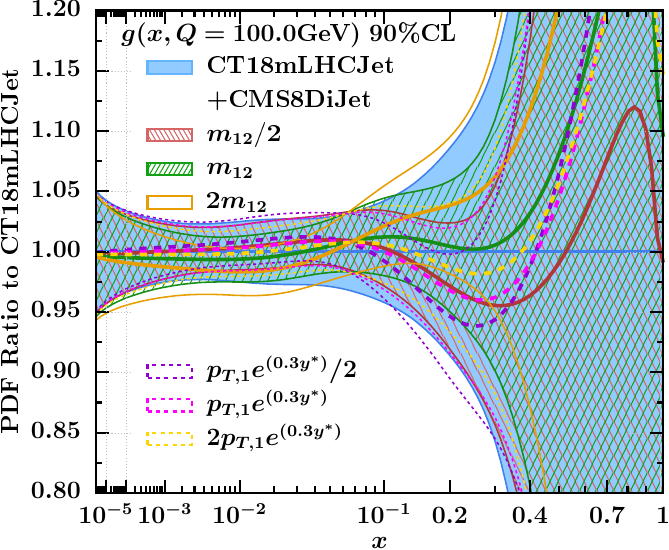}\\ (a) \hspace{3in}
 (b)
 	\caption{Gluon PDF ratios for CT18mLHCJet+ATL13IncJet and
          CT18mLHCJet+CMS8DiJet gluon PDFs over the best-fit of the
          CT18mLHCJet baseline PDFs for different scale choices.}
 	\label{Fig:ScaleVariation}
 \end{figure}

\rev{
For the single inclusive jet data, \texttt{APPLfast} grids with central scale choices $\mu_{F,R}^0 = p_T^j$ and 
 $H_T$ (scalar sum of the transverse
momenta of all jets) are available. In comparison, the scales $\mu_{F, R}^0=m_{12}$ are available for the dijet data, while another choice $\mu_{F, R}^0 =p_{T,1} e^{0.3 y^*}$ also exists for CMS 8 TeV dijet production~\cite{CMS:2017jfq}.
} 
For each choice of the central scale, we also varied it by a factor
$1/2$ or $2$, with the resulting changes in the PDFs illustrated in
Fig.~\ref{Fig:ScaleVariation}. The totality of variations of the QCD
scales changes the reduced $\chi^2/N_{\rm pt}$ by several units for
both inclusive jets and dijets~\cite{Ablat:2024Jet}, suggesting that
the control of missing higher-order corrections is still limited even
at NNLO. Furthermore, $\chi^2/N_{\rm pt}$ is affected comparably by
uncertainties due to the Monte Carlo integration in the NNLO
calculation and normalizations of systematic factors in the
experimental correlation matrices. Nevertheless, good fits can be
obtained with some scale choices, especially for dijet cross sections
with their more pronounced scale dependence.  While the impact of the
inclusive jet data on $g(x,Q)$ is relatively independent of the scale
choice used, the analogous pulls of the dijets are strongly correlated
with the scale choice, as shown in Fig.~\ref{Fig:ScaleVariation}.  For
this reason, as the final product, the study produced a combined fit,
titled ``CT18+IncJet''~\cite{Ablat:2024Jet}, with the inclusive jet
data sets only. Namely, the new ATLAS and CMS inclusive jet data sets
at 8 and 13 TeV were added to similar data sets at 7 and 8 TeV that
were already included in the CT18 baseline analysis.
\rev{The CT18+IncJet fit did not include the dijet data sets 
due to their larger scale uncertainty as well as tensions with some 
existing data sets~\cite{Ablat:2024Jet}.
}

In more detail, when fitting individual single-inclusive jet data
sets, the $p_T^j$ and $\hat{H}_T$ scale choices produce comparable
$\chi^2/N_{\rm pt}\approx 1-2$~\cite{Ablat:2024Jet}, with the exact
value depending on the various factors discussed already in the
context of CT18 fits to the LHC Run-1 jet data sets
\cite{Hou:2019efy}. On the other hand, the resulting gluon PDF changes
little among the various examined choices:
Fig.~\ref{Fig:ScaleVariation}(a) illustrates these changes based on
the fit to the ATLAS 13 TeV inclusive jet data
set~\cite{ATLAS:2017ble}. The final CT18+IncJet combination adopted
the scale $p_T^j$.  Figure~\ref{Fig:ScaleVariation}(b) illustrates
analogous changes in the gluon upon adding the CMS 8 TeV dijet
\cite{CMS:2017jfq} data set using several scale choices indicated in
the figure. In contrast to the inclusive jets, the impact of this new
dijet data on $g(x,Q)$ depends substantially on the choice of
scale. Other dijet sets also have pronounced scale dependence
\cite{Ablat:2024Jet}. Concerning the quality of fits, the upper
Table~\ref{Tab:newIncJetData} summarizes $\chi^2/N_{\rm pt}$ values
for the new inclusive jet data sets for the PDFs from the combined
CT18+IncJet and CT18 baseline fits.  The lower table lists the
$\chi^2/N_{\rm pt}$ values after adding each dijet data set one by one
to the CT18mLHCJet fit.  As we see, for the CMS 8 TeV dijet data set,
$\mu_{F,R} =p_{T,1} e^{0.3 y^*}$ gives a smaller $\chi^2/N_{\rm pt}$.

\begin{table}[t]
	\centering
	\begin{tabular}{ccc|cc}
		\hline Data set & Ref. & $N_{\rm pt}$ & CT18+IncJet &
                CT18 \\ \hline ATLAS 8 TeV IncJet &
                \cite{ATLAS:2017kux}& 171 & $1.76^{+0.20}_{-0.12}$ &
                $1.80^{+0.33}_{-0.16}$ \\ ATLAS 13 TeV IncJet &
                \cite{ATLAS:2017ble} &177 & $1.38^{+0.13}_{-0.10}$ &
                $1.39^{+0.20}_{-0.11}$ \\ CMS 13 TeV IncJet &
                \cite{CMS:2021yzl} &78 & $1.10^{+0.24}_{-0.17}$ &
                $1.11^{+0.30}_{-0.16}$ \\ \hline
	\end{tabular}
	\begin{tabular}{ccc|cc}
        \hline Data set & Ref. & $N_{\rm pt}$ &
        \multicolumn{2}{c}{CT18mLHCJet+ DiJet} \\ & & & $\mu_{F,R}^0
        =m_{12}$ & $\mu_{F,R}^0 =p_{T,1} e^{0.3 y^*}$ \\ \hline ATLAS
        7 TeV DiJet & \cite{ATLAS:2013jmu} &90 & $1.46$ & -- \\ CMS 7
        TeV DiJet & \cite{CMS:2012ftr} &54 & $1.78$ & --\\ CMS 8 TeV
        DiJet &\cite{CMS:2017jfq} &122 & $1.55$ & $1.12$ \\ ATLAS 13
        TeV DiJet & \cite{ATLAS:2017ble} &136 & $1.29$ & -- \\ \hline
	\end{tabular}

 \caption{Upper: The $\chi^2/N_{\rm pt}$ values for the indicated
   inclusive jet datasets included simultaneously in the CT18+IncJet
   fit, as well as the corresponding $\chi^2/N_{\rm pt} $ values for
   the CT18 NNLO PDFs. Lower: same, for the indicated dijet data sets
   upon adding them one-by-one to the CT18mLHCJet baseline
   fit.}\label{Tab:newIncJetData}
\end{table}

\subsection{Toward the global fit of the combined LHC Run-2 data sets \label{sec:nDYnTTnJet}}
The previous subsections summarized a series of preliminary NNLO fits
that explored the impact of the new LHC measurements in production of
lepton pairs (Sec.~\ref{sec:LHCDYdata}), top-quark pairs
(Sec.~\ref{sec:LHC13ttbar}), and hadronic jets
(Sec.~\ref{sec:LHCJets}). These explorations identified an optimal
selection of 13 data sets in the three categories of processes, with a
total of 776 data points, that form the extension of the CT18 global
data named ``CT18+nDYTTIncJet''. The ongoing work on this extension is
reported elsewhere \cite{Ablat:2024nhy}.

\begin{figure}[p]
\centering
\includegraphics[width=0.49\textwidth]{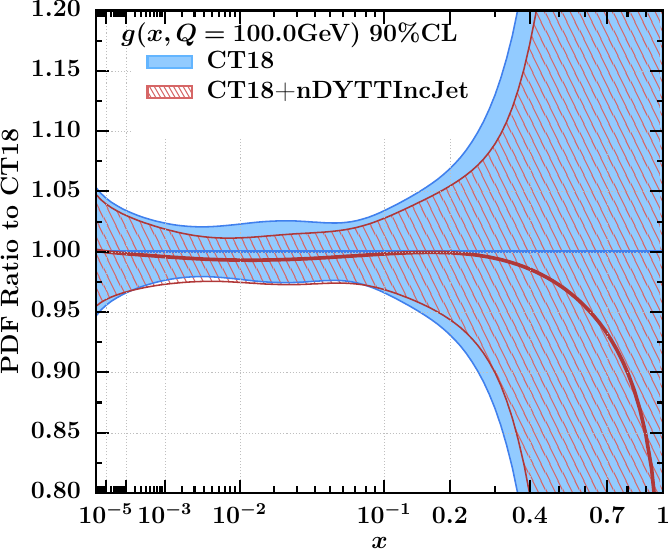}
\includegraphics[width=0.49\textwidth]{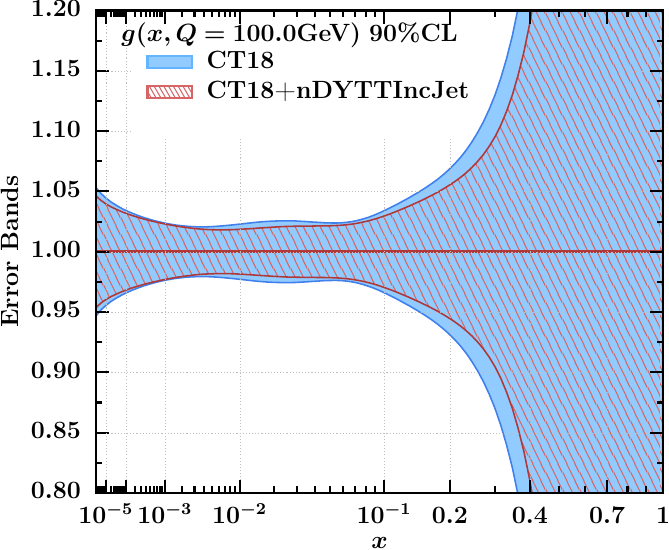}\\ (a) \hspace{3.5in}
(b)
\caption{(a) The ratio of the gluon PDF in the combined fit
  CT18+nDYTTIncJet NNLO to the respective CT18 PDF. (b) Comparison of
  the 90\% CL uncertainties on the gluon PDF in the CT18 and
  CT18+nDYTTIncJet NNLO fits.}\label{Fig:CT18+nDYTTIncJet}
\end{figure}

\begin{figure}[p]
\centering
\includegraphics[width=0.49\textwidth]{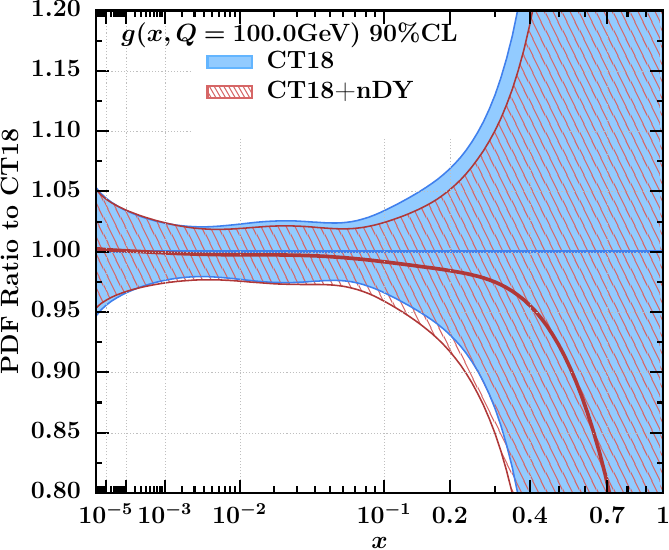}
\includegraphics[width=0.49\textwidth]{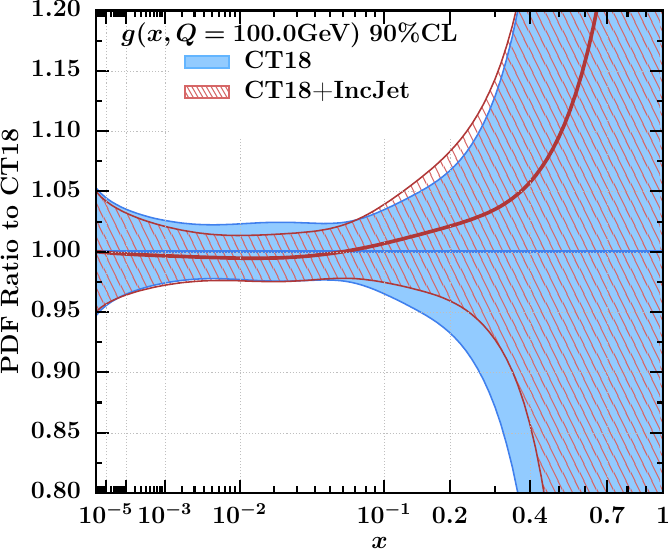}
\\ (a) \hspace{3.5in} (b)
	\caption{Ratios of the gluon PDFs in (a) CT18+nDY and (b)
          CT18+IncJet NNLO fits to the respective CT18
          PDF.}\label{fig:CT18+DY-CT18+incJet}
\end{figure}

Here we observe that CT18+nDYTTIncJet achieves an overall agreement
with all data sets with about the same total $\chi^2/N_{\rm pt}$ as in
the CT18 baseline fit. Also as in CT18, some residual tensions remain
among the data sets, reducing their net constraining power. While we
save the comparisons between the CT18 and CT18+nDYTTIncJet PDFs for
the full report, Fig.~\ref{Fig:CT18+nDYTTIncJet}(a) illustrates the
combined impact of the new data sets on the gluon PDF at $Q=100$~GeV,
plotted as the ratio to the CT18 NNLO gluon PDF.  The totality of the
LHC new data sets prefer a softer gluon PDF at $x > 0.3$. In
CT18+nDYTTIncJet, the LHC Run-2 $t\bar t$ cross sections are included
in their nTT2 combination, cf. Sec.~\ref{sec:LHC13ttbar}. We already
observed in Fig.~\ref{fig:nTT12}(b) that the nTT2 combination prefers
a softer gluon PDF at $x>0.1$. On the other hand, the separate fits
augmented by the new Drell-Yan data (CT18+nDY) and inclusive jet
(CT18+IncJet) data sets prefer a softer and harder gluon PDF at such
$x$, respectively, as illustrated in
Fig.~\ref{fig:CT18+DY-CT18+incJet}. In reflection of these pulls from
the new data sets, the central gluon PDF is reduced by about 1\% at
$x\approx 0.01$ and $Q\approx 100$ GeV, see
Fig.~\ref{Fig:CT18+nDYTTIncJet}(a).  The nominal gluon uncertainty is
mildly reduced at $x\gtrsim 10^{-3}$ according to
Fig.~\ref{Fig:CT18+nDYTTIncJet}(b).  Implications of the
CT18+nDYTTIncJet analysis for Higgs and other phenomenology will be
explored separately.
\clearpage

\section{Advancements in statistical techniques \label{sec:AdvancedStatistics}}
High accuracy and precision are paramount for the next generations of
collinear PDFs. Achieving such accuracy does not reduce to solely
increasing the statistics of experimental measurements and
perturbative order of theoretical calculations. The PDF uncertainty
also reflects the complexity of the global analysis and arises from
several groups of factors \cite{Kovarik:2019xvh}.  To provide the PDF
users with reliable uncertainty estimates, the CT group develops and
applies advanced statistical techniques for uncertainty quantification
(UQ) both \rev{during the global fit or using the published PDF error sets outside of the fit}. Such techniques are
discussed at length in the CT18 \cite{Hou:2019efy} and other
publications
\cite{Schmidt:2018hvu,Hou:2019gfw,Wang:2018heo,Hobbs:2019gob,
  Accardi:2021ysh, Jing:2023isu,Courtoy:2022ocu}. \rev{Among the advanced techniques employed during the fit, scans of fitted PDF values or parameters using Lagrange multipliers \cite{Stump:2001gu} assess constraints from individual data data sets and provide a more precise estimate of the uncertainty for the scanned parameter. It is also used to explore possible tensions among various data sets included in the global fit. We also developed a number of very fast techniques utilizing the published PDF error sets to estimate the sensitivity of unfitted data sets to the PDFs, such as \texttt{ePump}~\cite{Schmidt:2018hvu,Hou:2019gfw} and  {\it sensitivity}~\cite{Wang:2018heo,Hobbs:2019gob},  and to update the Hessian PDFs using such external data \cite{Schmidt:2018hvu,Hou:2019gfw}. In a related class, the hopscotch scan introduced in Ref.~\cite{Courtoy:2022ocu} realizes an in-depth assessment of the PDF uncertainty on a given QCD observable when both Hessian and Monte-Carlo error PDFs are available, like in the case of NNPDF ensembles.} We will now summarize
some recent advancements in this area.

\begin{figure}[p]
    \centering
    \includegraphics[width=0.465\textwidth]{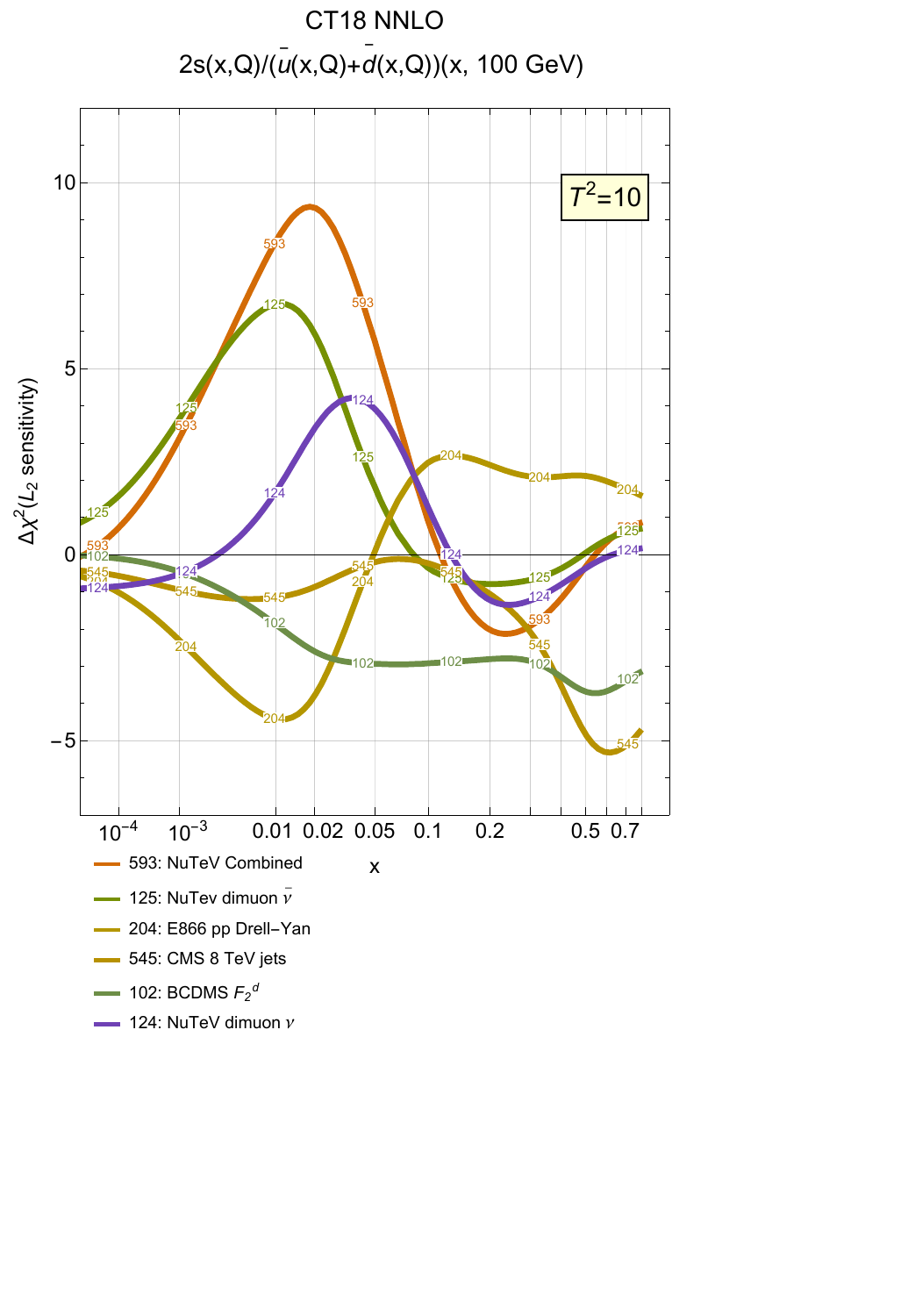}
    \includegraphics[width=0.465\textwidth]{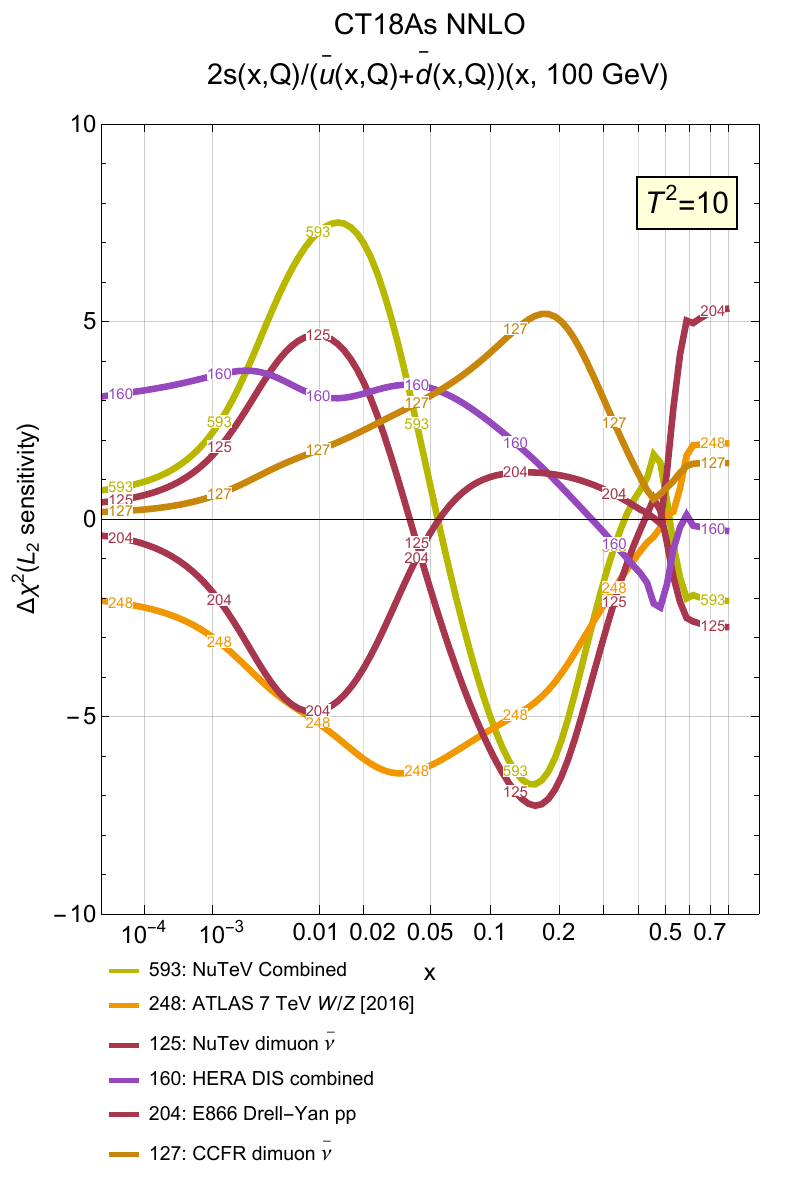}
    \includegraphics[width=0.465\textwidth]{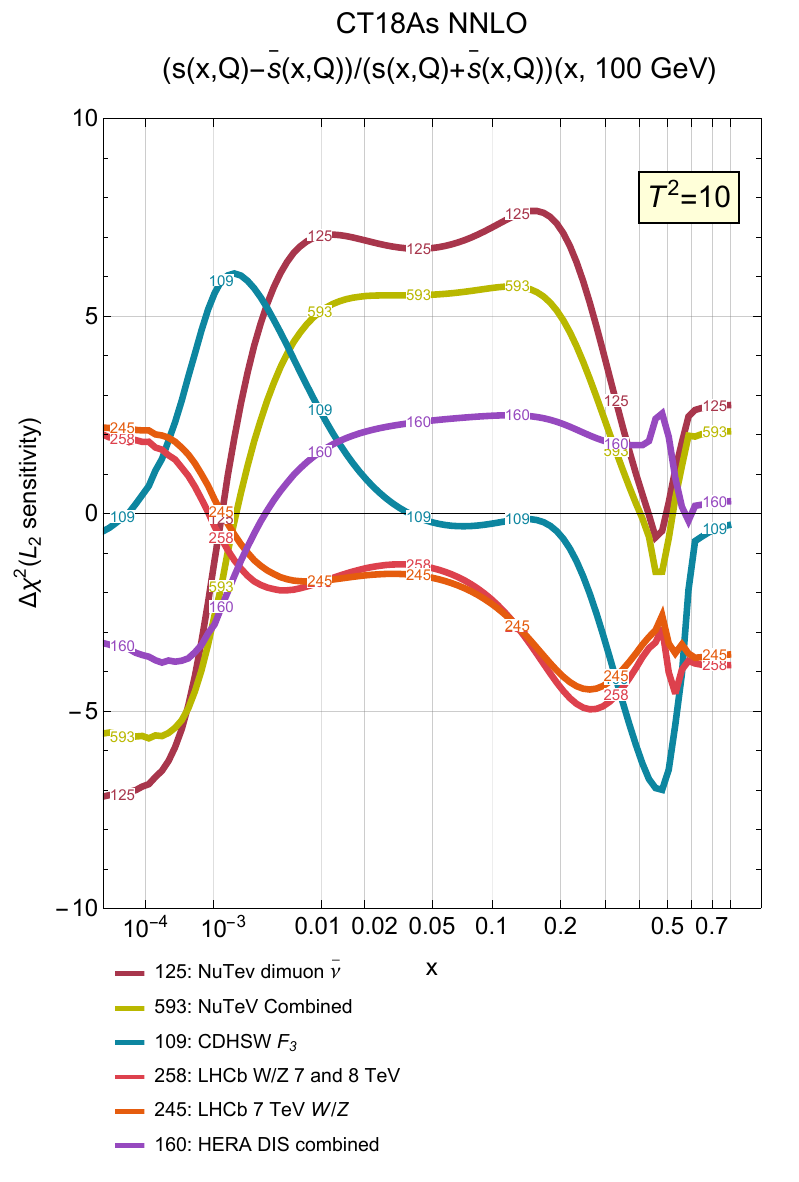}
    \includegraphics[width=0.465\textwidth]{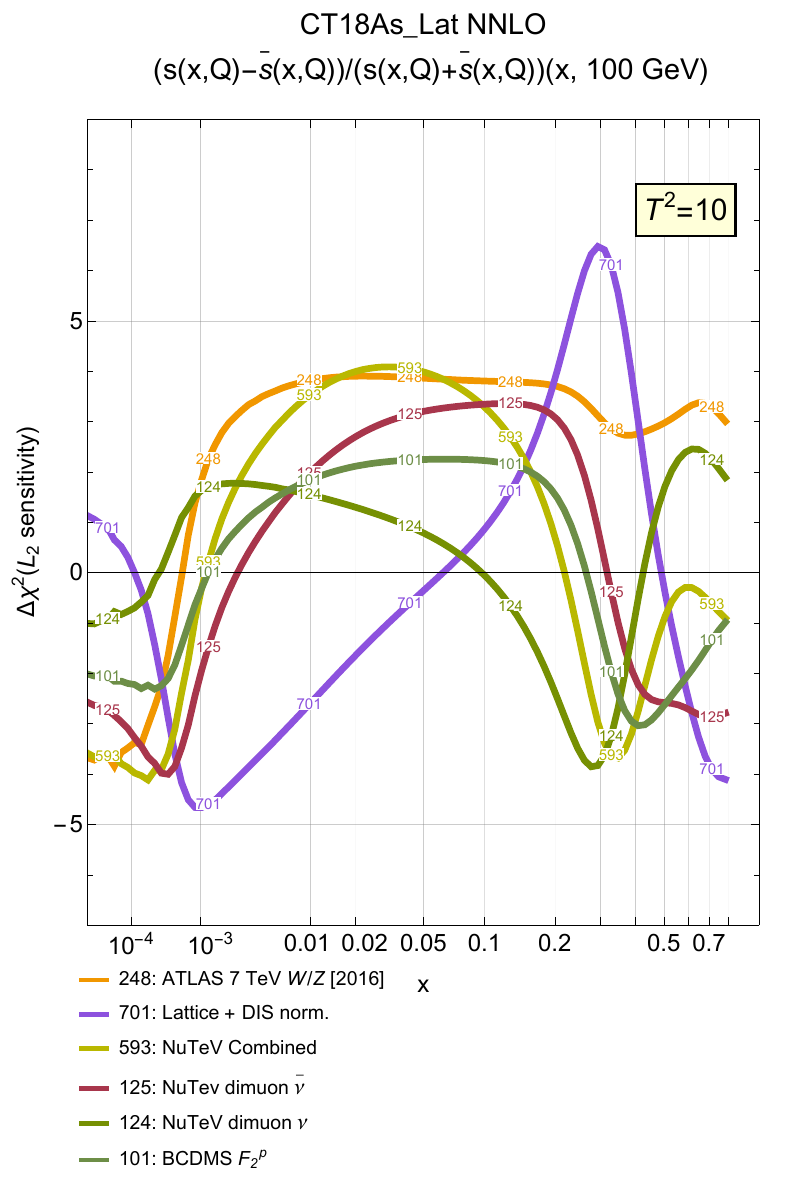}
    \caption{The $L_2$ sensitivities to the strangeness ratio
      $2s/(\bar{u}+\bar{d})$ at $Q=100~\GeV$ for CT18 and CT18As fits
      (upper) and similarly to the asymmetry $s_-$ for CT18As and
      CT18As\_Lat fits.}
    \label{fig:L2CT18AsLat}
\end{figure}

\subsection{Comparisons of PDF fits using $L_2$ sensitivity \label{sec:L2}}

To understand the factors driving the PDF behavior in the analyses by
various groups, the CTEQ-TEA group joined authors of the ATLAS'21 and
MSHT'20 studies to compare the experimental constraints in the fits
presented by the groups using a common metric called ``$L_2$
sensitivity'' \cite{Jing:2023isu}. This metric provides a fast
calculation to estimate the magnitudes and directions of statistical
pulls by individual experimental data sets, in particular on the
departure from the best fit, on the PDFs of a given flavor or
combination of flavors at the user-specified $Q$ and $x$ values. In
particular, the $L_2$ sensitivity renders much of the same information
as the Lagrange multiplier scans \cite{Stump:2001gu} about the pulls
and tensions of the data sets within the global fit, only now using
the tabulated PDF values and $\chi^2$ values of fitted experiments
obtained for each Hessian error set.  For a given data set $E$
included in the global fit, the $L_2$ sensitivity is constructed as a
product of the variation of $\chi^2$ for this experiment, or
$\chi^2_E$, and the cosine of the correlation angle between a
PDF-dependent quantity and
$\chi^2_E$~\cite{Wang:2018heo,Hobbs:2019gob}.  The $L_2$ sensitivity
is thus convenient for visualizing the pulls of the data sets on the
best-fit PDF without repeating the full fit. It extends and
streamlines a related technique in the \texttt{PDFSense} framework
\cite{Wang:2018heo} introduced in the course of the CT18 data
pre-selection.

Furthermore, the $L_2$ sensitivity provides a common metric to explore
experimental pulls across independent PDF analyses utilizing the
Hessian methodology.  The first comparison of this kind focused on
implications of nuclear dynamics affecting deuterium targets in the
contexts of the CTEQ-TEA and CTEQ-JLab NLO analyses
\cite{Accardi:2021ysh}.  It outlined the strategy for the more recent
and extensive investigation at NNLO in Ref.~\cite{Jing:2023isu}.  To
compare the Hessian PDF ensembles of CT18~\cite{Hou:2019efy},
MSHT20~\cite{Bailey:2020ooq} and ATLASpdf21~\cite{ATLAS:2021vod}, the
latter were produced with a common constant (global) tolerance
$T^2=10$.  Such a relatively small tolerance, which is different from
the ones made in the published PDF ensembles, was chosen to improve
the accuracy of the quasi-Gaussian approximation for $\chi^2$ that
motivates the concept of $L_2$ sensitivity and streamlines the
comparisons among diverse PDF ensembles. Then, the $L_2$ sensitivities
were plotted for the experiments fitted by each group, using the NNLO
accuracy for all groups, and for the approximate N$^3$LO in the case
of MSHT'20. An interactive website \cite{L2website} presents an
extensive collection of plots of $L_2$ sensitivities from this study,
as well as a C++ program \texttt{L2LHAexplorer} to plot $L_2$
sensitivities for other Hessian PDF ensembles available in the LHAPDF6
format \cite{LHAPDF6}.

Fig.~\ref{fig:L2CT18AsLat} illustrates how the $L_2$ sensitivities
elucidate the interplay of experimental constraints on the strangeness
and antistrangeness PDFs discussed in Sec.~\ref{sec:LHCDYdata}. Recall
the role of the ATLAS 7 TeV $W/Z$ data (ATL7WZ) in the CT18A NNLO fit,
which favor a larger $s+\bar{s}$ PDF around $x=0.02$ for $Q=2$
GeV. This preference is opposed by some DIS data sets, notably the
NuTeV SIDIS measurements~\cite{Mason:2006qa,NuTeV:2007uwm} and to some
degree the HERA I+II DIS combined cross section~\cite{H1:2015ubc}.
\rev{
We note that some datasets selected for the CT baseline, such as heavy-nuclear DIS measured by CCFR and NuTeV, involve corresponding nuclear corrections; the influence of these corrections on the $L_2$ sensitivities of key measurements was explored in Ref.~\cite{Accardi:2021ysh} alongside light-nuclear corrections for the deuteron.
}
The
$L_2$ sensitivities in the upper row of Fig.~\ref{fig:L2CT18AsLat}
quantify such opposing pulls on the ratio $2s/(\bar{u}+\bar{d})$ in
the CT18 and CT18As NNLO fits, with the positive (negative)
sensitivity indicating a preference for a smaller (larger) ratio at
the $x$ value on the horizontal axis. Focusing on $x=0.01-0.02$, we
see that, in the CT18 fit, the strongest preferences for a lower ratio
at these momentum fractions arise from the NuTeV combined (ID=593) and
anti-neutrino (ID=125) data sets. Even in CT18, the NuTeV preference
is opposed to some extent by the upward pull from the E866 $pp$
Drell-Yan cross section (ID=204). In CT18As, the upward pull is
strengthened by adding the ATL7WZ data.

In the lower row of Fig.~\ref{fig:L2CT18AsLat}, we see the analogous
plots for the asymmetry $s_-(x,Q)$ in the CT18As and CT18As\_Lat fits.
Allowing $s(x,Q_0)\neq \bar{s}(x,Q_0)$ in CT18As~\cite{Hou:2022onq}
(left) leads to a sizable asymmetry that results from the tradeoff
among several experiments. First, a non-zero difference $s-\bar s$ was
proposed long time ago to explain the anomalous measurement of the
weak mixing angle in neutrino DIS by NuTeV. The CT18As fit finds that
the NuTeV preference for the strangeness asymmetry is relatively mild,
and in fact the NuTeV antineutrino SIDIS (ID=125) opposes the even
stronger enhancement of $s-\bar s$ from the best-fit value preferred
by LHCb $W/Z$ production at 7 and 8 TeV (experiments 245 and
258). Inclusion of lattice constraints for $ 0.3 < x < 0.8$ in
CT18As\_Lat (right) changes the distribution of the pulls. The lattice
data indicate a strong preference for a smaller-than best-fit
strangeness asymmetry in the valence region, compensated by a negative
$L_2$ (positive pull) at small $x$ by means of the valence sum rule
enforcing the zero strangeness quantum number in the nucleons. Indeed,
we already saw in Fig.~\ref{fig:CT18AsLat} (right) that inclusion of
the lattice data in CT18As\_Lat substantially suppressed the
strangeness asymmetry at $x>0.3$~\cite{Hou:2022onq}.

These and many other comparisons of $L_2$ sensitivities are available
for ATLAS21, CT18, and MSHT'20 NNLO PDFs and MSHT approximate N$^3$LO
PDFs in Ref.~\cite{Jing:2023isu} and online at \cite{L2website}.  The
figures in these references plot the sensitivities either as the
dominant pulls from the experiments on a given PDF flavor (PDF
combination), or as the pulls on all PDF flavors (combination) from a
single experimental data set. For many experiments, the sensitivity
patterns or ``pulls'' on the PDFs are indistinguishable from those due
to statistical fluctuations in the data samples, reflecting a good
agreement of the experiment with the best-fit PDF model or a weak
constraint. For a small number of experiments, strong pulls reflect a
disagreement or a large influence from a given data set. The figures
on the website are formatted to emphasize the $L_2$ sensitivities from
such influential experiments and to identify the $x$ and $Q$ intervals
that are strongly affected.

Some patterns of the pulls are similar among the different global
analyses (CT and MSHT) or even the three groups (including the
non-global ATLAS fit). One such shared pattern is for the CMS 8 TeV
jet data set: in Ref.~\cite{Jing:2023isu}, it served as an example to
demonstrate that the $L_2$ sensitivities are in a good agreement with
the findings from the Lagrange Multiplier scans at various $x$
values. Other pulls differ substantially among CT and MSHT, e.g., for
the $R_s$ ratio and strangeness asymmetry discussed above. These
studies revealed interesting new aspects of the compared global fits,
such as an exceptional pull of the ATLAS 8 TeV data on $Z$-boson
transverse momentum on the gluon PDF in the MSHT'20 NNLO analysis,
differences in the impact of the E866 $pp$ and $pd$ data on the $\bar
d/\bar u$ ratio at $x>0.1$ in the CT and MSHT fits, and the
consistency of pulls of LHC $t\bar t$ production distributions on the
gluon PDFs in the fits of all three groups.  The information about the
pulls from the experiments is extracted at once over the whole $x$
range, in contrast to the LM scans that are usually limited to a few
kinematic points because of their strong demand on the computation
time.

To further illustrate the benefits of the $L_2$ sensitivity method,
Lucas Kotz \cite{Kotz:2024dfg} employed the open-source program
\texttt{L2LHAexplorer} \cite{L2website} to compute the sensitivities
of several processes available in the \texttt{xFitter}
framework~\cite{xFitterwebsite} to the CT18, CT18As, and MSHT20 NNLO
PDFs with $T^2=10$ (the same PDFs as the ones studied in
\cite{Jing:2023isu}).  For the processes that are implemented both in
\texttt{xFitter} and native CTEQ-TEA and MSHT fitting codes, such as
the HERA I+II inclusive DIS data set, comparisons of the resulting
$L_2$ sensitivities revealed some differences in the pulls on the
resulting PDFs due to the specific workings of the codes and settings
of the analyses, {\it e.g.} the adopted $\chi^2$ definitions. When
such differences are non-negligible compared to the PDF uncertainties,
they will need to be understood or eliminated via benchmarking studies
such as those done by the PDF4LHC working
group~\cite{PDF4LHCWorkingGroup:2022cjn}.  For the new data sets that
are already implemented only in \texttt{xFitter}, the $L_2$
sensitivities in Ref.~\cite{Kotz:2024dfg} provide an easy way to
project the likely impact of the candidate data sets on the PDFs,
should these data sets be implemented in the other fitting codes.  In
the provided examples, heavy-quark production at LHCb at small
transverse momenta seems to disfavor a gluon that turns strongly
negative at $x < 10^{-4}$ and $Q < 2$ GeV, while CMS 13 TeV inclusive
jet production data shows some sensitivity to $\bar u$ and $\bar d$ at
very high momentum fractions ($x > 0.5$), in addition to its usual
sensitivity to the gluon PDF in the kinematic region at $x > 0.01$
relevant for LHC Higgs production. The $L_2$ sensitivities hence guide
the optimal selection of the new data sets for the global fit.

\subsection{Tolerance, the likelihood ratio, and epistemic PDF uncertainty \label{sec:NewUQ}} 
Quantification of the PDF uncertainty presents a challenging problem
with particularly wide-ranging implications for precision tests of the
Standard Model, including the measurements of the $W$ boson mass,
electroweak mixing angle, and QCD coupling. PDF uncertainties arise
from several categories of sources that are broadly associated with
experimental, theoretical, parametrization, and methodological
factors~\cite{Kovarik:2019xvh}. Consistent combination of such
uncertainties, also accounting for their mutual correlations, drives
development of increasingly sophisticated statistical techniques, with
two mainstream approaches founded on the Hessian \cite{Pumplin:2001ct}
and stochastic sampling \cite{Giele:1998gw,Giele:2001mr}
methodologies. In the parlance of PDF fitters, the choice of tolerance
\cite{Pumplin:2002vw, Martin:2009iq,Lai:2010vv,Kovarik:2019xvh} or a
related criterion that defines the PDF uncertainty has lead to sizable
differences in the uncertainty estimates provided by different groups
even when they fitted a very similar data set in the context of a
joint benchmarking exercise \cite{PDF4LHCWorkingGroup:2022cjn}. Since
the publication of its first error PDF sets in the CTEQ6 analysis
\cite{Pumplin:2002vw}, the CTEQ-TEA group has been refining its
uncertainty quantification (UQ) methods with each generation of the
general-purpose PDFs. Its current two-tier tolerance prescription
implemented in CT18 NNLO PDFs \cite{Lai:2010vv,Gao:2013xoa} accounts
for both incomplete agreement (tensions) of the fitted experiments and
significant dependence on the chosen functional forms of PDFs (with
the latter contributing as much as a half of the total uncertainty
even in the regions with strong constraints from the data). The
upcoming release of CTEQ-TEA PDFs will further refine its PDF UQ.

What principle should guide the design of the uncertainty
quantification technique? It must be based on a solid foundation of
Bayesian statistics and account for drastic modifications in the
textbook strategies to reflect the large dimensionality of parametric
space in PDF fits. In the recent works \cite{Kovarik:2019xvh,
  Courtoy:2022ocu, Jing:2023isu}, we emphasized the high value of the
likelihood-ratio (Wilks) test for justifying the credibility intervals
on the PDFs. Given the theoretical predictions $T_1$ and $T_2$ based
on a pair of PDF ensembles, the likelihood-ratio test invokes the
Bayes theorem to update the ratio of Bayesian probabilities for $T_1$
and $T_2$ given the empirical data set $D$:
\begin{equation}
\frac{P(T_2|D)}{P(T_1|D)}= \frac{P(D|T_2)}{P(D|T_1)}
\frac{P(T_2)}{P(T_1)},
\end{equation}
with $P(D|T)$ the likelihood of $T$ for a given $D$, $P(T)$ the prior
probability of $T$, and $P(T|D)$ the posterior probability of $T$
given $D$.

The likelihood-ratio test provides a foundation to discriminate
between any pair of PDF solutions \cite{Soper:1994km} based on their
prior and likelihood probabilities. When two PDF solutions are
compared within a single fit and adopting identical functional forms
for their PDFs, i.e., within the same statistical model and using the
same prior, their posterior probabilities are distinguished solely
based on their likelihoods, i.e. their total $\chi^2$ values. From
this observation, we immediately justify the complementary techniques
of Lagrange Multiplier scans \cite{Stump:2001gu} and sensitivities
\cite{Hobbs:2019gob,Wang:2018heo,Accardi:2021ysh,Jing:2023isu} to
examine mutual consistency of the fitted data sets, which constitutes
an integral part of UQ in QCD global analyses and has direct
implications for the precision of the final PDFs. The above techniques
not only explore the impact of statistical uncertainties from each
individual experiment, but also elucidate the degree of agreement
among the individual experiments in the fitting process. When
eliminating substantial disagreements from the global data is not
feasible, the tolerance on the PDF uncertainty must be chosen to
account for them. The LM and $L_2$ sensitivity techniques contribute
to determination of the tolerance in a systematic way.

Alternatively, when the likelihoods of two models are the same, the
likelihood-ratio test posits unambiguously that the models are
distinguished only by their priors. An important case is when two PDF
sets differ only in their parametrization forms while rendering
exactly the same $\chi^2$ values and having no differences in their
priors. According to the likelihood-ratio test, such PDFs must be at
identical credibility levels, and Bayesian uncertainties must account
for statistical equivalence of such solutions. This logic usefully
guides the understanding of various sources of PDF uncertainties
either in the context of including PDF parametrization uncertainties
into the total CT uncertainty, justifying the $L_2$-sensitivity
studies~\cite{Jing:2023isu}, or in the analysis of possible biases in
sampling of PDF solutions with equivalent likelihoods and
priors~\cite{Courtoy:2022ocu}.

Such post-fit tests are increasingly important for representative
exploration of large-dimensional parametric spaces and will contribute
to defining PDF uncertainties. In that spirit, we will now distinguish
between two types of uncertainties, whose nomenclature is borrowed
from AI/ML: the aleatoric (stochastic) uncertainties, which are
irreducible for a given data set with a fixed number of scattering
events, and the epistemic ones, which may be reduced by improving
modeling and methodological choices, as well as by sufficiently
complete sampling over analysis workflows, parametrization forms, and
analysis settings. In the Monte-Carlo frameworks employed e.g. with
AI/ML models, the epistemic uncertainty can be further divided into
model and distributional components \cite{MalininGales:2018}, both
associated with the prior probability that we mentioned above. The
model uncertainty reflects the spread of parameters giving reasonable
description of the fitted data within the adopted model. Within the
PDF fits, it can be estimated for each model using a closure test on
pseudodata generated from some known ``truth'' distribution. The
distributional uncertainty arises from the mismatch between the
populations of solutions accessible within the considered models and
present in the real data.  This mismatch can then be remedied by
constructing a sufficiently representative class of models that
adequately sample the solutions consistent with the real data. Indeed,
in the language of Ref.~\cite{Courtoy:2022ocu}, ``distributional
uncertainty'' is referred to as the ``sampling uncertainty''.  While
challenging in general, estimation of the epistemic uncertainty is
facilitated by several representative sampling techniques such as the
hopscotch parameter exploration \cite{Courtoy:2022ocu} developed by
CTEQ-TEA authors. Furthermore, since the uncertainty due to the choice
of PDF parametrizations contributes a sizable part of the epistemic
uncertainty, several methods are developed within our group for
representative sampling over PDF parametrizations, as we will now
briefly review.

\subsection{Advanced polynomial and neural-network parametrizations \label{sec:Parametrizations}} 

Dependence on the functional forms of PDFs can be a leading source of
the epistemic uncertainty together with another source associated with
the selection of experiments within a given fitting methodology. The
presence of this uncertainty explains why the $\Delta \chi^2=1$
tolerance definition, associated with the aleatory experimental
uncertainty at 68\% probability level, is not sufficient on its own.
In the CT18 study~\cite{Hou:2019efy}, the parametrization uncertainty
was estimated by repeating the global fit with $>250$ functional forms
and then adjusting the tolerance on the final Hessian error ensemble
to approximately cover the spread of PDFs obtained with all candidate
forms.

\begin{figure}[t]
    \centering
    \includegraphics[width=0.57\textwidth]{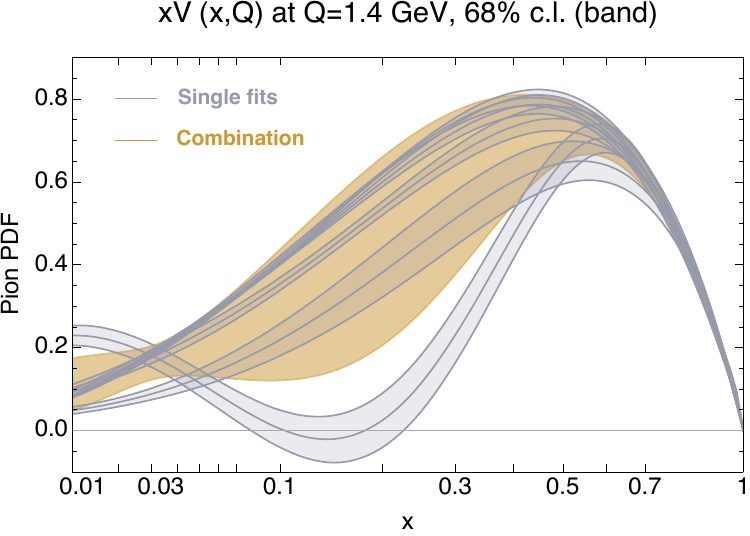}
    \includegraphics[width=0.39\textwidth]{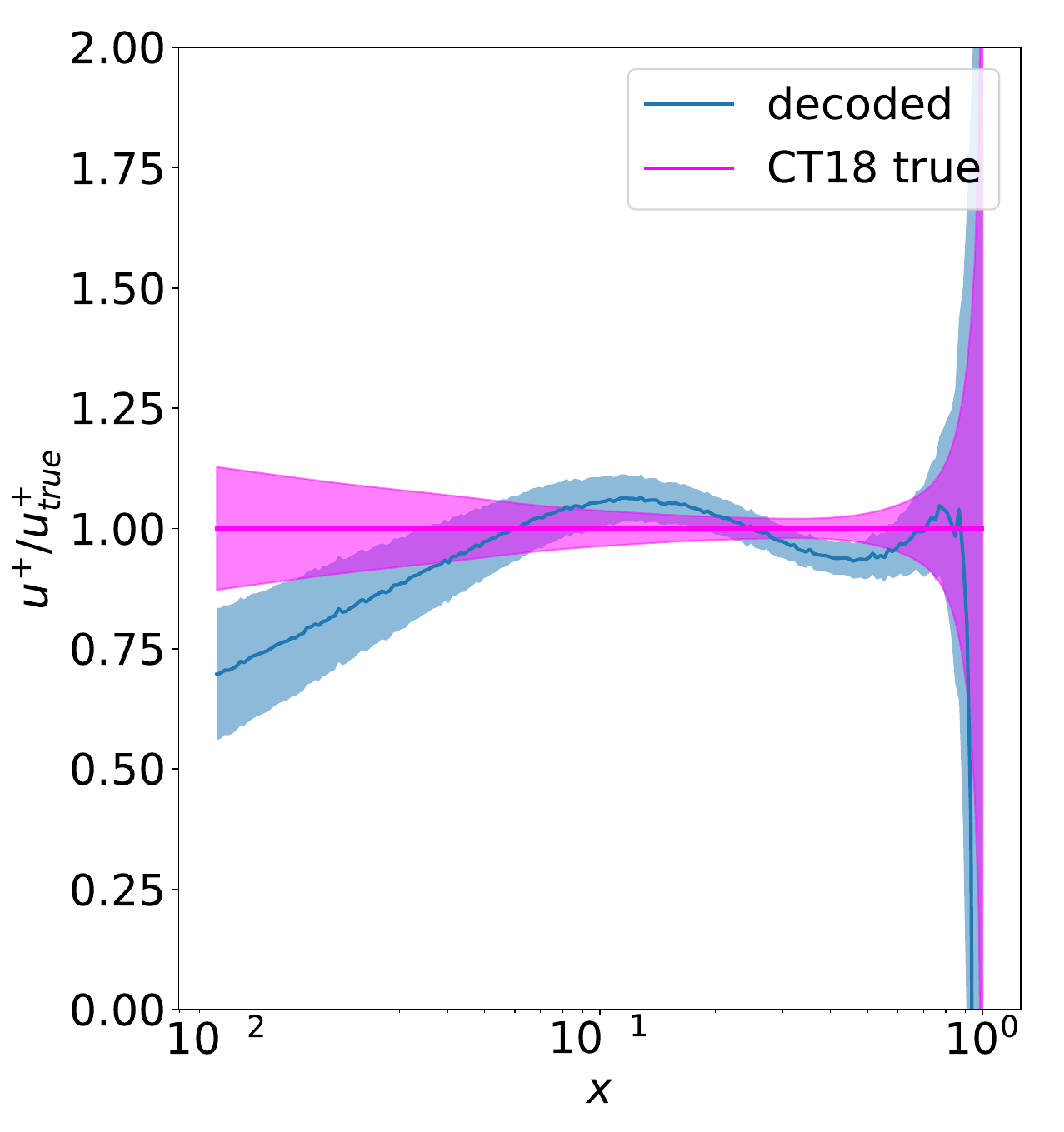}\\ \quad\hspace{50pt}
    (a) \hspace{3in} (b)
    \caption{(a) Single NLO fits (5 gray bands) and their combination
      using METAPDF (mustard band) for the valence PDF of the positive
      pion extracted in the Fant\^omas4QCD framework
      \cite{Kotz:2023pbu}.  (b) A reconstruction of the CT18 ``truth''
      combination $u_+(x,Q)=u(x,Q)+\bar u(x,Q)$ from its Mellin
      moments using a variational autoencoder inverse mapper
      (VAIM). From Ref.~\cite{Kriesten:2023uoi}.}
    \label{fig:fantomas}
\end{figure}

Along the same lines, and as a potential avenue to implement the
parametrization uncertainty in the upcoming CTEQ-TEA fits, a subgroup
of authors initiated a project Fantômas that streamlines generation
and exploration of versatile polynomial forms to parameterize the PDFs
using Bézier curves. Such forms can approximate arbitrary continuous
functions as a result of the Stone-Weierstrass approximation
theorem. A \textit{metamorph} is a generic two-component
parametrization introduced by this approach to reproduce the
asymptotic behaviors of the PDFs at $x\to 0$ and $x\to 1$, and to
control the flexibility of the PDFs at intermediate $x$ using
strategically placed \textit{control points}. \rev{In this approach, 
the polynomial component of the PDF parametrization is exactly computed 
from its values at several chosen points in $x$, or control points. 
These values of the polynomial itself serve as free parameters, which facilitates their mutual independence 
and streamlines generation of new functional forms for studies of the parametrization dependence.} 
The first physics application~\cite{Kotz:2023pbu} of this approach, for the global
analysis of parton distributions in a charged pion, shows an increased
uncertainty {\it w.r.t.} previous phenomenological analyses
\cite{Barry:2018ort,Barry:2021osv,Novikov:2020snp} of the pion PDFs
with fixed functional forms. In this study, the metamorph forms were
employed to generate over 100 candidate parametrizations, from which
the 5 most diverse forms were selected and combined into the final PDF
ensemble using the METAPDF technique \cite{Gao:2013bia}. All of the
final single Fant\^omas fits have a low chi-square value, ranging
between $\chi^2\in [\chi^2_{\rm min}, \,\chi^2_{\rm
    min}+\sqrt{2(N_{\rm pt}-N_{\rm par})}]$, with $N_{\rm par}/N_{\rm
  pt}\,\sim (1-3)\%$. The resulting PDFs, shown for the valence PDF of
the pion in Fig.~\ref{fig:fantomas}(a), display a variety of solutions
allowed by data as of now.  The final combination (in mustard in
Fig.~\ref{fig:fantomas}(a)) reflects the $68\%$ C.L. of the ensemble
made by the generating 5x100 MC replicas from the 5 selected
parametrization forms.
The Fantômas analysis, with its sampling over parametrizations,
demonstrates the role of methodological choices in the account for
epistemic uncertainties \cite{Kotz:2023pbu}. The Fantômas C++ module
is included into the \texttt{xFitter} and will be considered for
interpretable UQ in the upcoming CTEQ-TEA PDFs.

As a complementary alternative to the polynomial-based Fant\^omas
approach, a recent CT-adjacent analysis \cite{Kriesten:2023uoi}
explored the influence of model uncertainties in neural-network-based
PDF reconstructions. It considered a series of ML-based models for the
purpose of parameterizing the $x$ dependence of PDFs in a toy problem
(considering only $u$- and $d$-PDFs) extendable to the full flavor
basis of modern PDF analyses.  Specifically, a spectrum of
(variational) autoencoder models with an array of network topologies
was considered, focusing on imposing interpretability requirements on
the latent spaces that encode the PDFs learned during the
training. Given significant interest in the PDF-lattice
correspondence, Ref.~\cite{Kriesten:2023uoi} restricted loss surfaces
during model training to ensure that latent spaces encoded the
lattice-calculable Mellin moments. With this tractable latent space,
ML models could then be used generatively to predict the $x$
dependence of PDFs from their Mellin-space behavior with good
performance. Figure~\ref{fig:fantomas}(b) shows an example of such
reconstruction for $u(x,Q)+\bar u(x,Q)$. The VAIM technique employed
in this example achieves a reasonable agreement at $x>0.1$, where a
small number of Mellin moments is most restrictive. Such trained
models can function as a complementary statistical tool to explore
possible out-of-distribution behavior among various PDFs
parametrizations and facilitate UQ interpretations inside a basis of
separated aleatoric and epistemic errors. Further advances along these
lines, aimed to develop an explainable AI framework to interpret the
effects of internal assumptions in PDF analyses, were recently made in
\cite{Kriesten:2024are}.  We anticipate further development of both
polynomial and AI-based methods for parametrization uncertainty
quantification alongside the progress toward the next generation of
the CTEQ-TEA global fit.

\subsection{A Gaussian Mixture Model \label{sec:GMM}}

\begin{figure}
    \centering
    \includegraphics[width=0.48\textwidth]{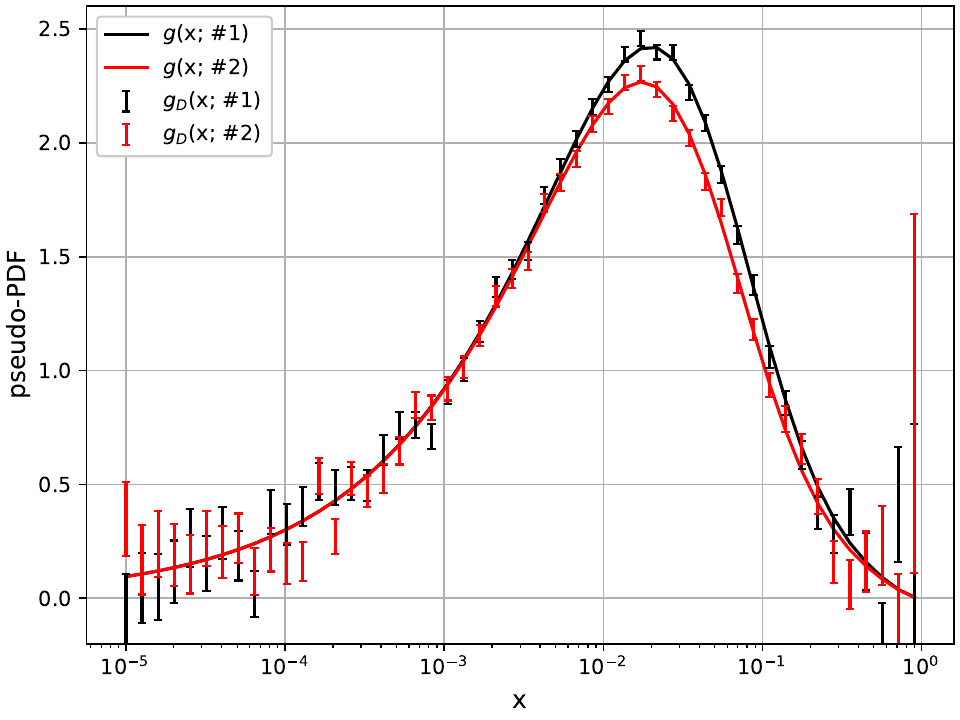}
    \includegraphics[width=0.50\textwidth]{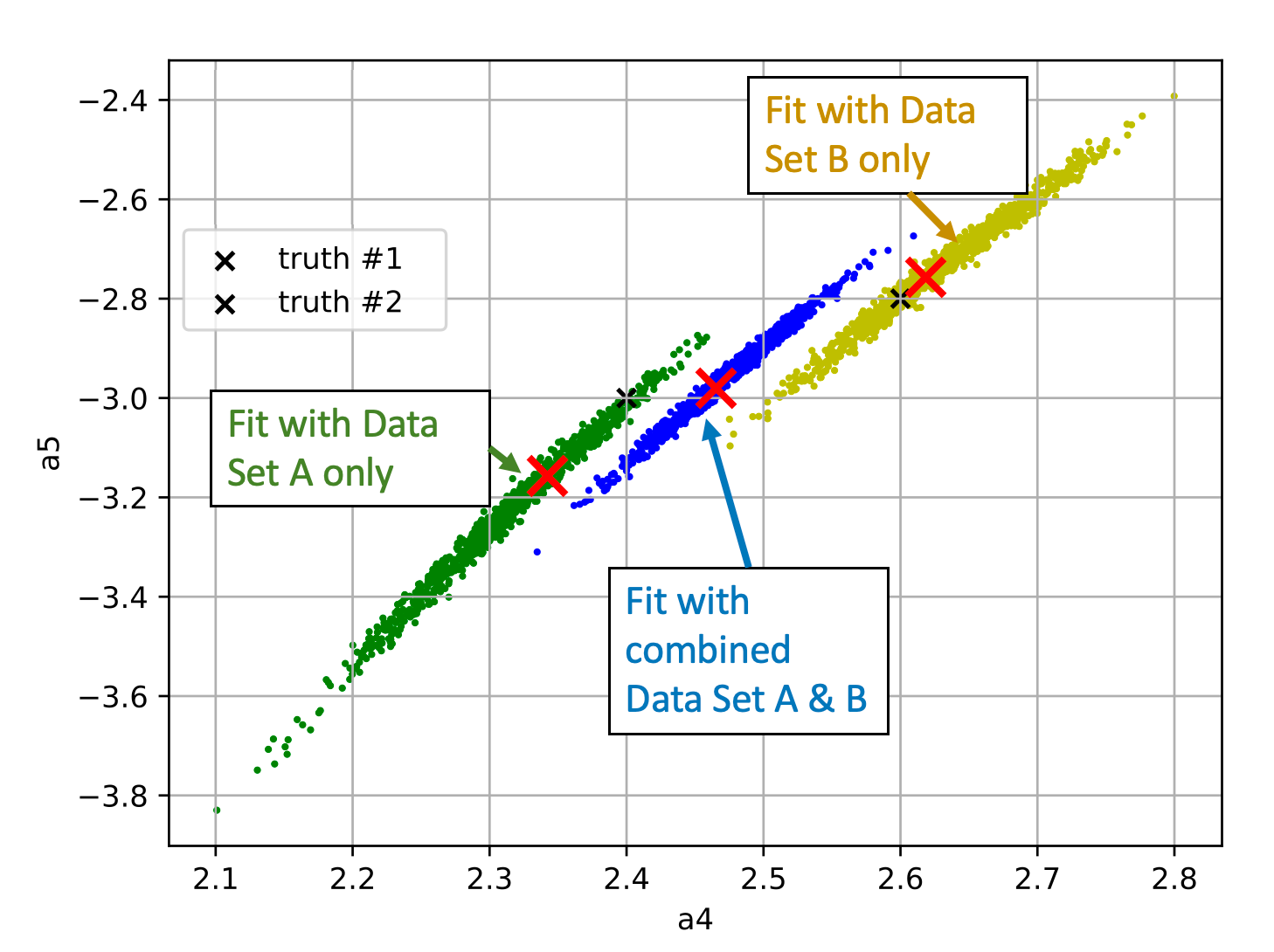}
    \\ (a)\hspace{0.5\textwidth}(b)\\ \quad\\ \includegraphics[width=0.49\textwidth]{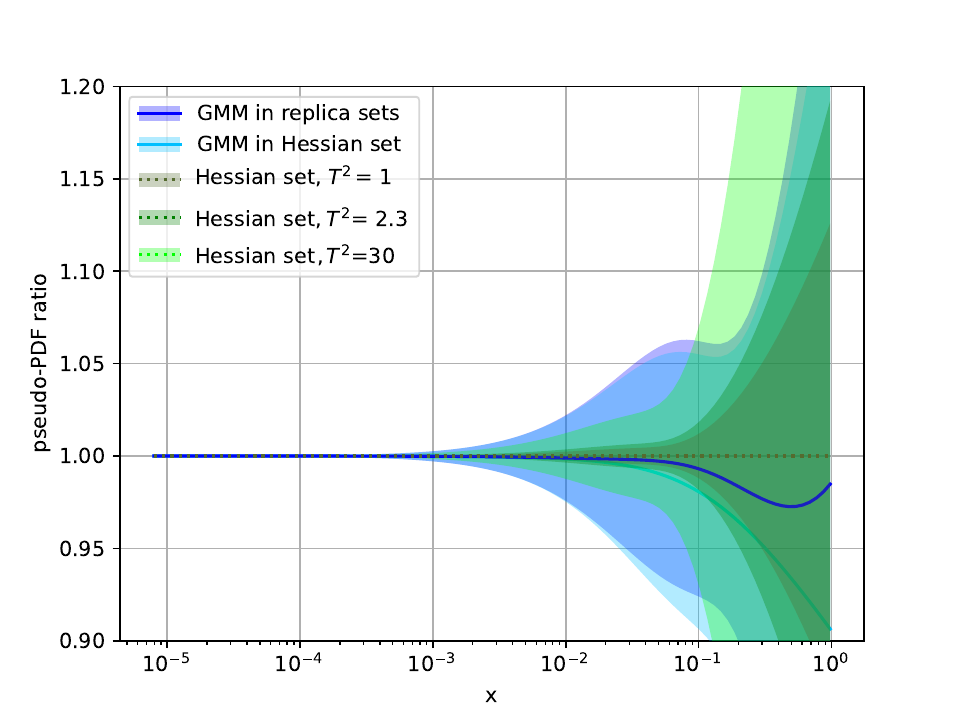}\\ (c)
    \caption{(a) Two pseudodata sets, $g_D(x; \# 1)$ (black) and
      $g_D(x; \# 2)$ (red), are in tension with each other. They are
      generated by random fluctuations around the solid curves $g(x;
      \# 1)$ and $g(x; \# 2)$.  (b) Distributions of pseudo-PDF
      parameters $a_4$ and $a_5$, fitted to data sets A (green) and B
      (yellow) individually using Monte-Carlo sampling, as well as in
      a combination of A and B using the product of likelihoods
      $P_A\left(D|T(\textbf{a})\right) \,
      P_B\left(D|T(\textbf{a})\right)$ (blue). (c) Comparison of the
      uncertainties obtained with the traditional Hessian technique
      with tolerances $T^2=1$, 2.3, and 30, and with the GMM analysis
      with $K=2$. Note that the Hessian tolerance criterion used here
      is simplistic and different from the more sophisticated
      tolerance prescriptions in realistic fits.  }
    \label{fig:GMM}
\end{figure}

As discussed in our recent study~\cite{Ablat:2023tiy},
Secs.~\ref{sec:LHCdata} and \ref{sec:L2}, some LHC precision data sets
are found to be in tension either among themselves (\emph{e.g.}, ATLAS
8 TeV $W$ boson~\cite{ATLAS:2019fgb} vs. CMS 13 TeV $Z$ boson
data~\cite{CMS:2019raw}) or with the pre-LHC ones (\emph{e.g.}, ATLAS
7 TeV $W,Z$ production (ATL7WZ)~\cite{ATLAS:2016nqi} vs.~NuTeV dimuon
SIDIS~\cite{NuTeV:2007uwm, Mason:2006qa}). Such data sets may pull the
PDFs in different directions, as illustrated in
Figs.~\ref{fig:CT18DYfit}, \ref{fig:CT18+DY-CT18+incJet}, and
\ref{fig:L2CT18AsLat}. There are several proposals for the estimation
of uncertainty when the data are in
tension~\cite{Tanabashi:2018oca,Erler:2020bif,Cowan:2018lhq,DAgostini:1999niu}. In
the context of PDF fits, tensions contribute to enlargement of the PDF
uncertainties estimated according to various tolerance criteria
developed over the years, \emph{e.g.}, in
\cite{Hou:2019efy,Bailey:2020ooq,Kovarik:2019xvh}.

Recently, a subgroup of CTEQ-TEA authors proposed a solution based on
a Gaussian Mixture Model (GMM)~\cite{Yan:2024yir} inspired by Bayesian
applications in astrophysics.  The GMM sheds light on model-averaging
in the presence of experimental tensions and predicts some features of
the uncertainties that may not be captured with the simplified $\Delta
\chi^2 = T^2$ tolerance.

We demonstrate the main idea in a toy example in Fig.~\ref{fig:GMM},
when two data sets A and B are in tension with one another and prefer
somewhat different underlying pseudo-PDFs, denoted as $g(x; \#1)$ and
$g(x; \#2)$ in Fig.~\ref{fig:GMM}(a), because of some unaccounted
systematic factor. In our toy example, pseudodata sets $A$ and $B$ are
generated by random Gaussian fluctuations around $g(x; \#1)$ and $g(x;
\#2)$, referred to ``truth \#1'' and ``truth \#2'' in the
discussion. Our goal is to estimate a single underlying objective
``truth'' pseudo-PDF from these imperfect inputs.

In Fig.~\ref{fig:GMM}(b), the tension among A and B is reflected in
the differences between the green and yellow regions populated by the
fitted parameters $\{a_4,a_5\}$ of the pseudo-PDF when A and B are
fitted individually. The distributions reflect the respective
likelihood distributions of A and B,
\begin{equation}
P_k(D|T(\textbf{a}))= \frac{\exp{-\frac{1}{2}\sum_{i,j=1}^{N_{{\rm
          pt},k}}\left(D_i-T_i(\textbf{a})\right)\,
    C_{k,ij}^{-1}\,\left(D_j-T_j(\textbf{a})\right)}}{\sqrt{(2\pi)^{N}\det(C_k)}},
\mbox{ with } k=A\mbox{ or B},
\end{equation}
which are not fully compatible.  Here $D_{i(j)}$ is a data point with
a covariance matrix $C_{k,ij}$ for data set $k=A$ or $B$,
$T_{i(j)}(\textbf{a})$ is a theoretical prediction for a parameter
vector $\textbf{a}$. In this case, we see that the confidence regions
indicated by green (for data set $A$) and yellow (for data set $B$)
enclose the values of truth \#1 and truth \#2. On the other hand, the
preferred region of a simultaneous fit of A and B with a shared vector
$\textbf{a}$, indicated by blue color in Fig.~\ref{fig:GMM}(b), does
not overlap with the yellow and green regions, and neither it encloses
the truth \#1 and \#2 values. So, the confidence region from the
simultaneous fit is too narrow.

As proposed in Ref.~\cite{Yan:2024yir}, a more realistic estimate of
the uncertainty on the combined data sets is possible by maximizing a
weighted sum of the likelihoods of A and B, constructed as
\begin{equation}
\prod_{k=A,B} \sum_{l=1}^{K}\omega_l\,P_{k}(D|T(\textbf{a}_l)),
\label{PDTGMM}
\end{equation}
where $\sum_l\omega_l=1$, and $K$ is a free parameter that can be
determined with the help of information criteria~\cite{Yan:2024yir}.
Figure~\ref{fig:GMM} (c) demonstrates such a combination for a
pseudo-PDF in the GMM model with $K=2$, as compared to the
conventional fit of likelihoods $P_A(D|T) P_B(D|T)$ and estimates of
the PDF uncertainties in the Hessian formalism with various constant
tolerances $\Delta \chi^2 = T^2$. The GMM approach produces a single
central fit that is similar to the standard one and an uncertainty
(blue band) that is more representative of the $x$ dependence of the
tension in the data sets.  As clearly illustrated in
Fig.~\ref{fig:GMM} (c), for $x$ between 0.01 and 0.1, the GMM
analysis, with $K=2$ in this case, predicts a larger uncertainty than
that from the the standard Hessian technique with tolerance $T^2=1$ or
even 30 (a very large value). This is in the $x$ region where two data
sets disagree in Fig.~\ref{fig:GMM}(a), and the conventional
uncertainty would be underestimated.  Outside of the region of
tension, the GMM uncertainty is closer to the Hessian estimates with
small $T^2$.  The toy example illustrates that a combination with the
GMM may capture the tensions of data sets without introducing an
explicit tolerance parameter.  We defer to Ref.~\cite{Yan:2024yir} for
a more detailed demonstration.

\section{PDFs at small momentum fractions \label{sec:smallx}}

In the kinematic region defined by both small $x\!\lesssim\!10^{-3}$
and small values of the momentum transfer, $Q\sim\mathcal{O}(\textrm{1
  GeV})$, deviations from DGLAP collinear factorization begin to
emerge due to the growth of logarithmic corrections
$\sim\!\alpha^m_s(Q^2)\log^n(1/x)$ described by the BFKL framework
\cite{Fadin:1975cb,Lipatov:1976zz,Kuraev:1976ge,Kuraev:1977fs,Balitsky:1978ic}. Eventually
at the smallest $x$ the growth of the scattering cross sections is
tamed by saturation of partonic
densities~\cite{Gribov:1983ivg,Golec-Biernat:1998zce,Morreale:2021pnn}.
Here the linear BFKL equation is superseded by the non-linear JIMWLK
one
\cite{Jalilian-Marian:1997qno,Jalilian-Marian:1997jhx,Weigert:2000gi,Iancu:2000hn,Ferreiro:2001qy}
that slows down the growth of scattering cross sections to satisfy
unitarity.  Various approaches have been proposed to compute these
small-$x$ effects in QCD phenomenology. In the PDF fits, possible
modifications in DIS at the smallest accessible $x$ where examined by
the NNPDF~\cite{Ball:2017otu} and
xFitter~\cite{xFitterDevelopersTeam:2018hym} authors based on a
realization of the NLO BFKL formalism
\cite{Altarelli:1999vw,Ball:1999sh,Altarelli:2000mh,Altarelli:2005ni,Bonvini:2016wki}
adopted in the respective fitting codes.  Alternatively, the CT18X
analysis~\cite{Hou:2019efy} showed that similar modifications in PDFs
can be obtained at pure NNLO by adopting a dynamic factorization scale
in DIS cross sections following a motivation from small-$x$ saturation
dynamics~\cite{Golec-Biernat:1998zce,Iancu:2002tr,Mueller:2002zm}. This
approach takes advantage of the scale dependence of a fixed-order
cross section to slow down the growth of PDFs and scattering cross
sections at $x\to 0$, as saturation predicts.

\begin{figure}[b]
    \centering
    \includegraphics[width=0.49\textwidth]{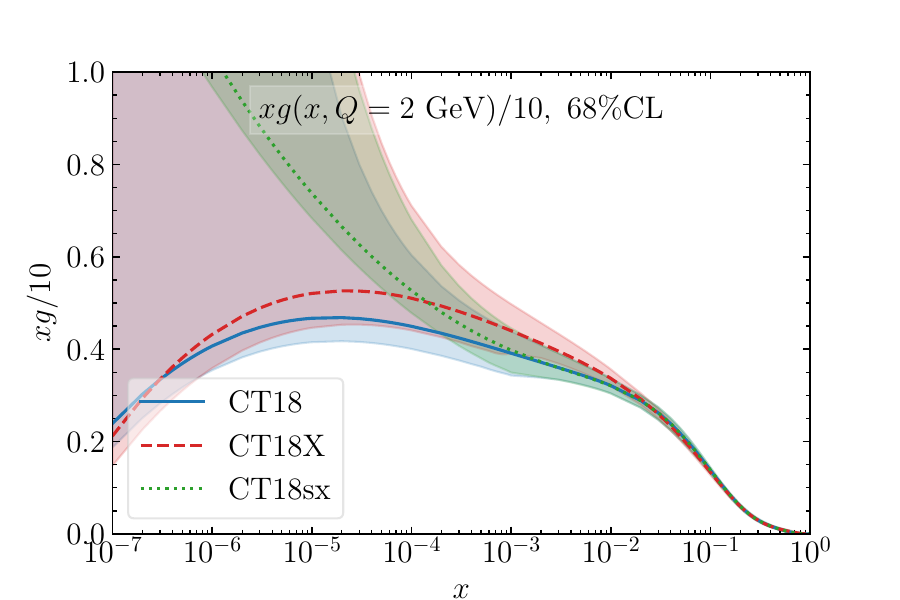}
    \includegraphics[width=0.49\textwidth]{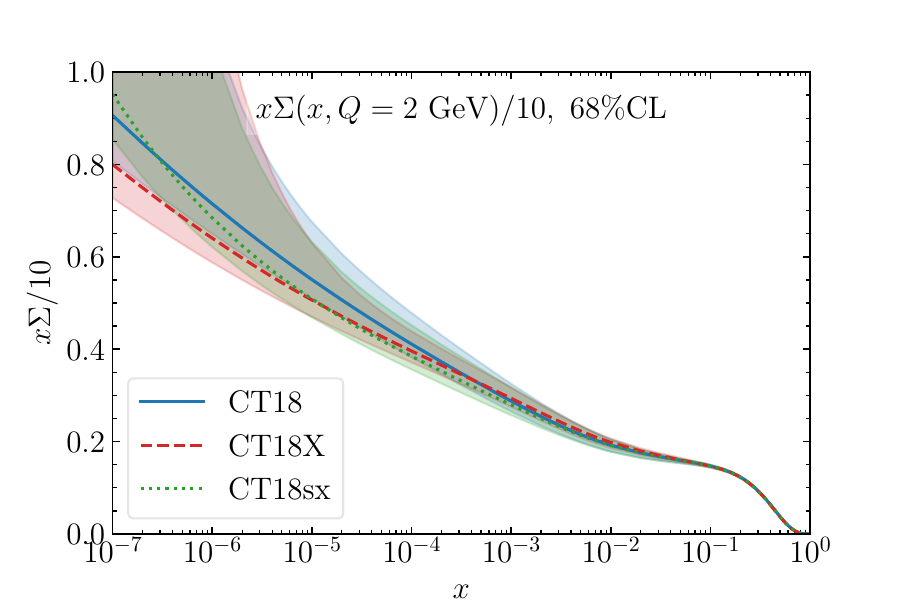}
    \caption{Comparison of the gluon and flavor singlet
      $\Sigma=\sum_i(q_i+\bar{q}_i)$ PDFs at $Q=2~\GeV$ among the
      CT18, CT18X and CT18sx PDF fits.}
    \label{fig:CT18sx}
\end{figure}

Availability of the PDFs with the BFKL or saturation components helps
to model possible experimental signatures in small-$x$ scattering.  As
an illustration, in Fig.~\ref{fig:CT18sx}, we compare PDFs for the
gluon and flavor-singlet light-quark combination,
$\Sigma\equiv\sum_{i}(q_i+\bar{q}_i)$, at $Q\!=\!2~\GeV$ for two
modified versions of CT18 NNLO PDFs: CT18X, obtained in a full global
fit with the saturation-motivated factorization scale $\mu_{\rm F,\,
  DIS}=0.8\,\sqrt{Q^2 + 0.3\mbox{ GeV}^2/x_B^{0.3}}$ dependent on
Bjorken $x$ in DIS; and CT18sx~\cite{Guzzi:2021fre}, obtained by
evolving the CT18 PDFs from the starting scale $Q_0=1.3~\GeV$ using
the code \texttt{APFEL}~\cite{Bertone:2013vaa} with NNLO DGLAP
splitting functions matched to the next-to-leading logarithm of
small-$x$ (NLLx) provided by
\texttt{HELL}~\cite{Bonvini:2016wki,Bonvini:2017ogt}.  At $10^{-5}
\lesssim x \lesssim 10^{-3}$, we see that both the resummed evolution
in CT18sx and the saturation model in CT18X enhance the gluon PDF,
while reducing the singlet quark PDF. At $x < 10^{-5}$, the NNLO+NLLx
splitting kernels evolve the CT18sx gluon faster than in CT18 at NNLO,
while CT18X eventually tames the small-$x$ growth of both the gluon
and singlet PDFs, as would be expected on the logic of saturation.
Since the CT18X PDFs are refitted with an alternative DIS scale, their
differences from CT18 are also seen at $x$ above $10^{-3}$.

\begin{figure}
    \centering
    \includegraphics[width=0.49\textwidth]{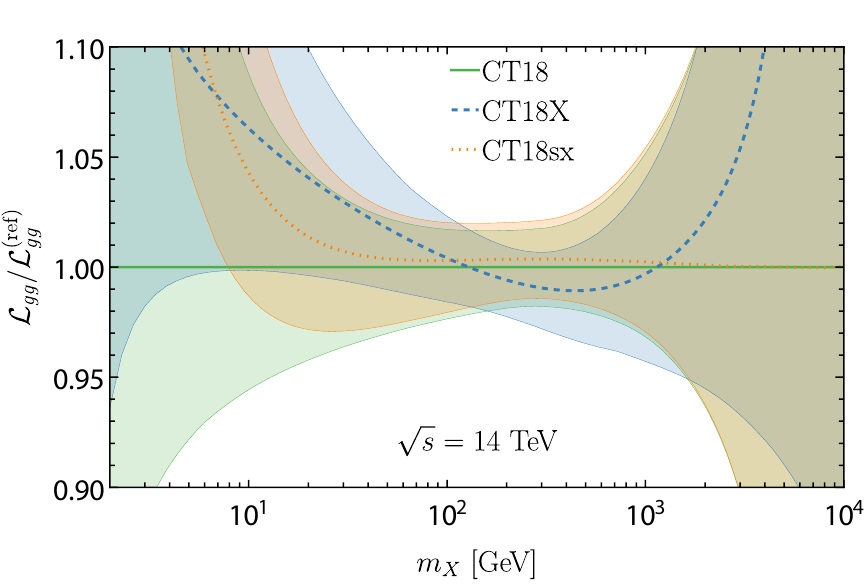}
    \includegraphics[width=0.49\textwidth]{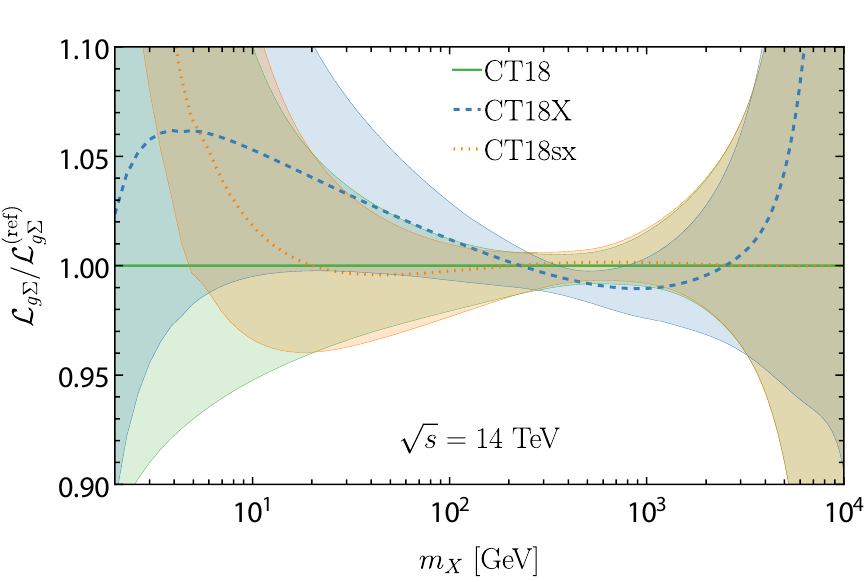}
    \includegraphics[width=0.49\textwidth]{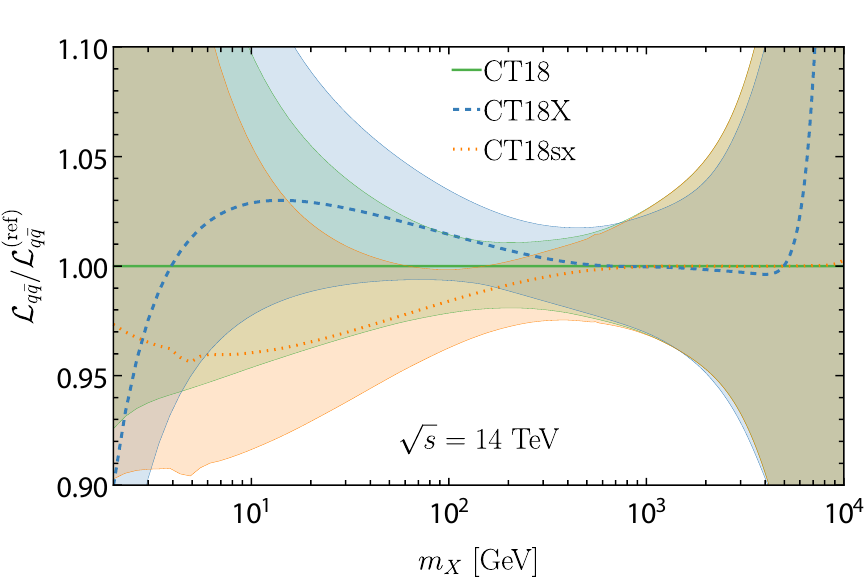}
    \includegraphics[width=0.49\textwidth]{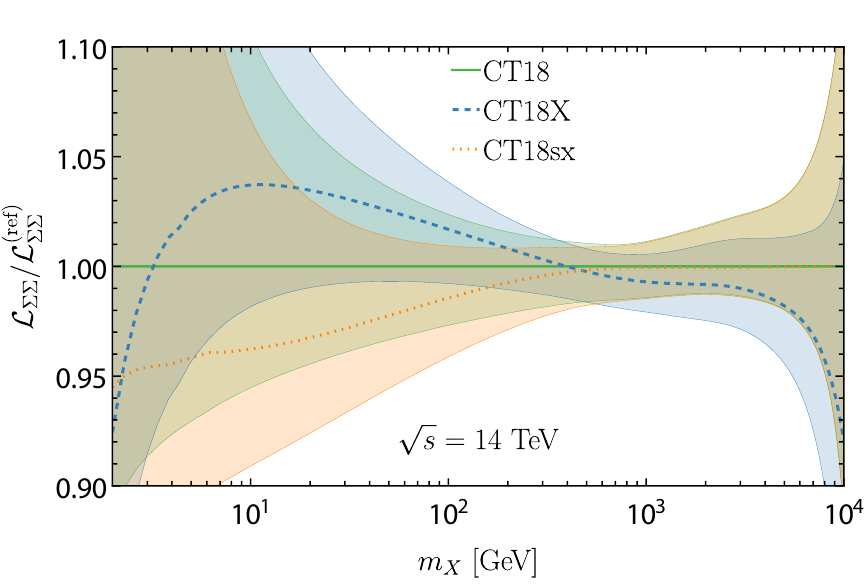}
    \caption{Similar to Fig.~\ref{fig:CT18sx}, but for parton-parton
      luminosities for the LHC 14 TeV.}
    \label{fig:lumi}
\end{figure}

The differences among the CT18, CT18X, and CT18sx propagate into
parton-parton luminosities at the LHC, which we show in
Fig.~\ref{fig:lumi} at $\sqrt{s}=14$ TeV as a function of the mass
$m_X$ of the final state. The luminosities are computed as in
\cite{Campbell:2006wx} and plotted as ratios to the CT18 ones without
restraining the rapidity of the final state. In particular, the
quark-antiquark luminosity is summed over active quark flavors,
$\mathcal{L}_{q\bar{q}}=\sum_{i}\mathcal{L}_{q_i\bar{q}_i}$, while the
flavor-singlet quark combination is defined as for
Fig.~\ref{fig:CT18sx}.  We see that the small-$x$ resummation in
CT18sx enhances the gluon-driven luminosities,
$\mathcal{L}_{gg,g\Sigma}$ and reduces the pure quark channels,
$\mathcal{L}_{q\bar{q},\Sigma\Sigma}$, for small invariant masses
$m_X\lesssim20~\GeV$, reflecting the shape of the respective PDF at
$x\sim m_X/\sqrt{s}\lesssim10^{-3}$.  In the large invariant-mass
region, the CT18sx parton luminosities naturally return to those of
CT18 due to the disappearance of the small-$x$ correction to DGLAP
evolution.  In comparison, the saturation and refitting of CT18X
enhance all luminosities at $3 \lesssim m_X \lesssim 300$
GeV. Saturation in CT18X predicts eventual suppression of quark
luminosities at $m_X \lesssim 3$ GeV.

In addition, in Ref.~\cite{Guzzi:2021fre} we found that both CT18X and
CT18sx enhance the structure function $F_2(x,Q^2)$ at small $x$ and
small $Q^2$. However, CT18sx enhances the longitudinal structure
function, $F_L(x,Q^2)$, as preferred by H1
measurements~\cite{H1:2013ktq}, while CT18X reduces $F_L(x,Q^2)$, as
preferred by results from ZEUS~\cite{ZEUS:2014thn}.

Our related work~\cite{Xie:2023suk} explored high-energy neutrino
cross sections and found that they can be enhanced by up to 20\% at
$E_\nu\!\sim\!10^{12}~\GeV$ due to the estimated contribution from the
small-$x$ resummation. Aside from this application, we have examined
these and related small-$x$ effects on heavy-quark
pair~\cite{Silvetti:2022hyc} or single-inclusive heavy-flavor
meson~\cite{Xie:2021ycd} production, which can be directly measured at
LHCb~\cite{LHCb:2013xam,LHCb:2015swx}. $D$-meson production in the
forward region, which is sensitive to the small-$x$ gluon and charm
PDFs, will provide an important source for the neutrinos measured at
FASER as well as other Forward Physics
Facilities~\cite{Anchordoqui:2021ghd,Feng:2022inv}; small-$x$ physics
will therefore be a relevant consideration to these programs as well.


\section{Parity violation and the high-$x$ sea \label{sec:PVsea}}

Detailed knowledge of the nucleon's sea PDFs at $x>0.3$ has been
elusive, yet it becomes increasingly relevant for high-luminosity
studies at large invariant masses at the LHC.
Poor knowledge of high-$x$ sea PDFs is attributable to their rapid
falloff relatively to the dominant valence PDFs as well as the subtle
behavior of flavor-symmetry breaking in the proton.
Unraveling the sea flavor dependence raises questions about the sign
and shape of $\bar{d}\!-\!\bar{u}$, relative magnitude of
$s\!+\!\bar{s}$, and possible signatures for $s\!-\!\bar{s}\!\neq\!0$
and its associated $x$-dependent shape. All these PDF features have
garnered attention recently, as they carry valuable information on how
the structure-forming properties of QCD break the
$\bar{u}\!=\!\bar{d}\!=\!s\!=\!\bar{s}$ flavor symmetry characteristic
of the strong interaction in the high-energy limit.
In addition, the high-$x$ behavior of the sea PDFs is pertinent to new
physics searches, as the SM backgrounds in a variety of searches for
new physics, including the forward-backward asymmetry $A_\mathrm{FB}$
of lepton pair production, can have significant sea-PDF
dependence~\cite{Fiaschi:2022wgl,Fu:2023rrs}.

In this context, parity-violating lepton scattering has been
highlighted as an additional process to unravel the flavor structure
of the nucleon sea at high $x$~\cite{Hobbs:2008mm,Brady:2011uy}. This
approach proceeds on the logic that the $\gamma Z$ interference
mechanism that dominates the parity-violating asymmetry,
$A^\mathrm{PV}$, effectively selects specific parton-level flavor and
charge currents in the target nucleon that are complementary to those
probed by inclusive DIS and other processes. Moreover, $A^\mathrm{PV}$
measured at energies below the electroweak scale might conceivably
augment information from, {\it e.g.}, collider or fixed-target
Drell-Yan process. In the latter category, this includes the
$\sigma^{pd}/(2 \sigma^{pp})$ deuteron-to-proton ratios measured by
the E866~\cite{NuSea:2001idv} and E906~\cite{SeaQuest:2021zxb}
experiments.

\begin{figure}[t!]
    \begin{center}  
        \includegraphics[width=0.47\textwidth,valign=t]{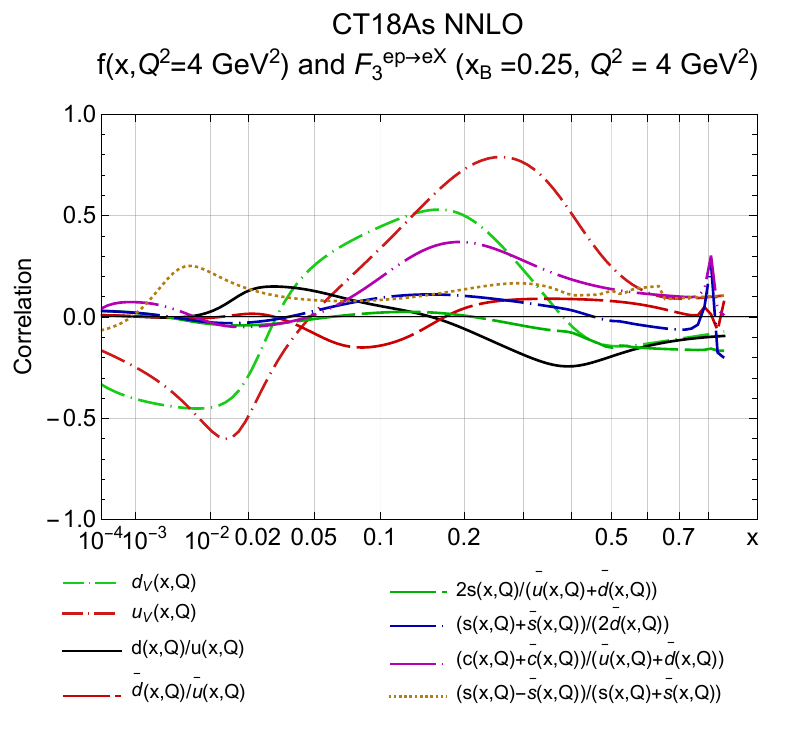}
        \includegraphics[width=0.51\textwidth,valign=t]{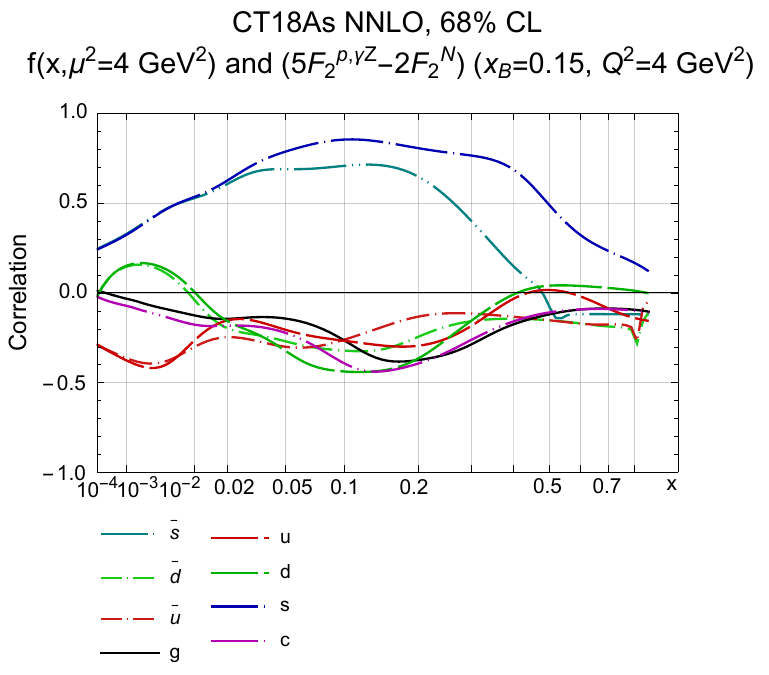}\\ (a)\hspace{0.5\textwidth}(b)\\ \quad\\
        \hspace{0.04\textwidth}\includegraphics[width=0.4\textwidth,valign=t]{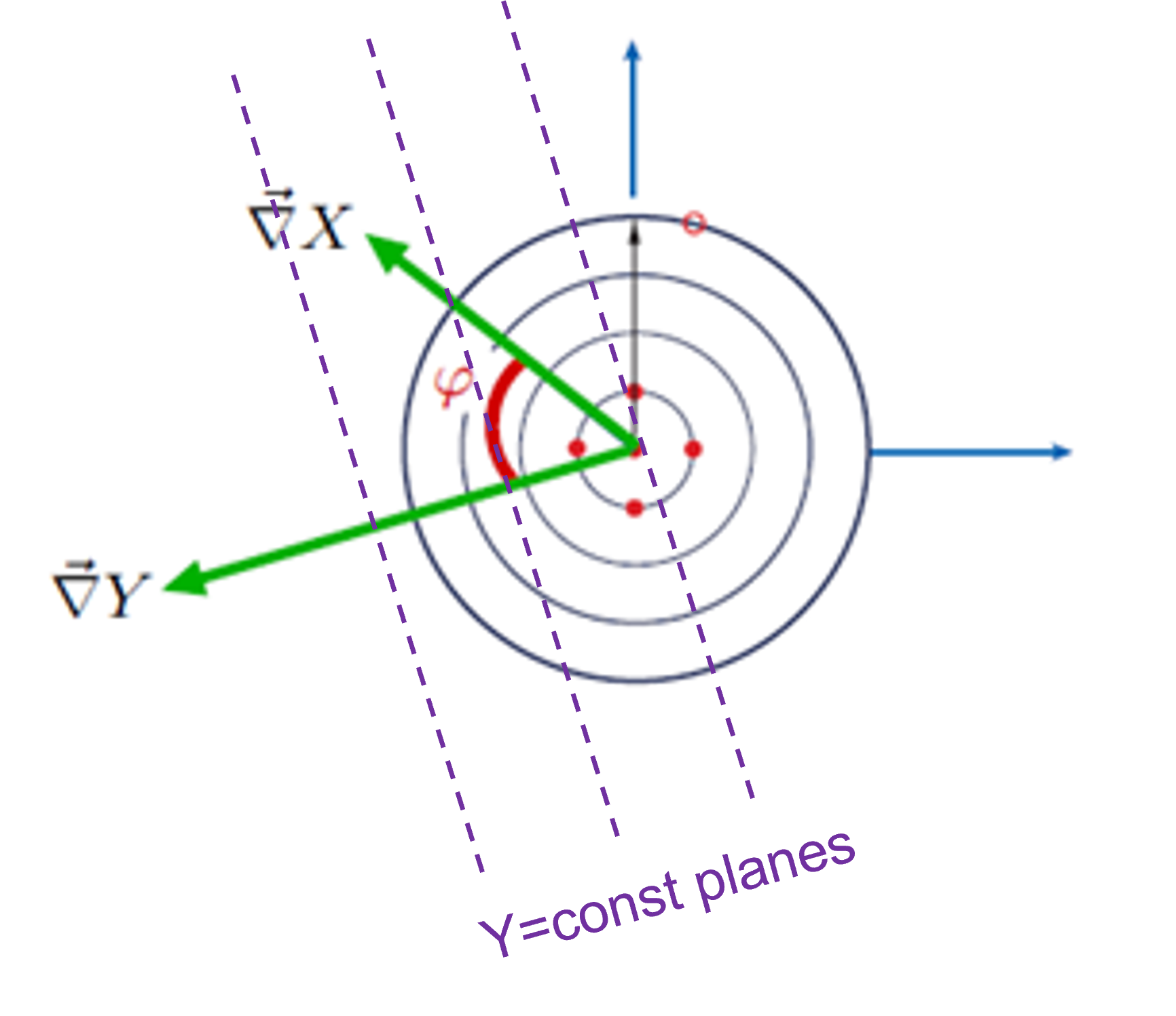}\hspace{0.04\textwidth}
        \includegraphics[width=0.51\textwidth,valign=t]{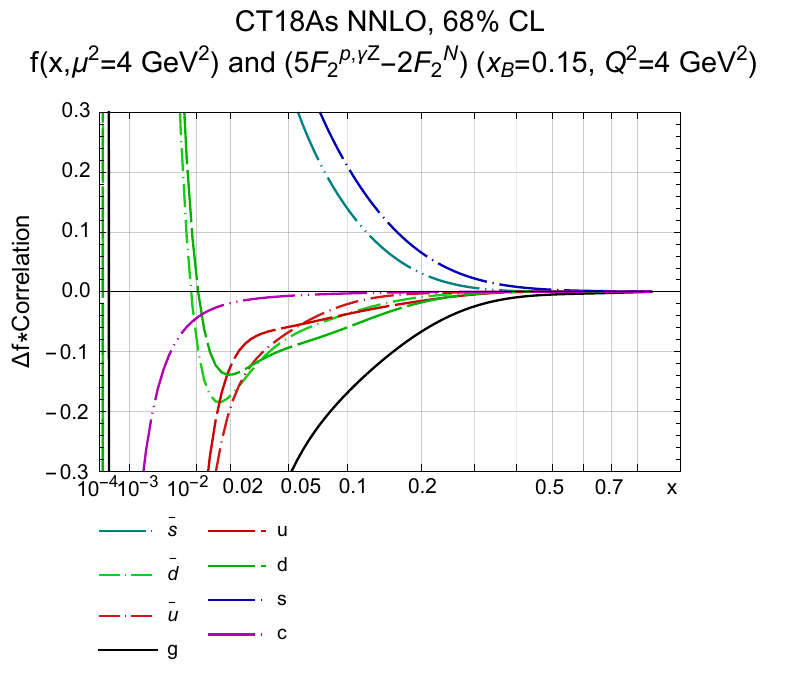}\\ (c)\hspace{0.5\textwidth}(d)
    \end{center}
    \caption{ When combined with other high-energy observables in a
      QCD global fit, parity-violating lepton scattering may access
      unique quark charge-flavor currents to help disentangle the
      flavor dependence of the nuclear sea at high $x$.  In the upper
      row, we plot the Hessian correlations of (a) several PDF flavor
      combinations with the parity-violating structure function
      $F_3(x_B,Q)$ and (b) the individual PDF flavors and a
      combination $5 F^{\gamma Z, p}_2(x_B,Q) - 2 F^N_2(x_B,Q)$ with a
      potential sensitivity to strangeness.  (c) We illustrate how the
      PDF-mediated sensitivity of $Y$ to $X$ can be estimated using
      the gradients computed with the Hessian error PDF sets. Using
      this technique, panel (d) estimates the sensitivity of
      constraints on PDFs $f(x,\mu^2 = 4\mbox{ GeV}^2)$ with indicated
      flavors to the combination $5 F^{\gamma Z, p}_2(x_B,Q) - 2
      F^N_2(x_B,Q)$ at $x_B=0.15$, $Q^2=4\mbox{ GeV}^2$ considered in
      (b).}
    \label{fig:PV}
\end{figure}

In a contribution to a recent whitepaper~\cite{Accardi:2023chb},
members of CT examined the potential for parity-violating electron
scattering at a 22~GeV upgrade of CEBAF at JLab to enhance knowledge
of the high-$x$ sea PDFs. In the absence of comprehensive pseudodata
with realistic projections for statistical and systematic
uncertainties, we explored the potential sensitivity through Hessian
correlations~\cite{Nadolsky:2008zw} of the PDFs with a number of
parity-violating structure function combinations entering
$A^\mathrm{PV}$.  The upper row in Fig.~\ref{fig:PV} plots examples of
such correlations explored at typical $x_B$ and $Q$ values accessible
at the JLab 22 GeV facility.  In particular, information on the
high-$x$ behavior of the $F_3$ structure function at modest $Q^2$
might increase accuracy of the $q\!-\!\bar{q}$ PDF combinations,
especially $u_V(x,Q)$, as shown in the upper left subfigure. When
taken together with information sensitive to the $u$- and $d$-PDFs,
such data may constrain the high-$x$ behavior of $\bar{u}$ and
$\bar{d}$.

In addition, the parity-violating $F^{\gamma Z, p}_2$ structure
function on the proton, when combined with inclusive scattering from
deuterium, could allow combinations like $5 F^{\gamma Z, p}_2 - 2
F^N_2$ to be extracted, for which we plot the PDF correlations in the
upper right subfigure of Fig.~\ref{fig:PV}. In this case, we observe
strong correlations with the independently parameterized $s$ and
$\bar{s}$ PDFs of the CT18As fit~\cite{Hou:2022onq}, confirming the
judgment \cite{SLACProposalE149bis} that such data could be
informative with respect to nucleon strangeness.

Crucially, achieving strong constraints on $s$ and $\bar s$ requires
good control over deuteron-structure
corrections~\cite{Accardi:2021ysh} to obtain the isoscalar nucleon
structure function, $F^N_2=(F_2^p+F_2^n)/2$. The ultimate PDF
constraining power of $A^\mathrm{PV}$ will closely depend on
experimental uncertainties. These observations can be refined using
the available techniques for the statistical exploration, such as the
Hessian PDF sensitivity \cite{Wang:2018heo,Hobbs:2019gob} already
discussed in Sec.~\ref{sec:L2}.

Working at the leading order in QCD and assuming the weak angle
satisfying $\sin^2\theta_w=1/4$, one finds \cite{Dalton:2023}
\begin{equation}
    3\left(5 F^{\gamma Z, p}_2(x,Q) - 2 F^N_2(x,Q)\right) =
    xs(x,Q)+x\bar s(x,Q), \label{ssFs}
\end{equation}
suggesting that the net strangeness can be determined from a
combination of the structure functions accessible in parity-conserving
DIS on deuterium and parity-violating DIS on proton. The relation
supports a strong Hessian correlation seen in Fig.~\ref{fig:PV}
between the left- and right-hand sides of Eq.~(\ref{ssFs}) obtained
for CT18As NNLO PDFs.

The connection gets more interesting if one takes the more precise
$\sin^2\theta_w=0.23121$ and adds NLO contributions to
Eq.~(\ref{ssFs}) to obtain
\begin{equation}
     3\left(5 F^{\gamma Z, p}_2 - 2 F^N_2\right) \approx 1.063
     \left(xs+x\bar s\right) + 0.25 \left(xu+x\bar u\right) + 0.063
     \left(xd+x\bar d\right) + \alpha_s(Q) F(q,g), \label{ssFs2}
\end{equation}
where $F(q,g)$ arises from higher-order contributions dependent both
on quark and gluon PDFs. We see that the PDF dependence of the
structure combination arises from the strangeness (anti)quark PDFs,
which are strongly correlated with the l.h.s.~yet are small, as well
as from larger PDFs of other flavors ($u, d, \dots$) that enter
Eq.~(\ref{ssFs2}) either directly with small numerical factors or
through the NLO contribution $F(q,g)$.
Ultimately, in the full NNLO calculation, we see in Fig.~\ref{fig:PV}
(b) that the Hessian correlation with these latter PDFs is weakly
negative, implying a minor anti-correlation.

To elucidate this interdependence in the case of the CT18As error PDF
set, we compute the sensitivity $S_{Y}(X)$, which estimates the change
$\delta Y(X)$ in the PDF $Y\equiv q,\ \bar q$, or $g$ on the
right-hand side of Eq.~(\ref{ssFs2}) when $X\equiv 5 F^{\gamma Z, p}_2
- 2 F^N_2$ increases by a 1$\sigma$ interval of the PDF uncertainty
from its central value. In the Hessian formalism, sensitivity
$S_{Y}(X)$ is a product of the PDF uncertainty $\Delta Y$ and
correlation $\cos\varphi$ between $X$ and $Y$:
\begin{equation}
    S_{Y}(X) = \vec\nabla Y \cdot \frac{\vec\nabla X}{\left|\vec\nabla
      X\right|} = \Delta Y \cos \varphi,
\end{equation}
as follows from the geometric picture depicted in
Fig.~\ref{fig:PV}(c). In Fig.~\ref{fig:PV}(d), we plot the
sensitivities of PDFs $Y=f_a(x,Q)$ for the specified flavors, at
$\mu=2$ GeV and $x$ shown on the horizontal axis, to the structure
function combination $X$ at $x_B=0.15$ and $Q=2$ GeV. We see that, for
the CT18As NNLO PDF set, an upward $1\sigma$ variation of $X$ results
in the increase of both $s$ and $\bar s$ across the whole $x$ range
and simultaneously in the compensating decreases of gluon, $u$ and
$\bar u$, and charm PDFs. It also leads to the decrease of down-type
(anti)quark PDFs at $x>0.01$ and their fast increase at $x<0.01$. The
absolute magnitudes of variations are comparable among the flavors.
We therefore conclude that, while $5 F^{\gamma Z, p}_2 - 2 F^N_2$ is
most strongly correlated with the small $s+\bar s$ according to
Fig.~\ref{fig:PV}(b), the net PDF sensitivity of this combination also
depends on the weakly anti-correlated, yet numerically significant,
other PDFs. Extraction of the (anti)strangeness PDF from PVDIS
requires simultaneous precision determination of the other
flavors. These estimates extend the insights in the whitepaper
\cite{Accardi:2023chb} and estimate quantitatively how a subtle
interplay among parity-violating observables, Drell-Yan data on
$\sigma^{pd}/(2 \sigma^{pp})$, and collider measurements of quantities
like $A_\mathrm{FB}$ may advance knowledge of the proton's high-$x$
sea in future experiments.


\section{CT18FC and nonperturbative charm in the proton \label{sec:FC}}

Control over the production and dynamics of massive quarks is
essential to the understanding of QCD. This applies to global analyses
of nucleon PDFs, for which the inclusion of heavy-quark mass effects
at NNLO has been shown to reduce perturbative scale uncertainties and
stabilize PDF extractions. (For a recent example, see
Refs.~\cite{Gao:2021fle,Guzzi:2011ew}, which explored an NNLO
general-mass prescription for charged and neutral-current DIS using
the CT default S-ACOT-$\chi$ scheme.) Such effects include the
contributions of heavy quarks in the running of $\alpha_s$,
perturbative parton-level matrix elements and DGLAP kernels, and lead
to the generation of heavy-quark PDFs through radiative QCD
processes. In addition, it was hypothesized shortly after the
formulation of QCD that heavy quarks --- particularly charm --- might
contribute to the proton's structure nonperturbatively near the
$Q_0\!\sim\!1.3$ GeV boundary of DGLAP evolution. This scenario has
usually been called {\it intrinsic charm}~\cite{Brodsky:1980pb}, and
an array of models have been proposed to describe the $x$ dependence
of the nonperturbative PDF. Problematically, the specific scale at
which these models should be mapped onto the factorization-based
formalism underlying PDF fits is not clear. As such, a number of PDF
analyses have attempted to determine a {\it fitted charm} (FC)
parametrization based on available high-energy data.

Nonperturbative charm has attracted interest in the last few years due
to the collection of potentially sensitive data at Run-1 and 2 at the
LHC. The FC studies included
Refs.~\cite{Pumplin:2007wg,Jimenez-Delgado:2014zga} as well as a
comprehensive CT14 IC analysis~\cite{Hou:2017khm} that addressed both
the theoretical and phenomenological aspects of the IC.

More recently, CT revisited fitted charm in light of the Run-1 LHC
data included in CT18, leading to the CT18FC analysis,
Ref.~\cite{Guzzi:2022rca}. Based on a systematic analysis of the
underlying QCD factorization framework and comprehensive sampling of
the various contributing PDF uncertainties, this work addressed, and
did not corroborate, the claim for a statistically significant
evidence of the intrinsic charm by the NNPDF
group~\cite{Ball:2022qks}. On the theory side, theoretical
interpretation of fitted charm parametrization in terms of intrinsic
charm of the nonperturbative models is not direct.  On the analysis
side, models with various magnitudes of the FC, including no fitted
charm at all, agree well with the global QCD data.  A separate study
\cite{Courtoy:2022ocu} showed, using the NNPDF fitting framework
itself, that the NNPDF conclusions about the IC/FC are strongly tied
to the specific methodology they use, notably assumptions about the
allowed shapes of gluon and other PDFs.  
\rev{Specifically, Ref.~\cite{Courtoy:2022ocu} emphasized that priors may have a strong impact on any type of the PDF ensemble, including MC/NN ones, and it provided quantitative estimates in the case of the NNPDF4.0 NNLO uncertainties using ``hopscotch'' PDF solutions constructed from the NNPDF Hessian eigenvector sets. The spread of hopscotch solutions reflects both experimental uncertainties and variations in methodology (e.g., sampling over hyperparameters). It is therefore consistent with an ensemble of PDFs that is larger than the NNPDF4.0 default ensemble generated by importance sampling of exclusively experimental errors for a fixed methodology. Some hopscotch solutions were found to be consistent with zero intrinsic charm, while having practically the same $\chi^2$ as the NNPDF4.0 central replica. According to the fundamental likelihood ratio test, such zero-IC solutions with the same $\chi^2$ must be included on the same footing as the NNPDF4.0 central replica, unless they are excluded by prior conditions such as integrability, positivity, and smoothness. If too stringent, the prior conditions may overconstrain the space of PDF solutions.  
}

\rev{
The central findings of the hopscotch analysis remain robust despite the rebuttal \cite{Ball:2022uon} from NNPDF authors. The arXiv version of Ref. \cite{Courtoy:2022ocu} includes responses to the key critiques raised by NNPDF.
In particular, most of 2330 hopscotch solutions available 
at \cite{HopscotchWebsite} are sufficiently smooth according to CTEQ-TEA criteria and mostly fall within NNPDF4.0 uncertainty bands. Only a fraction of these solutions was examined in \cite{Ball:2022uon} and declared to be overfitted upon failing the NNPDF smoothness test.}
An analysis
\cite{Harland-Lang:2024kvt} dedicated to closure testing of the MSHT
PDFs and comparisons against the NN approach arrives at a similar
conclusion that the NNPDF sensitivity to IC is strongly influenced by
the prior constraints, such as the positivity of the small-$x$ gluon.

On the experimental side, the strength of the preference for the
fitted charm is dependent on several factors.  When fitting the EMC
charm SIDIS data \cite{EuropeanMuon:1982xfn}, the quality of the fit
depends at least as much on the magnitude of the large-$x$ gluon as on
the magnitude of FC. With a harder gluon preferred by the CT and MSHT
fits, the EMC observation has little room for the FC or disfavors the
FC.  In another possible channel, $Z+c$ production at
LHCb~\cite{LHCb:2021stx}, the insufficiently controlled final-state
radiation strongly influences the possible room for the FC component
even at NNLO \cite{Hou:2017khm,Guzzi:2022rca}.  All these observations
consistently support the ambivalence of the existing data toward the
magnitude of the FC.

\begin{figure}
    \centering
    \includegraphics[width=0.4\textwidth]{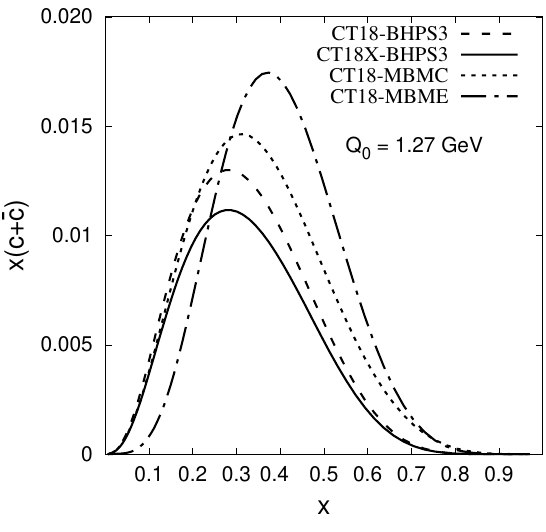}
    \includegraphics[width=0.54\textwidth]{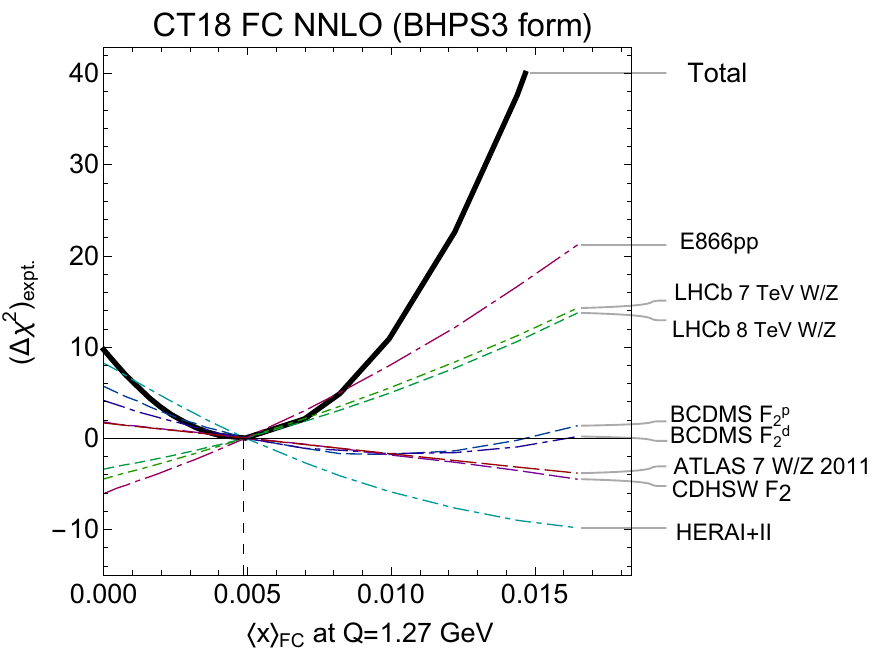}
    \caption{ From the recent CT18FC study~\cite{Guzzi:2022rca}, we
      plot the range of fitted charm input shapes at the evolution
      starting scale (left) as well as a $\Delta \chi^2$ profile
      (right) for one of these scenarios (CT18-BHPS3). While the fit
      has a shallow preference for $\langle x \rangle_\mathrm{FC}
      \equiv \langle x \rangle_{c^+} \sim 0.5\%$, this does not rise
      to a statistically significant signal according to default CT
      uncertainty conventions. Opposing pulls of fitted experiments
      are partly responsible for this weak sensitivity, as can be seen
      in the distinct $\langle x \rangle_\mathrm{FC}$ dependence of
      $(\Delta \chi^2)_\mathrm{expt.}$ of select experiments.  }
    \label{fig:charm}
\end{figure}

The final FC PDFs from the CT18FC study were released as a series of
parameterization scenarios for the high-$x$ FC, with a range of $x$
behaviors illustrated in Fig.~\ref{fig:charm} (left) at the initial
scale $Q_0\!\sim\!1.27$ GeV. The normalizations of the resulting FC
PDFs were then systematically varied over a broad range corresponding
to $\langle x \rangle_{c^+} = \int^1_0 dx x [c +\bar{c}](x) \in [0,
  0.02]$; as these normalizations were varied, the shape parameters
associated with the other PDF flavors were continuously refitted.

In this manner, CT18FC determined the sensitivity of the default CT
data set to fitted charm, finding a slight, $\Delta \chi^2\! \sim\!
10$-unit improvement in the fit for $\langle x \rangle_{c^+} \sim
0.5\%$ relative to a ``no FC'' scenario. Crucially, the depth of this
reduction in $\chi^2$, as shown in Fig.~\ref{fig:charm} (right), is
very shallow, and hence it can be concluded that the present data
comprising the CT baseline do not provide a statistically significant
signal for FC, in contrast to the findings in
Ref.~\cite{Ball:2022qks}. As another new advancement, for the first
time in the global QCD analysis some of the CT18FC scenarios included
the possible nonperturbative breaking of charm-anticharm symmetry,
$c_{-} =c-\bar{c}\neq 0$~\cite{Hobbs:2013bia} at the initial scale
$Q_0$. The data are even more ambiguous for the magnitude of $c_-$,
not corroborating the finding of evidence for $c_-(x,Q_0) \neq 0$
stated in a later study~\cite{NNPDF:2023tyk} based on the specific
NNPDF methodology.

For the outlined reasons, further theoretical advancements and data
from future LHC measurements, the EIC, or lattice-gauge theory could
be instrumental for distinguishing FC from other dynamical
contributions, assuming it has a phenomenologically significant
magnitude.


\section{The photon content of a neutron \label{sec:neutron}}

\begin{figure}
    \centering
    \includegraphics[width=0.49\textwidth]{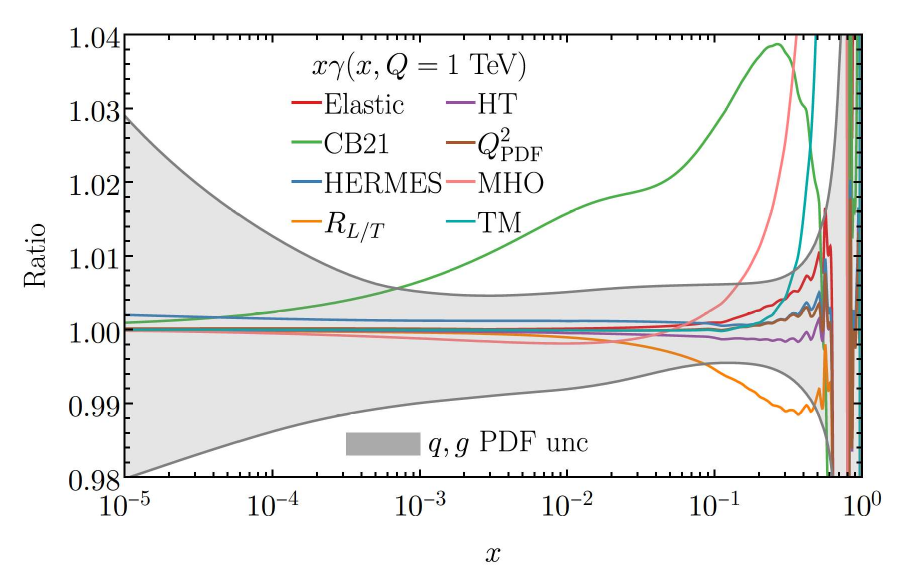}
    \includegraphics[width=0.49\textwidth]{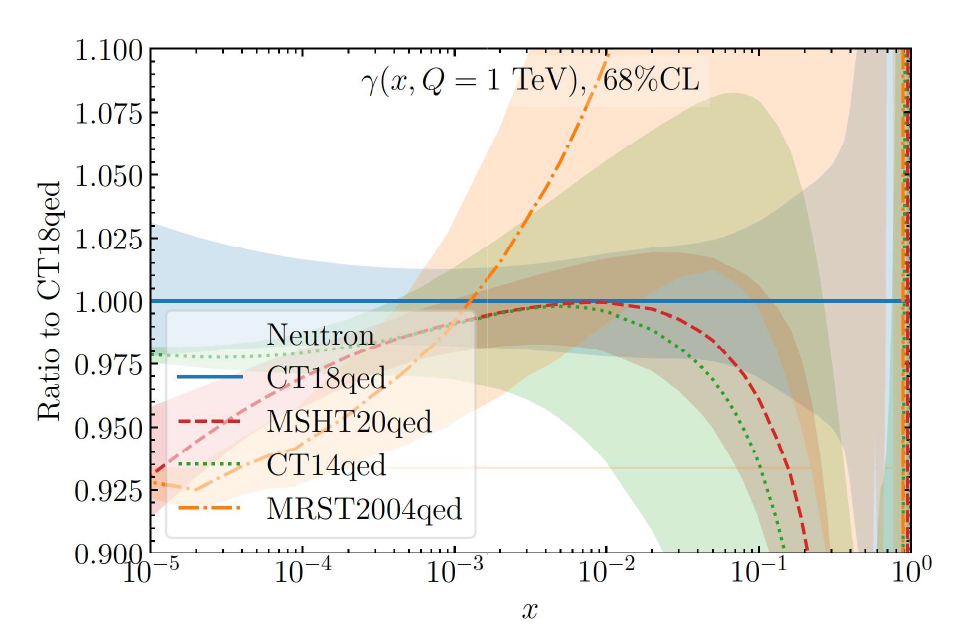}
    \caption{Left: Components of the uncertainty of the neutron's
      photon PDF at $Q=1~\TeV$ in the CT18qed
      framework~\cite{Xie:2023qbn}, arising from variations of quark
      and gluon PDFs, as well as low-$Q^2$ nonperturbative effects.
      Right: Comparison of the neutron's photon PDFs at $Q=1~\TeV$ as
      obtained in CT18qed~\cite{Xie:2023qbn},
      MSHT20qed~\cite{Cridge:2021pxm}, CT14qed~\cite{Schmidt:2015zda}
      and MRST2004qed~\cite{Martin:2004dh} NNLO analyses.}
    \label{fig:neutronQEDPDF}
\end{figure}
Precision predictions of QED effects in high-energy scattering off
nuclear targets used in the global fits require to know photon PDFs in
both protons and neutrons, as the two need not to be exactly the same.
Consideration of the photon component of the nucleon modifies PDFs or
quarks and gluons as well to the degree that may be appreciable at
N3LO QCD accuracy of the calculations.

Recently, as an extension of the CT18QED analysis of the photon PDF in
a proton~\cite{Xie:2021equ}, we have released an analogous study for
the neutron's photon content~\cite{Xie:2023qbn}. By leveraging the
LUXqed formalism~\cite{Manohar:2016nzj,Manohar:2017eqh}, the photon
PDF of a given hadron can be derived from DIS structure functions,
which in turn can be either directly measured in scattering
experiments or predicted using parametrizations of (anti)quark and
gluon PDFs in the QCD factorization framework.  In the CT18qed
approach~\cite{Xie:2021equ}, the photon PDF is initialized within the
LUXqed formalism at the starting scale \emph{e.g.} $Q=1.3~\GeV$ and
self-consistently determined at higher $Q$ via DGLAP evolution up to
NNLO QCD and NLO QED accuracy.

Fig.~\ref{fig:neutronQEDPDF} (left) illustrates that several sources
of uncertainty in the CT18qed photon PDF in the neutron at $Q=1~\TeV$
arise from variations of CT18 quark and gluon PDFs~\cite{Hou:2019efy}
as well as low-$Q^2$ factors. Similar plots of the photon PDF and
breakdowns of its uncertainties at other representative scales
($Q=1.3,10,100~\GeV$) can be found in Ref.~\cite{Xie:2023qbn}. The
low-$Q^2$ nonperturbative uncertainties include those from the elastic
electromagnetic form factors $G_{E,M}(Q^2)$
(Elastic)~\cite{Ye:2017gyb}; the structure functions determined by
CLAS~\cite{CLAS:2003iiq} and Christy-Bosted fits
(CB21)~\cite{Christy:2007ve,Bosted:2007xd} in the resonance region and
by the HERMES fit~\cite{HERMES:2011yno} in the low-$Q^2$ continuum
region; the longitudinal-transverse cross-section ratio,
$R_{L/T}$~\cite{E143:1998nvx}; higher-twist
(HT)~\cite{Accardi:2016qay,Abt:2016vjh} and target-mass (TM)
corrections; the matching scale $Q^2_{\rm PDF}$; and missing
higher-order (MHO) effects.  It turns out that the dominant
uncertainties come from low-$Q^2$ nonperturbative variations with the
associated effects potentially as large as 10\% at the starting scale
$Q_0$~\cite{Xie:2023qbn} and gradually decreasing at higher scales,
\emph{e.g.}, to 4\% at $Q=1~\TeV$ in the $x$ region of interest, as
shown in Fig.~\ref{fig:neutronQEDPDF} (left). In comparison, the
perturbative uncertainty induced by the quark-gluon PDFs only
increases slightly, while remaining below 3\% even up to $Q=1~\TeV$.

Fig.~\ref{fig:neutronQEDPDF} (right) compares the neutron's photon
PDFs at $Q=1~\TeV$ among CT18qed~\cite{Xie:2023qbn},
MSHT20qed~\cite{Cridge:2021pxm}, CT14qed~\cite{Schmidt:2015zda}, and
MRST2004qed~\cite{Martin:2004dh} frameworks. We see that the photon
PDFs of the second generation, CT18qed and MSHT20qed, agree very well
within the intermediate range $10^{-4}\lesssim x\lesssim0.1$.  In the
low-$x$ region $(x<10^{-4})$, the MSHT20qed gives a slightly smaller
photon, mainly driven by its smaller charge-weighted singlet PDF,
$\Sigma_e=\sum_{i}e_i^2(q_i+\bar{q}_i)$. In the large-$x$ region
$(x>0.1)$, the MSHT20qed also renders a smaller photon due to a
different treatment of the initial scale $Q_0$.  Similarly to the
MMHT2015qed~\cite{Harland-Lang:2019pla}, MSHT20qed takes the starting
scale $Q_0=1~\GeV$, as well as $Q_0^2$ rather than $Q_0^2/(1-z)$ in
the LUXqed formalism to determine the initialized photon, with both
choices reducing the photon PDF when $x>0.1$. In this comparison, we
do not show the NNPDF2.3qed photon PDF in the
neutron~\cite{Ball:2013hta}: it is substantially different from the
shown parametrizations and can be viewed in comparison figures in
Ref.~\cite{Xie:2023qbn}.

Intriguingly, the magnitude of the photon PDF and resulting
redistribution of the nucleon momentum among all partons depends on
the adopted treatment of low-energy dynamics and may even affect the
gluon PDF at the level that may be noticeable in N3LO computations for
Higgs boson production. This aspect is now being further investigated.
With the new CT18qed PDFs in the neutron, phenomenological predictions
in processes like $W$-boson production in neutrino-nucleus scattering
as well as the search of a axion-like particle (ALP) in a muon
beam-dump experiment show significant improvement due to reduction in
the photon PDF uncertainty and proper error estimation in the
large-$x$ region \cite{Xie:2023qbn}.


\section{Simultaneous SMEFT-PDF analyses \label{sec:SMEFT}}

\begin{figure}[tb]
\centering
\raisebox{0.35cm}{\includegraphics[width=0.49\textwidth,clip]{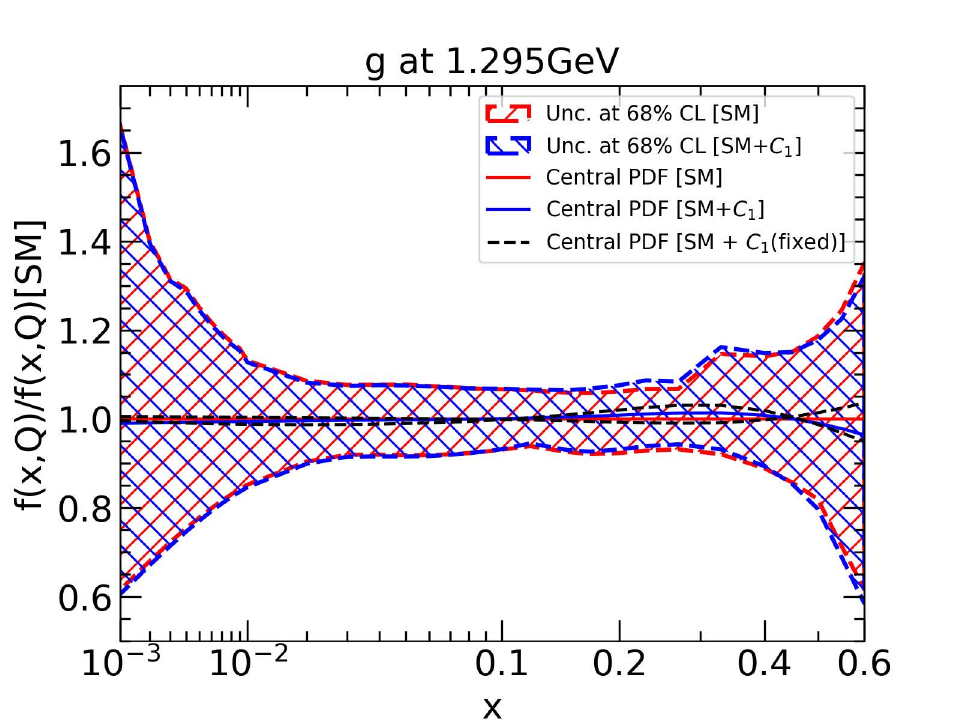}}
\includegraphics[width=0.49\textwidth,clip]{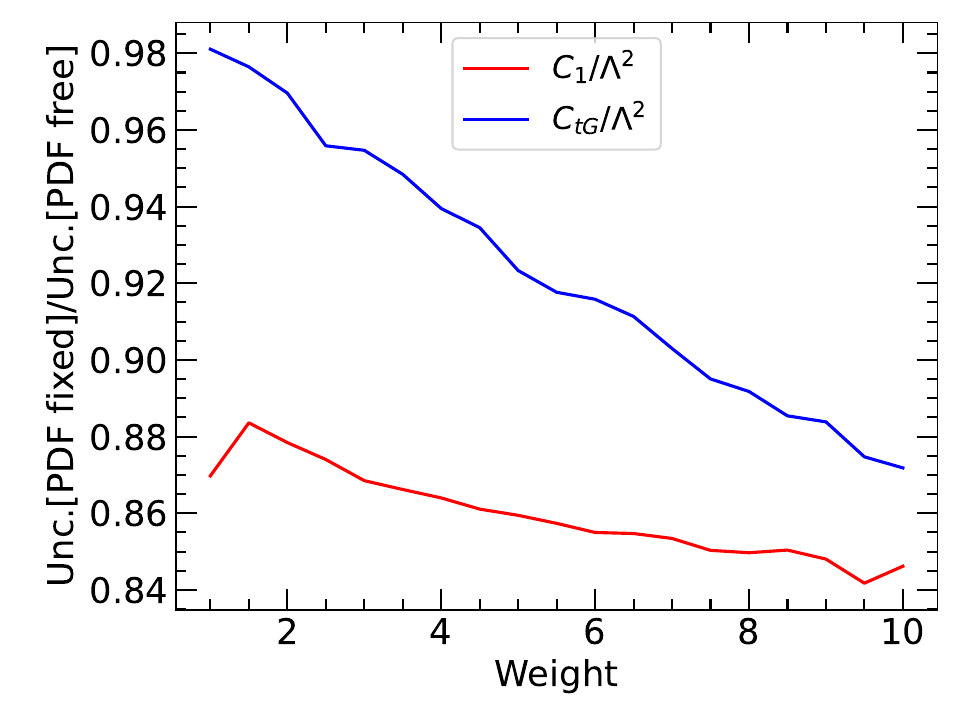}
\caption{ (Left) The fitted gluon PDF when jointly varied with the
  SMEFT $C_1$ contact-interaction coefficient (blue curve and band),
  relative to the corresponding SM-only fit (red curve and band)
  corresponding to $C_1 \!=\! 0$. The black-dashed curves indicate the
  central gluon PDF obtained in fits which fixed the $C_1$
  contact-interaction coefficient to its extreme allowed values at
  90\% C.L.
	(Right) While the indicated SMEFT-PDF correlations are mild,
  higher experimental precision in jet and $t\bar{t}$ data, simulated
  through increased weights for these data sets, cause larger
  underestimates in the $C^{(6)}_i$ uncertainties if these are not
  jointly fitted with PDFs.  }
\label{fig:eft}
\end{figure}
Searches for interactions beyond the Standard Model (BSM) have
historically proceeded through the formulation of specific theoretical
extensions that predicted experimental signatures amenable for
targeted testing against high-energy experiments.  More recently, the
lack of unambiguous signals for BSM physics at the LHC has motivated
the deployment of model-independent methods to identify subtle
deviations from SM predictions which may potentially lurk in the
global data set. These methods have often taken the form of effective
field theories (EFTs)~\cite{Wells:2015uba}, with a particularly
widely-adopted framework being SMEFT. The central assumption of SMEFT
is that the BSM takes the form of novel interactions entering at some
typical energy scale $\Lambda$ exceeding the reach of modern
experiments. At accessible energies, once heavy modes mediating these
interactions are integrated away, one is left with a basis of
higher-dimensional operators quantifying new pointlike interactions of
SM particles, in addition to the dimension-4 operators characterizing
the SM interactions. This basis allows a complete parametrization of
possible BSM signatures up to power-suppressed operators of higher
dimensionality (typically, up to dimension 6~\cite{Grzadkowski:2010es}
at the few-TeV scales).

Global fits of PDFs provide a traditional avenue for model-independent
searches of BSM physics involving QCD interactions. Recent proposals
aim to substantially expand their scope by adding new types of
measurements to constrain SMEFT parameters.  Taking the searches for
quark compositeness as an example, a resonant heavy state with mass
$\Lambda$ can be produced in the central rapidity region when both
initial-state partonic momentum fractions are of order
$\Lambda/\sqrt{s}$. However, in inclusive jet production, a contact
interaction may also enhance the rates of jets with highest $p_T$ in
the forward direction, where the same high scale $\Lambda$ can be
accessed in spacelike virtual exchanges as well as through
interference with the large QCD background \cite{Stump:2003yu}. Such
enhancements must be distinguished from uncertainties in sea PDFs at
high $x$, where the experimental constraints are still relatively
weak. However, the same PDFs affect a large range of $p_T$ and
rapidities, and hence it may be possible to pick out the BSM signals
from the overall PDF dependence by fitting diverse data sets over a
broad kinematic range.

This vision motivates the PDF-SMEFT program, which has enjoyed
significant development in the last several years, while running into
a potential circularity: the hadronic data used to constrain SMEFT
parametrizations are often independently analyzed in PDF fits,
suggesting the possibility that BSM signatures might be absorbed into
PDF fitting parameters and {\it vice
  versa}~\cite{Carrazza:2019sec,Hammou:2023heg}. A systematic approach
to resolving this circularity is to fit PDFs and the Wilson
coefficients of the SMEFT operators {\it simultaneously}.
Neglecting the interplay of PDF and EFT usually leads to a significant
overestimate of the EFT constraints.
A recent study \cite{Gao:2022srd} performed such an analysis within
the CT framework, specifically concentrating on a subset of jet and
$t\bar{t}$ production experiments.

This analysis employed a new method~\cite{Liu:2022plj} to encapsulate
the CT18 baseline PDFs in feed-forward neural networks to allow rapid
calculations of $\chi^2$ as the underlying PDF shapes were varied. On
this basis, SMEFT theoretical predictions could be added to SM
calculations for jet and $t\bar{t}$ processes. The resulting framework
then allowed efficient scans over the PDF shapes, SM input parameters
($\alpha_s$ and $m_c$), and the Wilson coefficients $C^{(6)}_i$ of key
SMEFT operator combinations at dimension 6. We direct interested
readers to Ref.~\cite{Gao:2022srd} for details, including an
exploration of the dependence of SM input parameters, an investigation
of the effect of linear and quadratic SMEFT contributions to $\chi^2$
limits, and studies of the interplay among various jet and $t\bar{t}$
data sets.

Ultimately, this study found only very minimal correlation between the
fitted PDFs and top- and jet-associated SMEFT coefficients. Though
mild, nonzero correlations were most pronounced for the gluon PDF at
high $x$, as shown in the left panel of Fig.~\ref{fig:eft}. Here, we
show the fitted gluon PDF under several scenarios in which SMEFT
coefficients are either simultaneously varied alongside the PDFs (blue
curves) or extracted under a SM-only assumption (red curves).  The
black curves are obtained by fitting PDFs with a nonzero $C_1$ SMEFT
coefficient fixed at the boundaries of its 90\% C.L.
Altogether, these results suggest that the risk associated with
fixed-PDF SMEFT analyses of jet and $t\bar{t}$ data is relatively
mild, although this first CT study only fitted 1-2 Wilson coefficients
with the PDFs. At the same time, enhanced experimental precision is
likely to make such fits more problematic. This can be visualized in
the right panel of Fig.~\ref{fig:eft}, which plots the size of the
SMEFT uncertainties for $C_1$ and the top-associated $C_{tG}$
coefficient extracted in fixed-PDF analyses relatively to those in
joint SMEFT-PDF fits.
The plotted curves show the behavior of this ratio as uncorrelated
experimental uncertainties are reduced, as simulated through increased
weights for the relevant combinations of data sets. In particular, a
ten-fold overweighting (corresponding to a factor $\sim$3 reduction in
experimental uncertainties for the jet and top data) leads to a
$\gtrsim\!10$-$15$\% underestimate in SMEFT uncertainties if
variability in the PDFs is not taken into account.  We point out that
the method proposed in Ref.~\cite{Liu:2022plj} and employed in
Ref.~\cite{Gao:2022srd}, resembles the \texttt{ePump} approach in the
Hessian representation, in that it closely tracks the PDF shapes of
the parent QCD analysis and does not allow a fully flexible variation
of the nonperturbative function forms of the PDFs while performing the
simultaneous PDF+EFT fits. Hence, we expect that a similar PDF+EFT fit
using a global analysis code with flexibly varying nonperturbative
functional forms for the PDFs, particularly in lightly constrained
regions of $x$ ({\it e.g.}, large $x$) which lack precision data,
could lead to even larger uncertainties of the EFT coefficients.
As such, joint SMEFT-PDF fits will be a precision activity of growing
importance in the HL-LHC era and a subject of ongoing study.


\section{Conclusions \label{sec:Conclusions}}
This article highlighted the array of recent developments within the
CT framework that emerged as we enter the high-precision LHC era as
well as plan for new studies at the Jefferson Laboratory and
Electron-Ion Collider.  These activities span a broad range of HEP
phenomenology and pave the way for the next-generation release of the
CT family of parton distributions.
The novel studies summarized in the main text fall into several
categories, including novel fits of recent, post-CT18 data;
explorations of statistical aspects of PDF determination and
uncertainty quantification; and investigations of high-profile
phenomenological questions of recent interest.

First, with respect to new data, we described the inclusion of a
series of post-CT18 LHC high-energy data sets from Drell-Yan pair,
jet, and $t\bar t$ production at energies ranging from 5.02 to 13 TeV.
The new fits will also include complementary data at fixed-target
kinematics, particularly SeaQuest (E906) \cite{SeaQuest:2021zxb}, and
altogether will enable more detailed studies of the flavor and $x$
dependence of the quark sea.
We expect that the new generation of CTEQ-TEA PDFs will reflect a
nontrivial interplay between the old and new data sets, subtleties in
multiloop perturbative QCD calculations, and possibly constraints
furnished by lattice QCD on less known PDF combinations such as the
strange quark sector.

In these studies, we have explored the mutual consistency of the new
data sets, as it has implications for the magnitude of PDF
uncertainties.  Focusing on the new vector boson production
observations, we examined several post-CT18 Drell-Yan data that
complement the ATLAS 7 TeV $W$/$Z$ (ATL7WZ) data included in the CT18Z
and CT18A fits. We found that, while some LHC Drell-Yan measurements
prefer an enhanced $s$-quark PDF at $x \lesssim 0.02$, others exhibit
less of such preference. Among them, the ATL7WZ data clearly prefer a
larger $s$-quark PDF at $10^{-2} \lesssim x \lesssim 0.1$ around 0.1,
causing some tension with the dimuon SIDIS and inclusive DIS data. We
also found that ATLAS 8 TeV $W$ data (ATL8W) stands out in that it
prefers a substantially larger {$\bar d$}-antiquark PDF at $x \sim
10^{-3}$.

We have also investigated the relationship between 13 TeV $t\bar{t}$
differential cross sections and the behavior of the gluon PDF,
particularly at high $x$. By adding $t\bar{t}$ differential
distributions into the fit one a time, we identified $t\bar{t}$ data
set combinations that are mutually consistent and deliver the
highest-impact experimental information.  In aggregate, we find that
the 13 TeV $t\bar{t}$ production prefers a softer gluon at high $x\!
\ge 0.1$.

In this article, we report on the first results of the ongoing study
\cite{Ablat:2024Jet} exploring analogous constraints from the
inclusive-jet and dijet differential distribution measurements at 8
and 13 TeV.  Though the $t\bar t$ and high-$p_T$ jet production at the
LHC largely overlap in the $x-Q$ plane, their matrix elements,
phase-space suppression, experimental and theoretical systematics are
different, and so are their constraints on the gluon PDF.  We found
that the post-CT18 inclusive jet data favor a larger $g$ PDF at
$x>0.1$, which is at odds with the preference of the post-CT18
Drell-Yan and top-quark data for a smaller gluon at comparable $x$
values. This behavior of inclusive jet data is quite robust against
scale variations in theory predictions. On the other hand, the impact
of dijet data substantially depends on the scale choice, especially in
the case of CMS8 TeV dijet data, causing varied pulls on the $g$ PDF
at $x>0.1$. Our ongoing study shows that the combination of the
post-CT18 Drell-Yan, top quark pair and inclusive-jet data can
nevertheless further constrain the gluon PDF at $x$ around $10^{-2}$,
relevant for Higgs boson production at the LHC, as well as at $x>0.1$.

After summarizing the impact of these new experiments on the PDFs, we
reviewed progress in statistical and numerical methods for PDF fits
while quantifying the associated uncertainties.
This activity represents a significant extension of the activity of
recent years to develop (fast) methods for PDF phenomenology,
including the Lagrange Multiplier approach and related studies;
\texttt{PDFSense} and $L_2$ sensitivity methods; \texttt{ePump}
Hessian updating tool; explorations of subtleties in sampling and
associated issues in Monte Carlo uncertainty determination; and
related benchmarking investigations to refine these calculations.
This effort reflects the high priority of achieving {\it robust} and
{\it replicable} PDF uncertainties to guarantee the statistical
interpretation of theoretical calculations for high-stakes
observables.
Within this context, we presented studies of new LHC data on PDFs
within the $L_2$ method \cite{Jing:2023isu,Kotz:2024dfg}; they guide
ongoing effort to include such data in the upcoming CT main release.
We highlighted a new numerical framework, Fant\^omas, to obtain a
flexibly generalizable basis to explore the PDF parametrization
dependence within the Hessian formalism \cite{Kotz:2023pbu}.
Complementing this activity, we also noted a recent effort to explore
PDFs in machine-learning approaches, including the study of PDF
parametrizations encapsulated by autoencoder networks with
interpretability constraints~\cite{Kriesten:2023uoi,Kriesten:2024are},
and work based on the application of Gaussian Mixture Model (GMM) for
analyzing the goodness of fits between data sets with various degrees
of tension \cite{Yan:2024yir}.

Remarkable progress has recently been made in global
analyses~\cite{McGowan:2022nag,NNPDF:2024nan,Cooper-Sarkar:2024crx} to incorporate partial
N$^3$LO DGLAP splitting functions and data-constrained estimates for
yet uncalculated N$^3$LO contributions. This approach, dubbed
approximate N$^3$LO (aN$^3$LO), agrees with the NNLO predictions
within the scale dependence uncertainty, but results in changes in the
PDFs that can lead to notable differences between the two orders, for
example, for Higgs boson production cross sections. A comparison of
such predictions for the $gg\to H^0$ channel with N$^3$LO
short-distance cross sections revealed that they differ more between
the MSHT and NNPDF aN$^3$LO PDFs than between their respective NNLO
PDFs, an indication that the results are still not under full control
at this accuracy level until more of the crucial components of the
global PDF fits are known to N$^3$LO.

Within the CTEQ-TEA group, we work on implementation of these critical
components for DGLAP evolution, DIS, and vector boson production. The
anticipated publication of the three-loop single-mass heavy-flavor
corrections to DIS \cite[ and references therein]{Ablinger:2024qxg}
and operator matrix elements~\cite[ and references
  therein]{Ablinger:2024xtt} in a practicable format will allow for a
more robust examination of N$^3$LO effects. In parallel, we have
continued to explore the impact of NLL resummation and saturation on
the gluon and other PDFs at the low-$x$ frontier of the accessible
kinematic region. The small-$x$ resummation corrections may alter the
PDFs across all $x$ via sum rules, especially in the case of the
gluon, in the ways that may not be fully captured by the full or
partial N3LO contributions.

Lastly, we discussed a series of phenomenological studies intimately
connected to PDF determination within the CT framework, or theoretical
predictions immediately derived from the CT PDFs themselves. In this
category, we highlighted new work to understand possible correlations
between the high-$x$ nucleon sea and low-energy parity-violating
measurements as might be undertaken at an upgraded 22 GeV JLab
facility; see \cite{Accardi:2023chb} and additional investigation in
Sec.~\ref{sec:PVsea}.  It is plausible on the basis of these
calculations that such data could extend sensitivity of the current
data set, especially if associated higher-twist and other corrections
can be controlled, and also complement measurements like the
forward-backward asymmetry ($A_\mathrm{FB}$) of the DY pairs produced
at the LHC.

Based on a systematic, representative exploration of various factors
contributing at NNLO accuracy, we critically examined the recent
claims of an experimental evidence for nonperturbative (intrinsic)
charm or its asymmetry. In this CT18FC NNLO study of fitted charm (FC)
in the nucleon \cite{Guzzi:2022rca}, for the first time in a global
PDF fit we allowed $c\! \neq\! \bar{c}$ in the nonperturbative
parametrization at the initial QCD scale to increase flexibility in
exploring possible scenarios. We found that the current global data
clearly have insufficient discriminating power to provide significant
evidence for fitted charm. This strong conclusion was independently
corroborated \cite{Courtoy:2022ocu} by a study of constraints on the
charm done directly within the NNPDF computational framework, \rev{with the key criticisms in the rebuttal \cite{Ball:2022uon} by NNPDF countered in the longer version of \cite{Courtoy:2022ocu} on arXiv}.  We
expect additional theoretical refinement of perturbative theory,
possible lattice constraints, and future experiments to eventually
shed light on the question.

Meanwhile, we also noted the first CTEQ-TEA study of the parton-level
photon PDF in the {\it neutron} \cite{Xie:2023qbn}, paralleling our
recent analogous study in the proton \cite{Xie:2021equ}. This analysis
highlighted a ground-up determination of QED-driven charge-symmetry
violation and opens the door to more precise phenomenology in the
neutrino sector and axionlike particle (ALP) searches.
On another footing, we summarized the conclusions of a first global
fit within the CT framework of allowing nonzero Wilson coefficients
within the SMEFT approach to parameterize possible signatures of new
physics in select jet and $t\bar{t}$ experiments
\cite{Gao:2022srd}. Correlations between fitted PDFs and SMEFT
parameters, which have been neglected in various past analyses, become
relevant as precision increases. Collectively, these studies drive the
multifaceted program of the development of the next-generation
CTEQ-TEA PDFs and leveraging them to ensure theoretical accuracy in
(B)SM phenomenology.

Finally, various advancements in the numerical implementation of
CTEQ-TEA PDFs are continuously made, including the release of
enhanced-precision LHAPDF grids, named CT18up, for the NNLO PDFs of
the CT18 family for precision and high-mass studies. 

\section*{Acknowledgments}

We thank O.~Gonz\'alez-Hern\'andez, B.~Kriesten, K. Rubin,
F.~I. Olness, and M. Ponce-Chavez for useful discussions and inputs.
The work of AA, SD and IS is supported by the National Natural Science
Foundation of China under Grants No.11965020 and No. 11847160. The
work of T.-J. Hou was supported by Natural Science Foundation of Hunan
province of China under Grant No. 2023JJ30496.  AC is supported by the
UNAM Grant No. DGAPA-PAPIIT (IN111222 and IN102225) and CONACyT Ciencia de Frontera
2019 No. 51244 (FORDECYT-PRONACES).
MG is supported by the National Science Foundation under Grants
No.~PHY-2112025 and No.~PHY-2412071.
The work of T.~J.~Hobbs at Argonne National Laboratory was supported
by the U.S.~Department of Energy, Office of Science, under Contract
No.~DE-AC02-06CH11357.
The work of HL is  partially supported
by the US National Science Foundation under grant PHY 1653405 ``CAREER: Constraining Parton Distribution Functions for New-Physics Searches'' and grant PHY~2209424. 
PMN was partially supported by the U.S. Department of Energy under
Grant No.~DE-SC0010129 and thanks the US DOE Institute for Nuclear
Theory at the University of Washington for its hospitality during the
completion of this work.
Research by CPY, KM, and KX at MSU was supported by 
the U.S. National Science Foundation under Grants No.~PHY-2310291 and PHY-2310497.
The work of KX was performed in part at the Aspen Center for Physics,
which is supported by National Science Foundation grant PHY-2210452.
This work used resources of high-performance computing clusters from
SMU M2/M3, MSU HPCC, KSU HPCs, as well as Pitt CRC.

\section*{Data availability statement}
The latest CT18up, CT18 FC, CT18 QED and other parton distributions discussed
in this article, as well as public computer programs developed by the CTEQ-TEA
group are available from the CTEQ-TEA website \cite{CTwebsite}.
We recommend the users to visit this website to find the latest releases of our PDFs.
The PDFs are provided as tabulated grids in two formats: as .pds files
for interpolation using a standalone interface available at \cite{CTwebsite},
and as LHAPDF6 files for interpolation using the LHAPDF library \cite{LHAPDF6}.
The counterpart CT18 NNLO grids from the 2019 release,
suitable when extra interpolation precision is not needed,
remain available both on the CTEQ-TEA \cite{CTwebsite} and LHAPDF
\cite{LHAPDF6} websites.


\begin{thebibliography}{100}

\bibitem{Hou:2019efy}
T.-J. Hou {\em et~al.}, ``{New CTEQ global analysis of quantum chromodynamics
  with high-precision data from the LHC},''
  \href{http://dx.doi.org/10.1103/PhysRevD.103.014013}{{\em Phys. Rev. D}
  {\bfseries 103} no.~1, (2021) 014013},
  \href{http://arxiv.org/abs/1912.10053}{{\ttfamily arXiv:1912.10053
  [hep-ph]}}.

\bibitem{Bailey:2020ooq}
S.~Bailey, T.~Cridge, L.~A. Harland-Lang, A.~D. Martin, and R.~S. Thorne,
  ``{Parton distributions from LHC, HERA, Tevatron and fixed target data:
  MSHT20 PDFs},'' \href{http://dx.doi.org/10.1140/epjc/s10052-021-09057-0}{{\em
  Eur. Phys. J. C} {\bfseries 81} no.~4, (2021) 341},
  \href{http://arxiv.org/abs/2012.04684}{{\ttfamily arXiv:2012.04684
  [hep-ph]}}.

\bibitem{NNPDF:2021njg}
{\bfseries NNPDF} Collaboration, R.~D. Ball {\em et~al.}, ``{The path to proton
  structure at 1\% accuracy},''
  \href{http://dx.doi.org/10.1140/epjc/s10052-022-10328-7}{{\em Eur. Phys. J.
  C} {\bfseries 82} no.~5, (2022) 428},
  \href{http://arxiv.org/abs/2109.02653}{{\ttfamily arXiv:2109.02653
  [hep-ph]}}.

\bibitem{H1:2021xxi}
{\bfseries H1, ZEUS} Collaboration, I.~Abt {\em et~al.}, ``{Impact of
  jet-production data on the next-to-next-to-leading-order determination of
  HERAPDF2.0 parton distributions},''
  \href{http://dx.doi.org/10.1140/epjc/s10052-022-10083-9}{{\em Eur. Phys. J.
  C} {\bfseries 82} no.~3, (2022) 243},
  \href{http://arxiv.org/abs/2112.01120}{{\ttfamily arXiv:2112.01120
  [hep-ex]}}.

\bibitem{ATLAS:2021vod}
{\bfseries ATLAS} Collaboration, G.~Aad {\em et~al.}, ``{Determination of the
  parton distribution functions of the proton using diverse ATLAS data from
  $pp$ collisions at $\sqrt{s} = 7$, 8 and 13~TeV},''
  \href{http://dx.doi.org/10.1140/epjc/s10052-022-10217-z}{{\em Eur. Phys. J.
  C} {\bfseries 82} no.~5, (2022) 438},
  \href{http://arxiv.org/abs/2112.11266}{{\ttfamily arXiv:2112.11266
  [hep-ex]}}.

\bibitem{Alekhin:2024bhs}
S.~Alekhin, M.~V. Garzelli, S.~O. Moch, and O.~Zenaiev, ``{NNLO PDFs driven by
  top-quark data},'' \href{http://arxiv.org/abs/2407.00545}{{\ttfamily
  arXiv:2407.00545 [hep-ph]}}.

\bibitem{Accardi:2023gyr}
A.~Accardi, X.~Jing, J.~F. Owens, and S.~Park, ``{Light quark and antiquark
  constraints from new electroweak data},''
  \href{http://dx.doi.org/10.1103/PhysRevD.107.113005}{{\em Phys. Rev. D}
  {\bfseries 107} no.~11, (2023) 113005},
  \href{http://arxiv.org/abs/2303.11509}{{\ttfamily arXiv:2303.11509
  [hep-ph]}}.

\bibitem{Sato:2019yez}
{\bfseries JAM} Collaboration, N.~Sato, C.~Andres, J.~J. Ethier, and
  W.~Melnitchouk, ``{Strange quark suppression from a simultaneous Monte Carlo
  analysis of parton distributions and fragmentation functions},''
  \href{http://dx.doi.org/10.1103/PhysRevD.101.074020}{{\em Phys. Rev. D}
  {\bfseries 101} no.~7, (2020) 074020},
  \href{http://arxiv.org/abs/1905.03788}{{\ttfamily arXiv:1905.03788
  [hep-ph]}}.

\bibitem{Moffat:2021dji}
{\bfseries Jefferson Lab Angular Momentum (JAM)} Collaboration, E.~Moffat,
  W.~Melnitchouk, T.~C. Rogers, and N.~Sato, ``{Simultaneous Monte~Carlo
  analysis of parton densities and fragmentation functions},''
  \href{http://dx.doi.org/10.1103/PhysRevD.104.016015}{{\em Phys. Rev. D}
  {\bfseries 104} no.~1, (2021) 016015},
  \href{http://arxiv.org/abs/2101.04664}{{\ttfamily arXiv:2101.04664
  [hep-ph]}}.

\bibitem{Vermaseren:2005qc}
J.~A.~M. Vermaseren, A.~Vogt, and S.~Moch, ``{The Third-order QCD corrections
  to deep-inelastic scattering by photon exchange},''
  \href{http://dx.doi.org/10.1016/j.nuclphysb.2005.06.020}{{\em Nucl. Phys. B}
  {\bfseries 724} (2005) 3--182},
  \href{http://arxiv.org/abs/hep-ph/0504242}{{\ttfamily arXiv:hep-ph/0504242}}.

\bibitem{Moch:2004xu}
S.~Moch, J.~A.~M. Vermaseren, and A.~Vogt, ``{The Longitudinal structure
  function at the third order},''
  \href{http://dx.doi.org/10.1016/j.physletb.2004.11.063}{{\em Phys. Lett. B}
  {\bfseries 606} (2005) 123--129},
  \href{http://arxiv.org/abs/hep-ph/0411112}{{\ttfamily arXiv:hep-ph/0411112}}.

\bibitem{Moch:2008fj}
S.~Moch, J.~A.~M. Vermaseren, and A.~Vogt, ``{Third-order QCD corrections to
  the charged-current structure function F(3)},''
  \href{http://dx.doi.org/10.1016/j.nuclphysb.2009.01.001}{{\em Nucl. Phys. B}
  {\bfseries 813} (2009) 220--258},
  \href{http://arxiv.org/abs/0812.4168}{{\ttfamily arXiv:0812.4168 [hep-ph]}}.

\bibitem{Davies:2016ruz}
J.~Davies, A.~Vogt, S.~Moch, and J.~A.~M. Vermaseren, ``{Non-singlet
  coefficient functions for charged-current deep-inelastic scattering to the
  third order in QCD},'' \href{http://dx.doi.org/10.22323/1.265.0059}{{\em PoS}
  {\bfseries DIS2016} (2016) 059},
  \href{http://arxiv.org/abs/1606.08907}{{\ttfamily arXiv:1606.08907
  [hep-ph]}}.

\bibitem{Blumlein:2022gpp}
J.~Bl\"umlein, P.~Marquard, C.~Schneider, and K.~Sch\"onwald, ``{The massless
  three-loop Wilson coefficients for the deep-inelastic structure functions
  F$_{2}$, F$_{L}$, xF$_{3}$ and g$_{1}$},''
  \href{http://dx.doi.org/10.1007/JHEP11(2022)156}{{\em JHEP} {\bfseries 11}
  (2022) 156}, \href{http://arxiv.org/abs/2208.14325}{{\ttfamily
  arXiv:2208.14325 [hep-ph]}}.

\bibitem{Ablinger:2024qxg}
J.~Ablinger, A.~Behring, J.~Bl\"umlein, A.~De~Freitas, A.~von Manteuffel,
  C.~Schneider, and K.~Schoenwald, ``{The three-loop single-mass heavy flavor
  corrections to deep-inelastic scattering},''
  \href{http://arxiv.org/abs/2407.02006}{{\ttfamily arXiv:2407.02006
  [hep-ph]}}.

\bibitem{Baglio:2022wzu}
J.~Baglio, C.~Duhr, B.~Mistlberger, and R.~Szafron, ``{Inclusive production
  cross sections at N$^{3}$LO},''
  \href{http://dx.doi.org/10.1007/JHEP12(2022)066}{{\em JHEP} {\bfseries 12}
  (2022) 066}, \href{http://arxiv.org/abs/2209.06138}{{\ttfamily
  arXiv:2209.06138 [hep-ph]}}.

\bibitem{Duhr:2020sdp}
C.~Duhr, F.~Dulat, and B.~Mistlberger, ``{Charged current Drell-Yan production
  at N$^{3}$LO},'' \href{http://dx.doi.org/10.1007/JHEP11(2020)143}{{\em JHEP}
  {\bfseries 11} (2020) 143}, \href{http://arxiv.org/abs/2007.13313}{{\ttfamily
  arXiv:2007.13313 [hep-ph]}}.

\bibitem{Duhr:2021vwj}
C.~Duhr and B.~Mistlberger, ``{Lepton-pair production at hadron colliders at
  N$^{3}$LO in QCD},'' \href{http://dx.doi.org/10.1007/JHEP03(2022)116}{{\em
  JHEP} {\bfseries 03} (2022) 116},
  \href{http://arxiv.org/abs/2111.10379}{{\ttfamily arXiv:2111.10379
  [hep-ph]}}.

\bibitem{Chen:2021vtu}
X.~Chen, T.~Gehrmann, N.~Glover, A.~Huss, T.-Z. Yang, and H.~X. Zhu,
  ``{Dilepton Rapidity Distribution in Drell-Yan Production to Third Order in
  QCD},'' \href{http://dx.doi.org/10.1103/PhysRevLett.128.052001}{{\em Phys.
  Rev. Lett.} {\bfseries 128} no.~5, (2022) 052001},
  \href{http://arxiv.org/abs/2107.09085}{{\ttfamily arXiv:2107.09085
  [hep-ph]}}.

\bibitem{Chen:2022lwc}
X.~Chen, T.~Gehrmann, N.~Glover, A.~Huss, T.-Z. Yang, and H.~X. Zhu,
  ``{Transverse mass distribution and charge asymmetry in W boson production to
  third order in QCD},''
  \href{http://dx.doi.org/10.1016/j.physletb.2023.137876}{{\em Phys. Lett. B}
  {\bfseries 840} (2023) 137876},
  \href{http://arxiv.org/abs/2205.11426}{{\ttfamily arXiv:2205.11426
  [hep-ph]}}.

\bibitem{Anastasiou:2015vya}
C.~Anastasiou, C.~Duhr, F.~Dulat, F.~Herzog, and B.~Mistlberger, ``{Higgs Boson
  Gluon-Fusion Production in QCD at Three Loops},''
  \href{http://dx.doi.org/10.1103/PhysRevLett.114.212001}{{\em Phys. Rev.
  Lett.} {\bfseries 114} (2015) 212001},
  \href{http://arxiv.org/abs/1503.06056}{{\ttfamily arXiv:1503.06056
  [hep-ph]}}.

\bibitem{Mistlberger:2018etf}
B.~Mistlberger, ``{Higgs boson production at hadron colliders at N$^{3}$LO in
  QCD},'' \href{http://dx.doi.org/10.1007/JHEP05(2018)028}{{\em JHEP}
  {\bfseries 05} (2018) 028}, \href{http://arxiv.org/abs/1802.00833}{{\ttfamily
  arXiv:1802.00833 [hep-ph]}}.

\bibitem{Dulat:2018bfe}
F.~Dulat, B.~Mistlberger, and A.~Pelloni, ``{Precision predictions at N$^3$LO
  for the Higgs boson rapidity distribution at the LHC},''
  \href{http://dx.doi.org/10.1103/PhysRevD.99.034004}{{\em Phys. Rev. D}
  {\bfseries 99} no.~3, (2019) 034004},
  \href{http://arxiv.org/abs/1810.09462}{{\ttfamily arXiv:1810.09462
  [hep-ph]}}.

\bibitem{Dreyer:2016oyx}
F.~A. Dreyer and A.~Karlberg, ``{Vector-Boson Fusion Higgs Production at Three
  Loops in QCD},'' \href{http://dx.doi.org/10.1103/PhysRevLett.117.072001}{{\em
  Phys. Rev. Lett.} {\bfseries 117} no.~7, (2016) 072001},
  \href{http://arxiv.org/abs/1606.00840}{{\ttfamily arXiv:1606.00840
  [hep-ph]}}.

\bibitem{Cieri:2018oms}
L.~Cieri, X.~Chen, T.~Gehrmann, E.~W.~N. Glover, and A.~Huss, ``{Higgs boson
  production at the LHC using the $q_T$ subtraction formalism at N$^3$LO
  QCD},'' \href{http://dx.doi.org/10.1007/JHEP02(2019)096}{{\em JHEP}
  {\bfseries 02} (2019) 096}, \href{http://arxiv.org/abs/1807.11501}{{\ttfamily
  arXiv:1807.11501 [hep-ph]}}.

\bibitem{Gehrmann:2018odt}
T.~Gehrmann, A.~Huss, J.~Niehues, A.~Vogt, and D.~M. Walker, ``{Jet production
  in charged-current deep-inelastic scattering to third order in QCD},''
  \href{http://dx.doi.org/10.1016/j.physletb.2019.03.003}{{\em Phys. Lett. B}
  {\bfseries 792} (2019) 182--186},
  \href{http://arxiv.org/abs/1812.06104}{{\ttfamily arXiv:1812.06104
  [hep-ph]}}.

\bibitem{Currie:2018fgr}
J.~Currie, T.~Gehrmann, E.~W.~N. Glover, A.~Huss, J.~Niehues, and A.~Vogt,
  ``{N$^{3}$LO corrections to jet production in deep inelastic scattering using
  the Projection-to-Born method},''
  \href{http://dx.doi.org/10.1007/JHEP05(2018)209}{{\em JHEP} {\bfseries 05}
  (2018) 209}, \href{http://arxiv.org/abs/1803.09973}{{\ttfamily
  arXiv:1803.09973 [hep-ph]}}.

\bibitem{Schmidt:2018hvu}
C.~Schmidt, J.~Pumplin, and C.-P. Yuan, ``{Updating and optimizing error parton
  distribution function sets in the Hessian approach},''
  \href{http://dx.doi.org/10.1103/PhysRevD.98.094005}{{\em Phys. Rev.}
  {\bfseries D98} no.~9, (2018) 094005},
\href{http://arxiv.org/abs/1806.07950}{{\ttfamily arXiv:1806.07950 [hep-ph]}}.

\bibitem{Hou:2019gfw}
T.-J. Hou, Z.~Yu, S.~Dulat, C.~Schmidt, and C.-P. Yuan, ``{Updating and
  optimizing error parton distribution function sets in the Hessian approach.
  II.},'' \href{http://dx.doi.org/10.1103/PhysRevD.100.114024}{{\em Phys. Rev.
  D} {\bfseries 100} no.~11, (2019) 114024},
  \href{http://arxiv.org/abs/1907.12177}{{\ttfamily arXiv:1907.12177
  [hep-ph]}}.

\bibitem{Wang:2018heo}
B.-T. Wang, T.~J. Hobbs, S.~Doyle, J.~Gao, T.-J. Hou, P.~M. Nadolsky, and F.~I.
  Olness, ``{Mapping the sensitivity of hadronic experiments to nucleon
  structure},'' \href{http://dx.doi.org/10.1103/PhysRevD.98.094030}{{\em Phys.
  Rev.} {\bfseries D98} no.~9, (2018) 094030},
\href{http://arxiv.org/abs/1803.02777}{{\ttfamily arXiv:1803.02777 [hep-ph]}}.

\bibitem{Hobbs:2019gob}
T.~J. Hobbs, B.-T. Wang, P.~M. Nadolsky, and F.~I. Olness, ``{Charting the
  coming synergy between lattice QCD and high-energy phenomenology},''
  \href{http://dx.doi.org/10.1103/PhysRevD.100.094040}{{\em Phys. Rev. D}
  {\bfseries 100} no.~9, (2019) 094040},
  \href{http://arxiv.org/abs/1904.00022}{{\ttfamily arXiv:1904.00022
  [hep-ph]}}.

\bibitem{Hou:2022onq}
T.-J. Hou, H.-W. Lin, M.~Yan, and C.-P. Yuan, ``{Impact of lattice strangeness
  asymmetry data in the CTEQ-TEA global analysis},''
  \href{http://dx.doi.org/10.1103/PhysRevD.107.076018}{{\em Phys. Rev. D}
  {\bfseries 107} no.~7, (2023) 076018},
  \href{http://arxiv.org/abs/2211.11064}{{\ttfamily arXiv:2211.11064
  [hep-ph]}}.

\bibitem{Courtoy:2022ocu}
A.~Courtoy, J.~Huston, P.~Nadolsky, K.~Xie, M.~Yan, and C.-P. Yuan, ``{Parton
  distributions need representative sampling},''
  \href{http://dx.doi.org/10.1103/PhysRevD.107.034008}{{\em Phys. Rev. D}
  {\bfseries 107} no.~3, (2023) 034008},
  \href{http://arxiv.org/abs/2205.10444}{{\ttfamily arXiv:2205.10444
  [hep-ph]}}.

\bibitem{Yan:2022pzl}
M.~Yan, T.-J. Hou, P.~Nadolsky, and C.-P. Yuan, ``{CT18 global PDF fit at
  leading order in QCD},''
  \href{http://dx.doi.org/10.1103/PhysRevD.107.116001}{{\em Phys. Rev. D}
  {\bfseries 107} no.~11, (2023) 116001},
  \href{http://arxiv.org/abs/2205.00137}{{\ttfamily arXiv:2205.00137
  [hep-ph]}}.

\bibitem{CTwebsite}
{\bfseries CTEQ-TEA} Collaboration, ``Cteq-tea research projects and results.''
  \protect{The CTEQ-TEA global analysis group},
  \url{https://cteq-tea.gitlab.io/}.

\bibitem{LHAPDF6}
\protect{LHAPDF library}. \url{https://lhapdf.hepforge.org/}.

\bibitem{Nagar:2019gij}
R.~Nagar, ``{Efficient interpolation and evolution of parton distribution
  functions},'' \href{http://dx.doi.org/10.22323/1.352.0022}{{\em PoS}
  {\bfseries DIS2019} (2019) 022},
  \href{http://arxiv.org/abs/1906.10059}{{\ttfamily arXiv:1906.10059
  [hep-ph]}}.

\bibitem{Diehl:2021gvs}
M.~Diehl, R.~Nagar, and F.~J. Tackmann, ``{ChiliPDF: Chebyshev interpolation
  for parton distributions},''
  \href{http://dx.doi.org/10.1140/epjc/s10052-022-10223-1}{{\em Eur. Phys. J.
  C} {\bfseries 82} no.~3, (2022) 257},
  \href{http://arxiv.org/abs/2112.09703}{{\ttfamily arXiv:2112.09703
  [hep-ph]}}.

\bibitem{Sitiwaldi:2023jjp}
{\bfseries CTEQ-TEA} Collaboration, I.~Sitiwaldi, K.~Xie, A.~Ablat, S.~Dulat,
  T.-J. Hou, and C.-.~P. Yuan, ``{Precision studies of the post-CT18 LHC
  Drell-Yan data in the CTEQ-TEA global analysis},''
  \href{http://dx.doi.org/10.1103/PhysRevD.108.034030}{{\em Phys. Rev. D}
  {\bfseries 108} no.~3, (2023) 034030},
  \href{http://arxiv.org/abs/2305.10733}{{\ttfamily arXiv:2305.10733
  [hep-ph]}}.

\bibitem{Ablat:2023tiy}
A.~Ablat, M.~Guzzi, K.~Xie, S.~Dulat, T.-J. Hou, I.~Sitiwaldi, and C.-P. Yuan,
  ``{Exploring the impact of high-precision top-quark pair production data on
  the structure of the proton at the LHC},''
  \href{http://dx.doi.org/10.1103/PhysRevD.109.054027}{{\em Phys. Rev. D}
  {\bfseries 109} no.~5, (2024) 054027},
  \href{http://arxiv.org/abs/2307.11153}{{\ttfamily arXiv:2307.11153
  [hep-ph]}}.

\bibitem{Ablat:2024Jet}
A.~Ablat, S.~Dulat, {\em et~al.}, ``{Exploring the impact of jet production
  data on the structure of the proton at the LHC}.'' In preparation, 2024.

\bibitem{ATLAS:2016nqi}
{\bfseries ATLAS} Collaboration, M.~Aaboud {\em et~al.}, ``{Precision
  measurement and interpretation of inclusive $W^+$ , $W^-$ and $Z/\gamma ^*$
  production cross sections with the ATLAS detector},''
  \href{http://dx.doi.org/10.1140/epjc/s10052-017-4911-9}{{\em Eur. Phys. J. C}
  {\bfseries 77} no.~6, (2017) 367},
  \href{http://arxiv.org/abs/1612.03016}{{\ttfamily arXiv:1612.03016
  [hep-ex]}}.

\bibitem{H1:2015ubc}
{\bfseries H1, ZEUS} Collaboration, H.~Abramowicz {\em et~al.}, ``{Combination
  of measurements of inclusive deep inelastic ${e^{\pm }p}$ scattering cross
  sections and QCD analysis of HERA data},''
  \href{http://dx.doi.org/10.1140/epjc/s10052-015-3710-4}{{\em Eur. Phys. J. C}
  {\bfseries 75} no.~12, (2015) 580},
  \href{http://arxiv.org/abs/1506.06042}{{\ttfamily arXiv:1506.06042
  [hep-ex]}}.

\bibitem{Mason:2006qa}
D.~A. Mason, \href{http://dx.doi.org/10.2172/879078}{{\em {Measurement of the
  strange - antistrange asymmetry at NLO in QCD from NuTeV dimuon data}}}.
\newblock PhD thesis, Oregon U., 2006.

\bibitem{NuTeV:2007uwm}
{\bfseries NuTeV} Collaboration, D.~Mason {\em et~al.}, ``{Measurement of the
  Nucleon Strange-Antistrange Asymmetry at Next-to-Leading Order in QCD from
  NuTeV Dimuon Data},''
  \href{http://dx.doi.org/10.1103/PhysRevLett.99.192001}{{\em Phys. Rev. Lett.}
  {\bfseries 99} (2007) 192001}.

\bibitem{ATLAS:2018pyl}
{\bfseries ATLAS} Collaboration, M.~Aaboud {\em et~al.}, ``{Measurements of $W$
  and $Z$ boson production in $pp$ collisions at $\sqrt{s}=5.02$ TeV with the
  ATLAS detector},''
  \href{http://dx.doi.org/10.1140/epjc/s10052-019-6622-x}{{\em Eur. Phys. J. C}
  {\bfseries 79} no.~2, (2019) 128},
  \href{http://arxiv.org/abs/1810.08424}{{\ttfamily arXiv:1810.08424
  [hep-ex]}}. [Erratum: Eur.Phys.J.C 79, 374 (2019)].

\bibitem{ATLAS:2019fgb}
{\bfseries ATLAS} Collaboration, G.~Aad {\em et~al.}, ``{Measurement of the
  cross-section and charge asymmetry of $W$ bosons produced in
  proton\textendash{}proton collisions at $\sqrt{s}=8~\text {TeV}$ with the
  ATLAS detector},''
  \href{http://dx.doi.org/10.1140/epjc/s10052-019-7199-0}{{\em Eur. Phys. J. C}
  {\bfseries 79} no.~9, (2019) 760},
  \href{http://arxiv.org/abs/1904.05631}{{\ttfamily arXiv:1904.05631
  [hep-ex]}}.

\bibitem{ATLAS:2017rue}
{\bfseries ATLAS} Collaboration, M.~Aaboud {\em et~al.}, ``{Measurement of the
  Drell-Yan triple-differential cross section in $pp$ collisions at $\sqrt{s} =
  8$ TeV},'' \href{http://dx.doi.org/10.1007/JHEP12(2017)059}{{\em JHEP}
  {\bfseries 12} (2017) 059}, \href{http://arxiv.org/abs/1710.05167}{{\ttfamily
  arXiv:1710.05167 [hep-ex]}}.

\bibitem{CMS:2019raw}
{\bfseries CMS} Collaboration, A.~M. Sirunyan {\em et~al.}, ``{Measurements of
  differential Z boson production cross sections in proton-proton collisions at
  $ \sqrt{s} $ = 13 TeV},''
  \href{http://dx.doi.org/10.1007/JHEP12(2019)061}{{\em JHEP} {\bfseries 12}
  (2019) 061}, \href{http://arxiv.org/abs/1909.04133}{{\ttfamily
  arXiv:1909.04133 [hep-ex]}}.

\bibitem{LHCb:2016zpq}
{\bfseries LHCb} Collaboration, R.~Aaij {\em et~al.}, ``{Measurement of forward
  $W\to e\nu$ production in $pp$ collisions at $\sqrt{s}=8\,$TeV},''
  \href{http://dx.doi.org/10.1007/JHEP10(2016)030}{{\em JHEP} {\bfseries 10}
  (2016) 030}, \href{http://arxiv.org/abs/1608.01484}{{\ttfamily
  arXiv:1608.01484 [hep-ex]}}.

\bibitem{LHCb:2021huf}
{\bfseries LHCb} Collaboration, R.~Aaij {\em et~al.}, ``{Precision measurement
  of forward $Z$ boson production in proton-proton collisions at $\sqrt{s} =
  13$ TeV},'' \href{http://dx.doi.org/10.1007/JHEP07(2022)026}{{\em JHEP}
  {\bfseries 07} (2022) 026}, \href{http://arxiv.org/abs/2112.07458}{{\ttfamily
  arXiv:2112.07458 [hep-ex]}}.

\bibitem{Carli:2010rw}
T.~Carli, D.~Clements, A.~Cooper-Sarkar, C.~Gwenlan, G.~P. Salam, F.~Siegert,
  P.~Starovoitov, and M.~Sutton, ``{A posteriori inclusion of parton density
  functions in NLO QCD final-state calculations at hadron colliders: The
  APPLGRID Project},''
  \href{http://dx.doi.org/10.1140/epjc/s10052-010-1255-0}{{\em Eur. Phys. J.}
  {\bfseries C66} (2010) 503--524},
\href{http://arxiv.org/abs/0911.2985}{{\ttfamily arXiv:0911.2985 [hep-ph]}}.

\bibitem{Campbell:2019dru}
J.~Campbell and T.~Neumann, ``{Precision Phenomenology with MCFM},''
  \href{http://dx.doi.org/10.1007/JHEP12(2019)034}{{\em JHEP} {\bfseries 12}
  (2019) 034}, \href{http://arxiv.org/abs/1909.09117}{{\ttfamily
  arXiv:1909.09117 [hep-ph]}}.

\bibitem{Isaacson:2022rts}
J.~Isaacson, Y.~Fu, and C.~P. Yuan, ``{ResBos2 and the CDF W Mass
  Measurement},'' \href{http://arxiv.org/abs/2205.02788}{{\ttfamily
  arXiv:2205.02788 [hep-ph]}}.

\bibitem{Czakon:2011xx}
M.~Czakon and A.~Mitov, ``{Top++: A Program for the Calculation of the Top-Pair
  Cross-Section at Hadron Colliders},''
  \href{http://dx.doi.org/10.1016/j.cpc.2014.06.021}{{\em Comput. Phys.
  Commun.} {\bfseries 185} (2014) 2930},
  \href{http://arxiv.org/abs/1112.5675}{{\ttfamily arXiv:1112.5675 [hep-ph]}}.

\bibitem{Alwall:2014hca}
J.~Alwall, R.~Frederix, S.~Frixione, V.~Hirschi, F.~Maltoni, O.~Mattelaer,
  H.~S. Shao, T.~Stelzer, P.~Torrielli, and M.~Zaro, ``{The automated
  computation of tree-level and next-to-leading order differential cross
  sections, and their matching to parton shower simulations},''
  \href{http://dx.doi.org/10.1007/JHEP07(2014)079}{{\em JHEP} {\bfseries 07}
  (2014) 079},
\href{http://arxiv.org/abs/1405.0301}{{\ttfamily arXiv:1405.0301 [hep-ph]}}.

\bibitem{Frederix:2018nkq}
R.~Frederix, S.~Frixione, V.~Hirschi, D.~Pagani, H.~S. Shao, and M.~Zaro,
  ``{The automation of next-to-leading order electroweak calculations},''
  \href{http://dx.doi.org/10.1007/JHEP11(2021)085}{{\em JHEP} {\bfseries 07}
  (2018) 185}, \href{http://arxiv.org/abs/1804.10017}{{\ttfamily
  arXiv:1804.10017 [hep-ph]}}. [Erratum: JHEP 11, 085 (2021)].

\bibitem{Carrazza:2020gss}
S.~Carrazza, E.~R. Nocera, C.~Schwan, and M.~Zaro, ``{PineAPPL: combining EW
  and QCD corrections for fast evaluation of LHC processes},''
  \href{http://dx.doi.org/10.1007/JHEP12(2020)108}{{\em JHEP} {\bfseries 12}
  (2020) 108}, \href{http://arxiv.org/abs/2008.12789}{{\ttfamily
  arXiv:2008.12789 [hep-ph]}}.

\bibitem{Nadolsky:2008zw}
P.~M. Nadolsky, H.-L. Lai, Q.-H. Cao, J.~Huston, J.~Pumplin, D.~Stump, W.-K.
  Tung, and C.-P. Yuan, ``{Implications of CTEQ global analysis for collider
  observables},'' \href{http://dx.doi.org/10.1103/PhysRevD.78.013004}{{\em
  Phys. Rev.} {\bfseries D78} (2008) 013004},
\href{http://arxiv.org/abs/0802.0007}{{\ttfamily arXiv:0802.0007 [hep-ph]}}.

\bibitem{CMS:2018adi}
{\bfseries CMS} Collaboration, A.~M. Sirunyan {\em et~al.}, ``{Measurements of
  $\mathrm{t\overline{t}}$ differential cross sections in proton-proton
  collisions at $\sqrt{s}=$ 13 TeV using events containing two leptons},''
  \href{http://dx.doi.org/10.1007/JHEP02(2019)149}{{\em JHEP} {\bfseries 02}
  (2019) 149}, \href{http://arxiv.org/abs/1811.06625}{{\ttfamily
  arXiv:1811.06625 [hep-ex]}}.

\bibitem{CMS:2021vhb}
{\bfseries CMS} Collaboration, A.~Tumasyan {\em et~al.}, ``{Measurement of
  differential $t \bar t$ production cross sections in the full kinematic range
  using lepton+jets events from proton-proton collisions at $\sqrt {s}$ =
  13\,\,TeV},'' \href{http://dx.doi.org/10.1103/PhysRevD.104.092013}{{\em Phys.
  Rev. D} {\bfseries 104} no.~9, (2021) 092013},
  \href{http://arxiv.org/abs/2108.02803}{{\ttfamily arXiv:2108.02803
  [hep-ex]}}.

\bibitem{ATLAS:2019hxz}
{\bfseries ATLAS} Collaboration, G.~Aad {\em et~al.}, ``{Measurements of
  top-quark pair differential and double-differential cross-sections in the
  $\ell$+jets channel with $pp$ collisions at $\sqrt{s}=13$ TeV using the ATLAS
  detector},'' \href{http://dx.doi.org/10.1140/epjc/s10052-019-7525-6}{{\em
  Eur. Phys. J. C} {\bfseries 79} no.~12, (2019) 1028},
  \href{http://arxiv.org/abs/1908.07305}{{\ttfamily arXiv:1908.07305
  [hep-ex]}}. [Erratum: Eur.Phys.J.C 80, 1092 (2020)].

\bibitem{ATLAS:2020ccu}
{\bfseries ATLAS} Collaboration, G.~Aad {\em et~al.}, ``{Measurements of
  top-quark pair single- and double-differential cross-sections in the
  all-hadronic channel in $pp$ collisions at $\sqrt{s}=13~\textrm{TeV}$ using
  the ATLAS detector},'' \href{http://dx.doi.org/10.1007/JHEP01(2021)033}{{\em
  JHEP} {\bfseries 01} (2021) 033},
  \href{http://arxiv.org/abs/2006.09274}{{\ttfamily arXiv:2006.09274
  [hep-ex]}}.

\bibitem{Czakon:2017dip}
M.~Czakon, D.~Heymes, and A.~Mitov, ``{fastNLO tables for NNLO top-quark pair
  differential distributions},''
  \href{http://arxiv.org/abs/1704.08551}{{\ttfamily arXiv:1704.08551
  [hep-ph]}}.

\bibitem{Czakon:2015owf}
M.~Czakon, D.~Heymes, and A.~Mitov, ``{High-precision differential predictions
  for top-quark pairs at the LHC},''
  \href{http://dx.doi.org/10.1103/PhysRevLett.116.082003}{{\em Phys. Rev.
  Lett.} {\bfseries 116} no.~8, (2016) 082003},
  \href{http://arxiv.org/abs/1511.00549}{{\ttfamily arXiv:1511.00549
  [hep-ph]}}.

\bibitem{Czakon:2014oma}
M.~Czakon and D.~Heymes, ``{Four-dimensional formulation of the sector-improved
  residue subtraction scheme},''
  \href{http://dx.doi.org/10.1016/j.nuclphysb.2014.11.006}{{\em Nucl. Phys. B}
  {\bfseries 890} (2014) 152--227},
  \href{http://arxiv.org/abs/1408.2500}{{\ttfamily arXiv:1408.2500 [hep-ph]}}.

\bibitem{Campbell:2015qma}
J.~M. Campbell, R.~K. Ellis, and W.~T. Giele, ``{A Multi-Threaded Version of
  MCFM},'' \href{http://dx.doi.org/10.1140/epjc/s10052-015-3461-2}{{\em Eur.
  Phys. J. C} {\bfseries 75} no.~6, (2015) 246},
  \href{http://arxiv.org/abs/1503.06182}{{\ttfamily arXiv:1503.06182
  [physics.comp-ph]}}.

\bibitem{Campbell:2012uf}
J.~M. Campbell and R.~K. Ellis, ``{Top-Quark Processes at NLO in Production and
  Decay},'' \href{http://dx.doi.org/10.1088/0954-3899/42/1/015005}{{\em J.
  Phys. G} {\bfseries 42} no.~1, (2015) 015005},
  \href{http://arxiv.org/abs/1204.1513}{{\ttfamily arXiv:1204.1513 [hep-ph]}}.

\bibitem{Grazzini:2017mhc}
M.~Grazzini, S.~Kallweit, and M.~Wiesemann, ``{Fully differential NNLO
  computations with MATRIX},''
  \href{http://dx.doi.org/10.1140/epjc/s10052-018-5771-7}{{\em Eur. Phys. J. C}
  {\bfseries 78} no.~7, (2018) 537},
  \href{http://arxiv.org/abs/1711.06631}{{\ttfamily arXiv:1711.06631
  [hep-ph]}}.

\bibitem{Catani:2019hip}
S.~Catani, S.~Devoto, M.~Grazzini, S.~Kallweit, and J.~Mazzitelli, ``{Top-quark
  pair production at the LHC: Fully differential QCD predictions at NNLO},''
  \href{http://dx.doi.org/10.1007/JHEP07(2019)100}{{\em JHEP} {\bfseries 07}
  (2019) 100}, \href{http://arxiv.org/abs/1906.06535}{{\ttfamily
  arXiv:1906.06535 [hep-ph]}}.

\bibitem{Catani:2007vq}
S.~Catani and M.~Grazzini, ``{An NNLO subtraction formalism in hadron
  collisions and its application to Higgs boson production at the LHC},''
  \href{http://dx.doi.org/10.1103/PhysRevLett.98.222002}{{\em Phys. Rev. Lett.}
  {\bfseries 98} (2007) 222002},
\href{http://arxiv.org/abs/hep-ph/0703012}{{\ttfamily arXiv:hep-ph/0703012
  [hep-ph]}}.

\bibitem{Pagani:2016caq}
D.~Pagani, I.~Tsinikos, and M.~Zaro, ``{The impact of the photon PDF and
  electroweak corrections on $t \bar{t}$ distributions},''
  \href{http://dx.doi.org/10.1140/epjc/s10052-016-4318-z}{{\em Eur. Phys. J. C}
  {\bfseries 76} no.~9, (2016) 479},
  \href{http://arxiv.org/abs/1606.01915}{{\ttfamily arXiv:1606.01915
  [hep-ph]}}.

\bibitem{Campbell:2016dks}
J.~M. Campbell, D.~Wackeroth, and J.~Zhou, ``{Study of weak corrections to
  Drell-Yan, top-quark pair, and dijet production at high energies with
  MCFM},'' \href{http://dx.doi.org/10.1103/PhysRevD.94.093009}{{\em Phys. Rev.
  D} {\bfseries 94} no.~9, (2016) 093009},
  \href{http://arxiv.org/abs/1608.03356}{{\ttfamily arXiv:1608.03356
  [hep-ph]}}.

\bibitem{ATLAS:2013pbc}
{\bfseries ATLAS} Collaboration, G.~Aad {\em et~al.}, ``{Measurement of the
  inclusive jet cross section in pp collisions at sqrt(s)=2.76 TeV and
  comparison to the inclusive jet cross section at sqrt(s)=7 TeV using the
  ATLAS detector},''
  \href{http://dx.doi.org/10.1140/epjc/s10052-013-2509-4}{{\em Eur. Phys. J. C}
  {\bfseries 73} no.~8, (2013) 2509},
  \href{http://arxiv.org/abs/1304.4739}{{\ttfamily arXiv:1304.4739 [hep-ex]}}.

\bibitem{CMS:2015jdl}
{\bfseries CMS} Collaboration, V.~Khachatryan {\em et~al.}, ``{Measurement of
  the inclusive jet cross section in pp collisions at $\sqrt{s} = 2.76\,\text
  {TeV}$},'' \href{http://dx.doi.org/10.1140/epjc/s10052-016-4083-z}{{\em Eur.
  Phys. J. C} {\bfseries 76} no.~5, (2016) 265},
  \href{http://arxiv.org/abs/1512.06212}{{\ttfamily arXiv:1512.06212
  [hep-ex]}}.

\bibitem{ATLAS:2017ble}
{\bfseries ATLAS} Collaboration, M.~Aaboud {\em et~al.}, ``{Measurement of
  inclusive jet and dijet cross-sections in proton-proton collisions at
  $\sqrt{s}=13$ TeV with the ATLAS detector},''
  \href{http://dx.doi.org/10.1007/JHEP05(2018)195}{{\em JHEP} {\bfseries 05}
  (2018) 195}, \href{http://arxiv.org/abs/1711.02692}{{\ttfamily
  arXiv:1711.02692 [hep-ex]}}.

\bibitem{CMS:2016jip}
{\bfseries CMS} Collaboration, V.~Khachatryan {\em et~al.}, ``{Measurement of
  the double-differential inclusive jet cross section in
  proton\textendash{}proton collisions at $\sqrt{s} = 13\,\text {TeV} $},''
  \href{http://dx.doi.org/10.1140/epjc/s10052-016-4286-3}{{\em Eur. Phys. J. C}
  {\bfseries 76} no.~8, (2016) 451},
  \href{http://arxiv.org/abs/1605.04436}{{\ttfamily arXiv:1605.04436
  [hep-ex]}}.

\bibitem{CMS:2021yzl}
{\bfseries CMS} Collaboration, A.~Tumasyan {\em et~al.}, ``{Measurement and QCD
  analysis of double-differential inclusive jet cross sections in proton-proton
  collisions at $ \sqrt{s} $ = 13 TeV},''
  \href{http://dx.doi.org/10.1007/JHEP02(2022)142}{{\em JHEP} {\bfseries 02}
  (2022) 142}, \href{http://arxiv.org/abs/2111.10431}{{\ttfamily
  arXiv:2111.10431 [hep-ex]}}. [Addendum: JHEP 12, 035 (2022)].

\bibitem{ATLAS:2013jmu}
{\bfseries ATLAS} Collaboration, G.~Aad {\em et~al.}, ``{Measurement of dijet
  cross sections in $pp$ collisions at 7 TeV centre-of-mass energy using the
  ATLAS detector},'' \href{http://dx.doi.org/10.1007/JHEP05(2014)059}{{\em
  JHEP} {\bfseries 05} (2014) 059},
  \href{http://arxiv.org/abs/1312.3524}{{\ttfamily arXiv:1312.3524 [hep-ex]}}.

\bibitem{CMS:2012ftr}
{\bfseries CMS} Collaboration, S.~Chatrchyan {\em et~al.}, ``{Measurements of
  Differential Jet Cross Sections in Proton-Proton Collisions at $\sqrt{s}=7$
  TeV with the CMS Detector},''
  \href{http://dx.doi.org/10.1103/PhysRevD.87.112002}{{\em Phys. Rev. D}
  {\bfseries 87} no.~11, (2013) 112002},
  \href{http://arxiv.org/abs/1212.6660}{{\ttfamily arXiv:1212.6660 [hep-ex]}}.
  [Erratum: Phys.Rev.D 87, 119902 (2013)].

\bibitem{CMS:2017jfq}
{\bfseries CMS} Collaboration, A.~M. Sirunyan {\em et~al.}, ``{Measurement of
  the triple-differential dijet cross section in proton-proton collisions at
  $\sqrt{s}=8\,\text {TeV} $ and constraints on parton distribution
  functions},'' \href{http://dx.doi.org/10.1140/epjc/s10052-017-5286-7}{{\em
  Eur. Phys. J. C} {\bfseries 77} no.~11, (2017) 746},
  \href{http://arxiv.org/abs/1705.02628}{{\ttfamily arXiv:1705.02628
  [hep-ex]}}.

\bibitem{CMS:2014nvq}
{\bfseries CMS} Collaboration, S.~Chatrchyan {\em et~al.}, ``{Measurement of
  the Ratio of Inclusive Jet Cross Sections using the Anti-$k_T$ Algorithm with
  Radius Parameters R=0.5 and 0.7 in pp Collisions at $\sqrt{s}=7$ TeV},''
  \href{http://dx.doi.org/10.1103/PhysRevD.90.072006}{{\em Phys. Rev. D}
  {\bfseries 90} no.~7, (2014) 072006},
  \href{http://arxiv.org/abs/1406.0324}{{\ttfamily arXiv:1406.0324 [hep-ex]}}.

\bibitem{ATLAS:2014riz}
{\bfseries ATLAS} Collaboration, G.~Aad {\em et~al.}, ``{Measurement of the
  inclusive jet cross-section in proton-proton collisions at $\sqrt{s}=7$ TeV
  using 4.5 fb$^{-1}$ of data with the ATLAS detector},''
  \href{http://dx.doi.org/10.1007/JHEP02(2015)153}{{\em JHEP} {\bfseries 02}
  (2015) 153}, \href{http://arxiv.org/abs/1410.8857}{{\ttfamily arXiv:1410.8857
  [hep-ex]}}. [Erratum: JHEP 09, 141 (2015)].

\bibitem{CMS:2016lna}
{\bfseries CMS} Collaboration, V.~Khachatryan {\em et~al.}, ``{Measurement and
  QCD analysis of double-differential inclusive jet cross sections in pp
  collisions at $ \sqrt{s}=8 $ TeV and cross section ratios to 2.76 and 7
  TeV},'' \href{http://dx.doi.org/10.1007/JHEP03(2017)156}{{\em JHEP}
  {\bfseries 03} (2017) 156}, \href{http://arxiv.org/abs/1609.05331}{{\ttfamily
  arXiv:1609.05331 [hep-ex]}}.

\bibitem{Britzger:2022lbf}
D.~Britzger {\em et~al.}, ``{NNLO interpolation grids for jet production at the
  LHC},'' \href{http://dx.doi.org/10.1140/epjc/s10052-022-10880-2}{{\em Eur.
  Phys. J. C} {\bfseries 82} no.~10, (2022) 930},
  \href{http://arxiv.org/abs/2207.13735}{{\ttfamily arXiv:2207.13735
  [hep-ph]}}.

\bibitem{Currie:2016bfm}
J.~Currie, E.~W.~N. Glover, and J.~Pires, ``{Next-to-Next-to Leading Order QCD
  Predictions for Single Jet Inclusive Production at the LHC},''
  \href{http://dx.doi.org/10.1103/PhysRevLett.118.072002}{{\em Phys. Rev.
  Lett.} {\bfseries 118} no.~7, (2017) 072002},
\href{http://arxiv.org/abs/1611.01460}{{\ttfamily arXiv:1611.01460 [hep-ph]}}.

\bibitem{Kluge:2006xs}
T.~Kluge, K.~Rabbertz, and M.~Wobisch,
  \href{http://dx.doi.org/10.1142/9789812706706_0110}{``{FastNLO: Fast pQCD
  calculations for PDF fits},''} pp.~483--486.
\newblock 2006.
\newblock \href{http://arxiv.org/abs/hep-ph/0609285}{{\ttfamily
  arXiv:hep-ph/0609285 [hep-ph]}}.
\newblock
\url{http://lss.fnal.gov/cgi-bin/find_paper.pl?conf-06-352}.
\newblock

\bibitem{Wobisch:2011ij}
{\bfseries fastNLO} Collaboration, M.~Wobisch, D.~Britzger, T.~Kluge,
  K.~Rabbertz, and F.~Stober, ``{Theory-Data Comparisons for Jet Measurements
  in Hadron-Induced Processes},''
  \href{http://arxiv.org/abs/1109.1310}{{\ttfamily arXiv:1109.1310 [hep-ph]}}.

\bibitem{Britzger:2012bs}
{\bfseries fastNLO} Collaboration, D.~Britzger, K.~Rabbertz, F.~Stober, and
  M.~Wobisch, \href{http://dx.doi.org/10.3204/DESY-PROC-2012-02/165}{``New
  features in version 2 of the fastnlo project,''} in {\em 20th International
  Workshop on Deep-Inelastic Scattering and Related Subjects}, pp.~217--221.
\newblock 2012.
\newblock \href{http://arxiv.org/abs/1208.3641}{{\ttfamily arXiv:1208.3641
  [hep-ph]}}.

\bibitem{PloughshareProject}
M.~Sutton and B.~Patawah, ``Ploughshare: for all your interpolation grid
  needs.'' \url{https://ploughshare.web.cern.ch/ploughshare/}.

\bibitem{Czakon:2019tmo}
M.~Czakon, A.~van Hameren, A.~Mitov, and R.~Poncelet, ``{Single-jet inclusive
  rates with exact color at $ \mathcal{O} $ ($ {\alpha}_s^4 $)},''
  \href{http://dx.doi.org/10.1007/JHEP10(2019)262}{{\em JHEP} {\bfseries 10}
  (2019) 262}, \href{http://arxiv.org/abs/1907.12911}{{\ttfamily
  arXiv:1907.12911 [hep-ph]}}.

\bibitem{Chen:2022tpk}
X.~Chen, T.~Gehrmann, E.~W.~N. Glover, A.~Huss, and J.~Mo, ``{NNLO QCD
  corrections in full colour for jet production observables at the LHC},''
  \href{http://dx.doi.org/10.1007/JHEP09(2022)025}{{\em JHEP} {\bfseries 09}
  (2022) 025}, \href{http://arxiv.org/abs/2204.10173}{{\ttfamily
  arXiv:2204.10173 [hep-ph]}}.

\bibitem{Cridge:2023ozx}
T.~Cridge, L.~A. Harland-Lang, and R.~S. Thorne, ``{The impact of LHC jet and
  Z$p_T$ data at up to approximate N${}^3$LO order in the MSHT global PDF
  fit},'' \href{http://dx.doi.org/10.1140/epjc/s10052-024-12771-0}{{\em Eur.
  Phys. J. C} {\bfseries 84} no.~4, (2024) 446},
  \href{http://arxiv.org/abs/2312.12505}{{\ttfamily arXiv:2312.12505
  [hep-ph]}}.

\bibitem{ATLAS:2017kux}
{\bfseries ATLAS} Collaboration, M.~Aaboud {\em et~al.}, ``{Measurement of the
  inclusive jet cross-sections in proton-proton collisions at $ \sqrt{s}=8 $
  TeV with the ATLAS detector},''
  \href{http://dx.doi.org/10.1007/JHEP09(2017)020}{{\em JHEP} {\bfseries 09}
  (2017) 020}, \href{http://arxiv.org/abs/1706.03192}{{\ttfamily
  arXiv:1706.03192 [hep-ex]}}.

\bibitem{Ablat:2024nhy}
A.~Ablat {\em et~al.}, ``{The upcoming CTEQ-TEA parton distributions in a
  nutshell},'' \href{http://arxiv.org/abs/2408.11131}{{\ttfamily
  arXiv:2408.11131 [hep-ph]}}.

\bibitem{Kovarik:2019xvh}
K.~Kova\v{r}\'\i{}k, P.~M. Nadolsky, and D.~E. Soper, ``{Hadronic structure in
  high-energy collisions},''
  \href{http://dx.doi.org/10.1103/RevModPhys.92.045003}{{\em Rev. Mod. Phys.}
  {\bfseries 92} no.~4, (2020) 045003},
  \href{http://arxiv.org/abs/1905.06957}{{\ttfamily arXiv:1905.06957
  [hep-ph]}}.

\bibitem{Accardi:2021ysh}
A.~Accardi, T.~J. Hobbs, X.~Jing, and P.~M. Nadolsky, ``{Deuterium scattering
  experiments in CTEQ global QCD analyses: a comparative investigation},''
  \href{http://dx.doi.org/10.1140/epjc/s10052-021-09318-y}{{\em Eur. Phys. J.
  C} {\bfseries 81} no.~7, (2021) 603},
  \href{http://arxiv.org/abs/2102.01107}{{\ttfamily arXiv:2102.01107
  [hep-ph]}}.

\bibitem{Jing:2023isu}
X.~Jing {\em et~al.}, ``{Quantifying the interplay of experimental constraints
  in analyses of parton distributions},''
  \href{http://dx.doi.org/10.1103/PhysRevD.108.034029}{{\em Phys. Rev. D}
  {\bfseries 108} no.~3, (2023) 034029},
  \href{http://arxiv.org/abs/2306.03918}{{\ttfamily arXiv:2306.03918
  [hep-ph]}}.

\bibitem{Stump:2001gu}
D.~Stump, J.~Pumplin, R.~Brock, D.~Casey, J.~Huston, J.~Kalk, H.-L. Lai, and
  W.-K. Tung, ``{Uncertainties of predictions from parton distribution
  functions. 1. The Lagrange multiplier method},''
  \href{http://dx.doi.org/10.1103/PhysRevD.65.014012}{{\em Phys. Rev.}
  {\bfseries D65} (2001) 014012},
\href{http://arxiv.org/abs/hep-ph/0101051}{{\ttfamily arXiv:hep-ph/0101051
  [hep-ph]}}.

\bibitem{L2website}
\protect{The online plotter of ATLAS21, CT18, and MSHT20 sensitivities},
  \url{https://metapdf.hepforge.org/L2}.

\bibitem{Kotz:2024dfg}
L.~Kotz, ``{A study of experimental sensitivities to proton parton
  distributions with xFitter},''
  \href{http://arxiv.org/abs/2401.11350}{{\ttfamily arXiv:2401.11350
  [hep-ph]}}.

\bibitem{xFitterwebsite}
``The {xFitter} project is an open source {QCD} fit framework ready to extract
  {PDFs} and assess the impact of new data.''
  \url{https://www.xfitter.org/xFitter/}.

\bibitem{PDF4LHCWorkingGroup:2022cjn}
{\bfseries PDF4LHC Working Group} Collaboration, R.~D. Ball {\em et~al.},
  ``{The PDF4LHC21 combination of global PDF fits for the LHC Run III},''
  \href{http://dx.doi.org/10.1088/1361-6471/ac7216}{{\em J. Phys. G} {\bfseries
  49} no.~8, (2022) 080501}, \href{http://arxiv.org/abs/2203.05506}{{\ttfamily
  arXiv:2203.05506 [hep-ph]}}.

\bibitem{Pumplin:2001ct}
J.~Pumplin, D.~Stump, R.~Brock, D.~Casey, J.~Huston, J.~Kalk, H.-L. Lai, and
  W.-K. Tung, ``{Uncertainties of predictions from parton distribution
  functions. 2. The Hessian method},''
  \href{http://dx.doi.org/10.1103/PhysRevD.65.014013}{{\em Phys. Rev.}
  {\bfseries D65} (2001) 014013},
\href{http://arxiv.org/abs/hep-ph/0101032}{{\ttfamily arXiv:hep-ph/0101032
  [hep-ph]}}.

\bibitem{Giele:1998gw}
W.~T. Giele and S.~Keller, ``{Implications of hadron collider observables on
  parton distribution function uncertainties},''
  \href{http://dx.doi.org/10.1103/PhysRevD.58.094023}{{\em Phys. Rev. D}
  {\bfseries 58} (1998) 094023},
  \href{http://arxiv.org/abs/hep-ph/9803393}{{\ttfamily arXiv:hep-ph/9803393}}.

\bibitem{Giele:2001mr}
W.~T. Giele, S.~A. Keller, and D.~A. Kosower, ``{Parton Distribution Function
  Uncertainties},'' \href{http://arxiv.org/abs/hep-ph/0104052}{{\ttfamily
  arXiv:hep-ph/0104052}}.

\bibitem{Pumplin:2002vw}
J.~Pumplin, D.~R. Stump, J.~Huston, H.~L. Lai, P.~M. Nadolsky, and W.-K. Tung,
  ``{New generation of parton distributions with uncertainties from global QCD
  analysis},'' \href{http://dx.doi.org/10.1088/1126-6708/2002/07/012}{{\em
  JHEP} {\bfseries 07} (2002) 012},
\href{http://arxiv.org/abs/hep-ph/0201195}{{\ttfamily arXiv:hep-ph/0201195
  [hep-ph]}}.

\bibitem{Martin:2009iq}
A.~D. Martin, W.~J. Stirling, R.~S. Thorne, and G.~Watt, ``{Parton
  distributions for the LHC},''
  \href{http://dx.doi.org/10.1140/epjc/s10052-009-1072-5}{{\em Eur. Phys. J.}
  {\bfseries C63} (2009) 189--285},
\href{http://arxiv.org/abs/0901.0002}{{\ttfamily arXiv:0901.0002 [hep-ph]}}.

\bibitem{Lai:2010vv}
H.-L. Lai, M.~Guzzi, J.~Huston, Z.~Li, P.~M. Nadolsky, J.~Pumplin, and C.~P.
  Yuan, ``{New parton distributions for collider physics},''
  \href{http://dx.doi.org/10.1103/PhysRevD.82.074024}{{\em Phys. Rev.}
  {\bfseries D82} (2010) 074024},
\href{http://arxiv.org/abs/1007.2241}{{\ttfamily arXiv:1007.2241 [hep-ph]}}.

\bibitem{Gao:2013xoa}
J.~Gao, M.~Guzzi, J.~Huston, H.-L. Lai, Z.~Li, P.~Nadolsky, J.~Pumplin,
  D.~Stump, and C.-P. Yuan, ``{CT10 next-to-next-to-leading order global
  analysis of QCD},'' \href{http://dx.doi.org/10.1103/PhysRevD.89.033009}{{\em
  Phys. Rev.} {\bfseries D89} no.~3, (2014) 033009},
\href{http://arxiv.org/abs/1302.6246}{{\ttfamily arXiv:1302.6246 [hep-ph]}}.

\bibitem{Soper:1994km}
D.~E. Soper and J.~C. Collins, ``{Issues in the determination of parton
  distribution functions},''
  \href{http://arxiv.org/abs/hep-ph/9411214}{{\ttfamily arXiv:hep-ph/9411214}}.

\bibitem{MalininGales:2018}
A.~Malinin and M.~Gales, ``Predictive uncertainty estimation via prior
  networks,'' in {\em Proceedings of the 32nd International Conference on
  Neural Information Processing Systems}, NIPS'18, p.~7047–7058.
\newblock Curran Associates Inc., Red Hook, NY, USA, 2018.

\bibitem{Kotz:2023pbu}
L.~Kotz, A.~Courtoy, P.~Nadolsky, F.~Olness, and M.~Ponce-Chavez, ``{Analysis
  of parton distributions in a pion with B\'ezier parametrizations},''
  \href{http://dx.doi.org/10.1103/PhysRevD.109.074027}{{\em Phys. Rev. D}
  {\bfseries 109} no.~7, (2024) 074027},
  \href{http://arxiv.org/abs/2311.08447}{{\ttfamily arXiv:2311.08447
  [hep-ph]}}.

\bibitem{Barry:2018ort}
P.~C. Barry, N.~Sato, W.~Melnitchouk, and C.-R. Ji, ``{First Monte Carlo Global
  QCD Analysis of Pion Parton Distributions},''
  \href{http://dx.doi.org/10.1103/PhysRevLett.121.152001}{{\em Phys. Rev.
  Lett.} {\bfseries 121} no.~15, (2018) 152001},
  \href{http://arxiv.org/abs/1804.01965}{{\ttfamily arXiv:1804.01965
  [hep-ph]}}.

\bibitem{Barry:2021osv}
{\bfseries Jefferson Lab Angular Momentum (JAM)} Collaboration, P.~C. Barry,
  C.-R. Ji, N.~Sato, and W.~Melnitchouk, ``{Global QCD Analysis of Pion Parton
  Distributions with Threshold Resummation},''
  \href{http://dx.doi.org/10.1103/PhysRevLett.127.232001}{{\em Phys. Rev.
  Lett.} {\bfseries 127} no.~23, (2021) 232001},
  \href{http://arxiv.org/abs/2108.05822}{{\ttfamily arXiv:2108.05822
  [hep-ph]}}.

\bibitem{Novikov:2020snp}
I.~Novikov {\em et~al.}, ``{Parton Distribution Functions of the Charged Pion
  Within The xFitter Framework},''
  \href{http://dx.doi.org/10.1103/PhysRevD.102.014040}{{\em Phys. Rev. D}
  {\bfseries 102} no.~1, (2020) 014040},
  \href{http://arxiv.org/abs/2002.02902}{{\ttfamily arXiv:2002.02902
  [hep-ph]}}.

\bibitem{Gao:2013bia}
J.~Gao and P.~Nadolsky, ``{A meta-analysis of parton distribution functions},''
  \href{http://dx.doi.org/10.1007/JHEP07(2014)035}{{\em JHEP} {\bfseries 07}
  (2014) 035}, \href{http://arxiv.org/abs/1401.0013}{{\ttfamily arXiv:1401.0013
  [hep-ph]}}.

\bibitem{Kriesten:2023uoi}
B.~Kriesten and T.~J. Hobbs, ``{Learning PDFs through Interpretable Latent
  Representations in Mellin Space},''
  \href{http://arxiv.org/abs/2312.02278}{{\ttfamily arXiv:2312.02278
  [hep-ph]}}.

\bibitem{Kriesten:2024are}
B.~Kriesten, J.~Gomprecht, and T.~J. Hobbs, ``{Explainable AI classification
  for parton density theory},''
  \href{http://arxiv.org/abs/2407.03411}{{\ttfamily arXiv:2407.03411
  [hep-ph]}}.

\bibitem{Tanabashi:2018oca}
{\bfseries Particle Data Group} Collaboration, M.~Tanabashi {\em et~al.},
  ``{Review of Particle Physics},''
\href{http://dx.doi.org/10.1103/PhysRevD.98.030001}{{\em Phys. Rev.} {\bfseries
  D98} no.~3, (2018) 030001}.

\bibitem{Erler:2020bif}
J.~Erler and R.~Ferro-Hern\'andez, ``{Alternative to the application of PDG
  scale factors},''
  \href{http://dx.doi.org/10.1140/epjc/s10052-020-8115-3}{{\em Eur. Phys. J. C}
  {\bfseries 80} no.~6, (2020) 541},
  \href{http://arxiv.org/abs/2004.01219}{{\ttfamily arXiv:2004.01219
  [physics.data-an]}}.

\bibitem{Cowan:2018lhq}
G.~Cowan, ``{Statistical Models with Uncertain Error Parameters},''
  \href{http://dx.doi.org/10.1140/epjc/s10052-019-6644-4}{{\em Eur. Phys. J. C}
  {\bfseries 79} no.~2, (2019) 133},
  \href{http://arxiv.org/abs/1809.05778}{{\ttfamily arXiv:1809.05778
  [physics.data-an]}}.

\bibitem{DAgostini:1999niu}
G.~D'Agostini, ``{Sceptical combination of experimental results: General
  considerations and application to epsilon-prime / epsilon},''
  \href{http://arxiv.org/abs/hep-ex/9910036}{{\ttfamily arXiv:hep-ex/9910036}}.

\bibitem{Yan:2024yir}
M.~Yan, T.-J. Hou, Z.~Li, K.~Mohan, and C.-P. Yuan, ``{A generalized
  statistical model for fits to parton distributions},''
  \href{http://arxiv.org/abs/2406.01664}{{\ttfamily arXiv:2406.01664
  [hep-ph]}}.

\bibitem{Fadin:1975cb}
V.~S. Fadin, E.~A. Kuraev, and L.~N. Lipatov, ``{On the Pomeranchuk Singularity
  in Asymptotically Free Theories},''
  \href{http://dx.doi.org/10.1016/0370-2693(75)90524-9}{{\em Phys. Lett. B}
  {\bfseries 60} (1975) 50--52}.

\bibitem{Lipatov:1976zz}
L.~N. Lipatov, ``{Reggeization of the Vector Meson and the Vacuum Singularity
  in Nonabelian Gauge Theories},'' {\em Sov. J. Nucl. Phys.} {\bfseries 23}
  (1976) 338--345.

\bibitem{Kuraev:1976ge}
E.~A. Kuraev, L.~N. Lipatov, and V.~S. Fadin, ``{Multi - Reggeon Processes in
  the Yang-Mills Theory},'' {\em Sov. Phys. JETP} {\bfseries 44} (1976)
  443--450.

\bibitem{Kuraev:1977fs}
E.~A. Kuraev, L.~N. Lipatov, and V.~S. Fadin, ``{The Pomeranchuk Singularity in
  Nonabelian Gauge Theories},'' {\em Sov. Phys. JETP} {\bfseries 45} (1977)
  199--204.

\bibitem{Balitsky:1978ic}
I.~I. Balitsky and L.~N. Lipatov, ``{The Pomeranchuk Singularity in Quantum
  Chromodynamics},'' {\em Sov. J. Nucl. Phys.} {\bfseries 28} (1978) 822--829.

\bibitem{Gribov:1983ivg}
L.~V. Gribov, E.~M. Levin, and M.~G. Ryskin, ``{Semihard Processes in QCD},''
  \href{http://dx.doi.org/10.1016/0370-1573(83)90022-4}{{\em Phys. Rept.}
  {\bfseries 100} (1983) 1--150}.

\bibitem{Golec-Biernat:1998zce}
K.~J. Golec-Biernat and M.~Wusthoff, ``{Saturation effects in deep inelastic
  scattering at low $Q^2$ and its implications on diffraction},''
  \href{http://dx.doi.org/10.1103/PhysRevD.59.014017}{{\em Phys. Rev. D}
  {\bfseries 59} (1998) 014017},
  \href{http://arxiv.org/abs/hep-ph/9807513}{{\ttfamily arXiv:hep-ph/9807513}}.

\bibitem{Morreale:2021pnn}
A.~Morreale and F.~Salazar, ``{Mining for Gluon Saturation at Colliders},''
  \href{http://dx.doi.org/10.3390/universe7080312}{{\em Universe} {\bfseries 7}
  no.~8, (2021) 312}, \href{http://arxiv.org/abs/2108.08254}{{\ttfamily
  arXiv:2108.08254 [hep-ph]}}.

\bibitem{Jalilian-Marian:1997qno}
J.~Jalilian-Marian, A.~Kovner, A.~Leonidov, and H.~Weigert, ``{The BFKL
  equation from the Wilson renormalization group},''
  \href{http://dx.doi.org/10.1016/S0550-3213(97)00440-9}{{\em Nucl. Phys. B}
  {\bfseries 504} (1997) 415--431},
  \href{http://arxiv.org/abs/hep-ph/9701284}{{\ttfamily arXiv:hep-ph/9701284}}.

\bibitem{Jalilian-Marian:1997jhx}
J.~Jalilian-Marian, A.~Kovner, A.~Leonidov, and H.~Weigert, ``{The Wilson
  renormalization group for low x physics: Towards the high density regime},''
  \href{http://dx.doi.org/10.1103/PhysRevD.59.014014}{{\em Phys. Rev. D}
  {\bfseries 59} (1998) 014014},
  \href{http://arxiv.org/abs/hep-ph/9706377}{{\ttfamily arXiv:hep-ph/9706377}}.

\bibitem{Weigert:2000gi}
H.~Weigert, ``{Unitarity at small Bjorken x},''
  \href{http://dx.doi.org/10.1016/S0375-9474(01)01668-2}{{\em Nucl. Phys. A}
  {\bfseries 703} (2002) 823--860},
  \href{http://arxiv.org/abs/hep-ph/0004044}{{\ttfamily arXiv:hep-ph/0004044}}.

\bibitem{Iancu:2000hn}
E.~Iancu, A.~Leonidov, and L.~D. McLerran, ``{Nonlinear gluon evolution in the
  color glass condensate. 1.},''
  \href{http://dx.doi.org/10.1016/S0375-9474(01)00642-X}{{\em Nucl. Phys. A}
  {\bfseries 692} (2001) 583--645},
  \href{http://arxiv.org/abs/hep-ph/0011241}{{\ttfamily arXiv:hep-ph/0011241}}.

\bibitem{Ferreiro:2001qy}
E.~Ferreiro, E.~Iancu, A.~Leonidov, and L.~McLerran, ``{Nonlinear gluon
  evolution in the color glass condensate. 2.},''
  \href{http://dx.doi.org/10.1016/S0375-9474(01)01329-X}{{\em Nucl. Phys. A}
  {\bfseries 703} (2002) 489--538},
  \href{http://arxiv.org/abs/hep-ph/0109115}{{\ttfamily arXiv:hep-ph/0109115}}.

\bibitem{Ball:2017otu}
R.~D. Ball, V.~Bertone, M.~Bonvini, S.~Marzani, J.~Rojo, and L.~Rottoli,
  ``{Parton distributions with small-x resummation: evidence for BFKL dynamics
  in HERA data},'' \href{http://dx.doi.org/10.1140/epjc/s10052-018-5774-4}{{\em
  Eur. Phys. J. C} {\bfseries 78} no.~4, (2018) 321},
  \href{http://arxiv.org/abs/1710.05935}{{\ttfamily arXiv:1710.05935
  [hep-ph]}}.

\bibitem{xFitterDevelopersTeam:2018hym}
{\bfseries xFitter Developers' Team} Collaboration, H.~Abdolmaleki {\em
  et~al.}, ``{Impact of low-$x$ resummation on QCD analysis of HERA data},''
  \href{http://dx.doi.org/10.1140/epjc/s10052-018-6090-8}{{\em Eur. Phys. J. C}
  {\bfseries 78} no.~8, (2018) 621},
  \href{http://arxiv.org/abs/1802.00064}{{\ttfamily arXiv:1802.00064
  [hep-ph]}}.

\bibitem{Altarelli:1999vw}
G.~Altarelli, R.~D. Ball, and S.~Forte, ``{Resummation of singlet parton
  evolution at small x},''
  \href{http://dx.doi.org/10.1016/S0550-3213(00)00032-8}{{\em Nucl. Phys. B}
  {\bfseries 575} (2000) 313--329},
  \href{http://arxiv.org/abs/hep-ph/9911273}{{\ttfamily arXiv:hep-ph/9911273}}.

\bibitem{Ball:1999sh}
R.~D. Ball and S.~Forte, ``{The Small x behavior of Altarelli-Parisi splitting
  functions},'' \href{http://dx.doi.org/10.1016/S0370-2693(99)01013-8}{{\em
  Phys. Lett. B} {\bfseries 465} (1999) 271--281},
  \href{http://arxiv.org/abs/hep-ph/9906222}{{\ttfamily arXiv:hep-ph/9906222}}.

\bibitem{Altarelli:2000mh}
G.~Altarelli, R.~D. Ball, and S.~Forte, ``{Small x resummation and HERA
  structure function data},''
  \href{http://dx.doi.org/10.1016/S0550-3213(01)00023-2}{{\em Nucl. Phys. B}
  {\bfseries 599} (2001) 383--423},
  \href{http://arxiv.org/abs/hep-ph/0011270}{{\ttfamily arXiv:hep-ph/0011270}}.

\bibitem{Altarelli:2005ni}
G.~Altarelli, R.~D. Ball, and S.~Forte, ``{Perturbatively stable resummed small
  x evolution kernels},''
  \href{http://dx.doi.org/10.1016/j.nuclphysb.2006.01.046}{{\em Nucl. Phys. B}
  {\bfseries 742} (2006) 1--40},
  \href{http://arxiv.org/abs/hep-ph/0512237}{{\ttfamily arXiv:hep-ph/0512237}}.

\bibitem{Bonvini:2016wki}
M.~Bonvini, S.~Marzani, and T.~Peraro, ``{Small-$x$ resummation from HELL},''
  \href{http://dx.doi.org/10.1140/epjc/s10052-016-4445-6}{{\em Eur. Phys. J. C}
  {\bfseries 76} no.~11, (2016) 597},
  \href{http://arxiv.org/abs/1607.02153}{{\ttfamily arXiv:1607.02153
  [hep-ph]}}.

\bibitem{Iancu:2002tr}
E.~Iancu, K.~Itakura, and L.~McLerran, ``{Geometric scaling above the
  saturation scale},''
  \href{http://dx.doi.org/10.1016/S0375-9474(02)01010-2}{{\em Nucl. Phys. A}
  {\bfseries 708} (2002) 327--352},
  \href{http://arxiv.org/abs/hep-ph/0203137}{{\ttfamily arXiv:hep-ph/0203137}}.

\bibitem{Mueller:2002zm}
A.~H. Mueller and D.~N. Triantafyllopoulos, ``{The Energy dependence of the
  saturation momentum},''
  \href{http://dx.doi.org/10.1016/S0550-3213(02)00581-3}{{\em Nucl. Phys. B}
  {\bfseries 640} (2002) 331--350},
  \href{http://arxiv.org/abs/hep-ph/0205167}{{\ttfamily arXiv:hep-ph/0205167}}.

\bibitem{Guzzi:2021fre}
M.~Guzzi {\em et~al.}, ``{NNLO constraints on proton PDFs from the SeaQuest and
  STAR experiments and other developments in the CTEQ-TEA global analysis},''
  \href{http://dx.doi.org/10.21468/SciPostPhysProc.8.005}{{\em SciPost Phys.
  Proc.} {\bfseries 8} (2022) 005},
  \href{http://arxiv.org/abs/2108.06596}{{\ttfamily arXiv:2108.06596
  [hep-ph]}}.

\bibitem{Bertone:2013vaa}
{\bfseries APFEL} Collaboration, V.~Bertone, S.~Carrazza, and J.~Rojo,
  ``{APFEL: A PDF Evolution Library with QED corrections},''
  \href{http://dx.doi.org/10.1016/j.cpc.2014.03.007}{{\em Comput. Phys.
  Commun.} {\bfseries 185} (2014) 1647--1668},
  \href{http://arxiv.org/abs/1310.1394}{{\ttfamily arXiv:1310.1394 [hep-ph]}}.

\bibitem{Bonvini:2017ogt}
M.~Bonvini, S.~Marzani, and C.~Muselli, ``{Towards parton distribution
  functions with small-$x$ resummation: HELL 2.0}''
  \href{http://dx.doi.org/10.1007/JHEP12(2017)117}{{\em JHEP} {\bfseries 12}
  (2017) 117}, \href{http://arxiv.org/abs/1708.07510}{{\ttfamily
  arXiv:1708.07510 [hep-ph]}}.

\bibitem{Campbell:2006wx}
J.~M. Campbell, J.~W. Huston, and W.~J. Stirling, ``{Hard Interactions of
  Quarks and Gluons: A Primer for LHC Physics},''
  \href{http://dx.doi.org/10.1088/0034-4885/70/1/R02}{{\em Rept. Prog. Phys.}
  {\bfseries 70} (2007) 89},
\href{http://arxiv.org/abs/hep-ph/0611148}{{\ttfamily arXiv:hep-ph/0611148
  [hep-ph]}}.

\bibitem{H1:2013ktq}
{\bfseries H1} Collaboration, V.~Andreev {\em et~al.}, ``{Measurement of
  inclusive $e p$ cross sections at high $Q^2$ at $\sqrt s =$ 225 and 252 GeV
  and of the longitudinal proton structure function $F_L$ at HERA},''
  \href{http://dx.doi.org/10.1140/epjc/s10052-014-2814-6}{{\em Eur. Phys. J. C}
  {\bfseries 74} no.~4, (2014) 2814},
  \href{http://arxiv.org/abs/1312.4821}{{\ttfamily arXiv:1312.4821 [hep-ex]}}.

\bibitem{ZEUS:2014thn}
{\bfseries ZEUS} Collaboration, H.~Abramowicz {\em et~al.}, ``{Deep inelastic
  cross-section measurements at large y with the ZEUS detector at HERA},''
  \href{http://dx.doi.org/10.1103/PhysRevD.90.072002}{{\em Phys. Rev. D}
  {\bfseries 90} no.~7, (2014) 072002},
  \href{http://arxiv.org/abs/1404.6376}{{\ttfamily arXiv:1404.6376 [hep-ex]}}.

\bibitem{Xie:2023suk}
{\bfseries CTEQ-TEA} Collaboration, K.~Xie, J.~Gao, T.~J. Hobbs, D.~R. Stump,
  and C.-P. Yuan, ``{High-energy neutrino deep inelastic scattering cross
  sections},'' \href{http://dx.doi.org/10.1103/PhysRevD.109.113001}{{\em Phys.
  Rev. D} {\bfseries 109} no.~11, (2024) 113001},
  \href{http://arxiv.org/abs/2303.13607}{{\ttfamily arXiv:2303.13607
  [hep-ph]}}.

\bibitem{Silvetti:2022hyc}
F.~Silvetti and M.~Bonvini, ``{Differential heavy quark pair production at
  small x},'' \href{http://dx.doi.org/10.1140/epjc/s10052-023-11326-z}{{\em
  Eur. Phys. J. C} {\bfseries 83} no.~4, (2023) 267},
  \href{http://arxiv.org/abs/2211.10142}{{\ttfamily arXiv:2211.10142
  [hep-ph]}}.

\bibitem{Xie:2021ycd}
K.~Xie, J.~M. Campbell, and P.~M. Nadolsky, ``{A general-mass scheme for prompt
  charm production at hadron colliders},''
  \href{http://dx.doi.org/10.21468/SciPostPhysProc.8.084}{{\em SciPost Phys.
  Proc.} {\bfseries 8} (2022) 084},
  \href{http://arxiv.org/abs/2108.03741}{{\ttfamily arXiv:2108.03741
  [hep-ph]}}.

\bibitem{LHCb:2013xam}
{\bfseries LHCb} Collaboration, R.~Aaij {\em et~al.}, ``{Prompt charm
  production in pp collisions at sqrt(s)=7 TeV},''
  \href{http://dx.doi.org/10.1016/j.nuclphysb.2013.02.010}{{\em Nucl. Phys. B}
  {\bfseries 871} (2013) 1--20},
  \href{http://arxiv.org/abs/1302.2864}{{\ttfamily arXiv:1302.2864 [hep-ex]}}.

\bibitem{LHCb:2015swx}
{\bfseries LHCb} Collaboration, R.~Aaij {\em et~al.}, ``{Measurements of prompt
  charm production cross-sections in $pp$ collisions at $ \sqrt{s}=13 $ TeV},''
  \href{http://dx.doi.org/10.1007/JHEP03(2016)159}{{\em JHEP} {\bfseries 03}
  (2016) 159}, \href{http://arxiv.org/abs/1510.01707}{{\ttfamily
  arXiv:1510.01707 [hep-ex]}}. [Erratum: JHEP 09, 013 (2016), Erratum: JHEP 05,
  074 (2017)].

\bibitem{Anchordoqui:2021ghd}
L.~A. Anchordoqui {\em et~al.}, ``{The Forward Physics Facility: Sites,
  experiments, and physics potential},''
  \href{http://dx.doi.org/10.1016/j.physrep.2022.04.004}{{\em Phys. Rept.}
  {\bfseries 968} (2022) 1--50},
  \href{http://arxiv.org/abs/2109.10905}{{\ttfamily arXiv:2109.10905
  [hep-ph]}}.

\bibitem{Feng:2022inv}
J.~L. Feng {\em et~al.}, ``{The Forward Physics Facility at the High-Luminosity
  LHC},'' \href{http://dx.doi.org/10.1088/1361-6471/ac865e}{{\em J. Phys. G}
  {\bfseries 50} no.~3, (2023) 030501},
  \href{http://arxiv.org/abs/2203.05090}{{\ttfamily arXiv:2203.05090
  [hep-ex]}}.

\bibitem{Fiaschi:2022wgl}
J.~Fiaschi, F.~Giuli, F.~Hautmann, S.~Moch, and S.~Moretti, ``{Z'-boson
  dilepton searches and the high-x quark density},''
  \href{http://dx.doi.org/10.1016/j.physletb.2023.137915}{{\em Phys. Lett. B}
  {\bfseries 841} (2023) 137915},
  \href{http://arxiv.org/abs/2211.06188}{{\ttfamily arXiv:2211.06188
  [hep-ph]}}.

\bibitem{Fu:2023rrs}
Y.~Fu, R.~Brock, D.~Hayden, and C.-P. Yuan, ``{Probing Parton distribution
  functions at large x via Drell-Yan Forward-Backward Asymmetry},''
  \href{http://arxiv.org/abs/2307.07839}{{\ttfamily arXiv:2307.07839
  [hep-ph]}}.

\bibitem{Hobbs:2008mm}
T.~Hobbs and W.~Melnitchouk, ``{Finite-$Q^2$ corrections to parity-violating
  DIS},'' \href{http://dx.doi.org/10.1103/PhysRevD.77.114023}{{\em Phys. Rev.
  D} {\bfseries 77} (2008) 114023},
  \href{http://arxiv.org/abs/0801.4791}{{\ttfamily arXiv:0801.4791 [hep-ph]}}.

\bibitem{Brady:2011uy}
L.~T. Brady, A.~Accardi, T.~J. Hobbs, and W.~Melnitchouk, ``{Next-to leading
  order analysis of target mass corrections to structure functions and
  asymmetries},'' \href{http://dx.doi.org/10.1103/PhysRevD.84.074008}{{\em
  Phys. Rev. D} {\bfseries 84} (2011) 074008},
  \href{http://arxiv.org/abs/1108.4734}{{\ttfamily arXiv:1108.4734 [hep-ph]}}.
  [Erratum: Phys.Rev.D 85, 039902 (2012)].

\bibitem{NuSea:2001idv}
{\bfseries NuSea} Collaboration, R.~S. Towell {\em et~al.}, ``{Improved
  measurement of the anti-d / anti-u asymmetry in the nucleon sea},''
  \href{http://dx.doi.org/10.1103/PhysRevD.64.052002}{{\em Phys. Rev. D}
  {\bfseries 64} (2001) 052002},
  \href{http://arxiv.org/abs/hep-ex/0103030}{{\ttfamily arXiv:hep-ex/0103030}}.

\bibitem{SeaQuest:2021zxb}
{\bfseries SeaQuest} Collaboration, J.~Dove {\em et~al.}, ``{The asymmetry of
  antimatter in the proton},''
  \href{http://dx.doi.org/10.1038/s41586-022-04707-z}{{\em Nature} {\bfseries
  590} no.~7847, (2021) 561--565},
  \href{http://arxiv.org/abs/2103.04024}{{\ttfamily arXiv:2103.04024
  [hep-ph]}}. [Erratum: Nature 604, E26 (2022)].

\bibitem{Accardi:2023chb}
A.~Accardi {\em et~al.}, ``{Strong Interaction Physics at the Luminosity
  Frontier with 22 GeV Electrons at Jefferson Lab},''
  \href{http://arxiv.org/abs/2306.09360}{{\ttfamily arXiv:2306.09360
  [nucl-ex]}}.

\bibitem{SLACProposalE149bis}
M.~Dalton, ``\protect{DIS-PARITY: Parity violation in Deep Inelastic Electron
  Scattering}.'' \protect{SLAC proposal E149bis},
  \url{https://www.slac.stanford.edu/grp/rd/epac/Proposal/E149-bis.pdf}, May,
  1993.

\bibitem{Dalton:2023}
M.~Dalton. \protect{talk at the "Science at the Luminosity Frontier: Jefferson
  Lab at 22 GeV" Workshop},
  \url{https://www.jlab.org/conference/luminosity22gev}, January, 2023.

\bibitem{Gao:2021fle}
J.~Gao, T.~J. Hobbs, P.~M. Nadolsky, C.~Sun, and C.-P. Yuan, ``{General
  heavy-flavor mass scheme for charged-current DIS at NNLO and beyond},''
  \href{http://dx.doi.org/10.1103/PhysRevD.105.L011503}{{\em Phys. Rev. D}
  {\bfseries 105} no.~1, (2022) L011503},
  \href{http://arxiv.org/abs/2107.00460}{{\ttfamily arXiv:2107.00460
  [hep-ph]}}.

\bibitem{Guzzi:2011ew}
M.~Guzzi, P.~M. Nadolsky, H.-L. Lai, and C.-P. Yuan, ``{General-Mass Treatment
  for Deep Inelastic Scattering at Two-Loop Accuracy},''
  \href{http://dx.doi.org/10.1103/PhysRevD.86.053005}{{\em Phys. Rev.}
  {\bfseries D86} (2012) 053005},
\href{http://arxiv.org/abs/1108.5112}{{\ttfamily arXiv:1108.5112 [hep-ph]}}.

\bibitem{Brodsky:1980pb}
S.~J. Brodsky, P.~Hoyer, C.~Peterson, and N.~Sakai, ``{The Intrinsic Charm of
  the Proton},''
\href{http://dx.doi.org/10.1016/0370-2693(80)90364-0}{{\em Phys. Lett.}
  {\bfseries B93} (1980) 451--455}.

\bibitem{Pumplin:2007wg}
J.~Pumplin, H.-L. Lai, and W.-K. Tung, ``{The Charm Parton Content of the
  Nucleon},'' \href{http://dx.doi.org/10.1103/PhysRevD.75.054029}{{\em Phys.
  Rev.} {\bfseries D75} (2007) 054029},
\href{http://arxiv.org/abs/hep-ph/0701220}{{\ttfamily arXiv:hep-ph/0701220
  [hep-ph]}}.

\bibitem{Jimenez-Delgado:2014zga}
P.~Jimenez-Delgado, T.~J. Hobbs, J.~T. Londergan, and W.~Melnitchouk, ``{New
  limits on intrinsic charm in the nucleon from global analysis of parton
  distributions},''
  \href{http://dx.doi.org/10.1103/PhysRevLett.114.082002}{{\em Phys. Rev.
  Lett.} {\bfseries 114} no.~8, (2015) 082002},
\href{http://arxiv.org/abs/1408.1708}{{\ttfamily arXiv:1408.1708 [hep-ph]}}.

\bibitem{Hou:2017khm}
T.-J. Hou, S.~Dulat, J.~Gao, M.~Guzzi, J.~Huston, P.~Nadolsky, C.~Schmidt,
  J.~Winter, K.~Xie, and C.~P. Yuan, ``{CT14 Intrinsic Charm Parton
  Distribution Functions from CTEQ-TEA Global Analysis},''
  \href{http://dx.doi.org/10.1007/JHEP02(2018)059}{{\em JHEP} {\bfseries 02}
  (2018) 059},
\href{http://arxiv.org/abs/1707.00657}{{\ttfamily arXiv:1707.00657 [hep-ph]}}.

\bibitem{Guzzi:2022rca}
M.~Guzzi, T.~J. Hobbs, K.~Xie, J.~Huston, P.~Nadolsky, and C.-P. Yuan, ``{The
  persistent nonperturbative charm enigma},''
  \href{http://dx.doi.org/10.1016/j.physletb.2023.137975}{{\em Phys. Lett. B}
  {\bfseries 843} (2023) 137975},
  \href{http://arxiv.org/abs/2211.01387}{{\ttfamily arXiv:2211.01387
  [hep-ph]}}.

\bibitem{Ball:2022qks}
{\bfseries NNPDF} Collaboration, R.~D. Ball, A.~Candido, J.~Cruz-Martinez,
  S.~Forte, T.~Giani, F.~Hekhorn, K.~Kudashkin, G.~Magni, and J.~Rojo,
  ``{Evidence for intrinsic charm quarks in the proton},''
  \href{http://dx.doi.org/10.1038/s41586-022-04998-2}{{\em Nature} {\bfseries
  608} no.~7923, (2022) 483--487},
  \href{http://arxiv.org/abs/2208.08372}{{\ttfamily arXiv:2208.08372
  [hep-ph]}}.

\bibitem{Ball:2022uon}
{\bfseries NNPDF} Collaboration, R.~D. Ball, J.~Cruz-Martinez, L.~Del~Debbio,
  S.~Forte, Z.~Kassabov, E.~R. Nocera, J.~Rojo, R.~Stegeman, and M.~Ubiali,
  ``{Response to ''Parton distributions need representative sampling''},''
  \href{http://arxiv.org/abs/2211.12961}{{\ttfamily arXiv:2211.12961
  [hep-ph]}}.

\bibitem{HopscotchWebsite}
``2022 hopscotch scans of the lhc cross sections, supplementary material.''
  \url{https://ct.hepforge.org/PDFs/2022hopscotch/}.

\bibitem{Harland-Lang:2024kvt}
L.~A. Harland-Lang, T.~Cridge, and R.~S. Thorne, ``{A Stress Test of Global PDF
  Fits: Closure Testing the MSHT PDFs and a First Direct Comparison to the
  Neural Net Approach},'' \href{http://arxiv.org/abs/2407.07944}{{\ttfamily
  arXiv:2407.07944 [hep-ph]}}.

\bibitem{EuropeanMuon:1982xfn}
{\bfseries European Muon} Collaboration, J.~J. Aubert {\em et~al.},
  ``{Production of charmed particles in 250-GeV $\mu^+$ - iron interactions},''
  \href{http://dx.doi.org/10.1016/0550-3213(83)90174-8}{{\em Nucl. Phys. B}
  {\bfseries 213} (1983) 31--64}.

\bibitem{LHCb:2021stx}
{\bfseries LHCb} Collaboration, R.~Aaij {\em et~al.}, ``{Study of Z Bosons
  Produced in Association with Charm in the Forward Region},''
  \href{http://dx.doi.org/10.1103/PhysRevLett.128.082001}{{\em Phys. Rev.
  Lett.} {\bfseries 128} no.~8, (2022) 082001},
  \href{http://arxiv.org/abs/2109.08084}{{\ttfamily arXiv:2109.08084
  [hep-ex]}}.

\bibitem{Hobbs:2013bia}
T.~J. Hobbs, J.~T. Londergan, and W.~Melnitchouk, ``{Phenomenology of
  nonperturbative charm in the nucleon},''
  \href{http://dx.doi.org/10.1103/PhysRevD.89.074008}{{\em Phys. Rev.}
  {\bfseries D89} no.~7, (2014) 074008},
\href{http://arxiv.org/abs/1311.1578}{{\ttfamily arXiv:1311.1578 [hep-ph]}}.

\bibitem{NNPDF:2023tyk}
{\bfseries NNPDF} Collaboration, R.~D. Ball, A.~Candido, J.~Cruz-Martinez,
  S.~Forte, T.~Giani, F.~Hekhorn, G.~Magni, E.~R. Nocera, J.~Rojo, and
  R.~Stegeman, ``{Intrinsic charm quark valence distribution of the proton},''
  \href{http://dx.doi.org/10.1103/PhysRevD.109.L091501}{{\em Phys. Rev. D}
  {\bfseries 109} no.~9, (2024) L091501},
  \href{http://arxiv.org/abs/2311.00743}{{\ttfamily arXiv:2311.00743
  [hep-ph]}}.

\bibitem{Xie:2023qbn}
{\bfseries CTEQ-TEA} Collaboration, K.~Xie, B.~Zhou, and T.~J. Hobbs, ``{The
  photon content of the neutron},''
  \href{http://dx.doi.org/10.1007/JHEP04(2024)022}{{\em JHEP} {\bfseries 04}
  (2024) 022}, \href{http://arxiv.org/abs/2305.10497}{{\ttfamily
  arXiv:2305.10497 [hep-ph]}}.

\bibitem{Cridge:2021pxm}
T.~Cridge, L.~A. Harland-Lang, A.~D. Martin, and R.~S. Thorne, ``{QED parton
  distribution functions in the MSHT20 fit},''
  \href{http://dx.doi.org/10.1140/epjc/s10052-022-10028-2}{{\em Eur. Phys. J.
  C} {\bfseries 82} no.~1, (2022) 90},
  \href{http://arxiv.org/abs/2111.05357}{{\ttfamily arXiv:2111.05357
  [hep-ph]}}.

\bibitem{Schmidt:2015zda}
C.~Schmidt, J.~Pumplin, D.~Stump, and C.-P. Yuan, ``{CT14QED parton
  distribution functions from isolated photon production in deep inelastic
  scattering},'' \href{http://dx.doi.org/10.1103/PhysRevD.93.114015}{{\em Phys.
  Rev.} {\bfseries D93} no.~11, (2016) 114015},
\href{http://arxiv.org/abs/1509.02905}{{\ttfamily arXiv:1509.02905 [hep-ph]}}.

\bibitem{Martin:2004dh}
A.~D. Martin, R.~G. Roberts, W.~J. Stirling, and R.~S. Thorne, ``{Parton
  distributions incorporating QED contributions},''
  \href{http://dx.doi.org/10.1140/epjc/s2004-02088-7}{{\em Eur. Phys. J.}
  {\bfseries C39} (2005) 155--161},
\href{http://arxiv.org/abs/hep-ph/0411040}{{\ttfamily arXiv:hep-ph/0411040
  [hep-ph]}}.

\bibitem{Xie:2021equ}
{\bfseries CTEQ-TEA} Collaboration, K.~Xie, T.~J. Hobbs, T.-J. Hou, C.~Schmidt,
  M.~Yan, and C.-P. Yuan, ``{Photon PDF within the CT18 global analysis},''
  \href{http://dx.doi.org/10.1103/PhysRevD.105.054006}{{\em Phys. Rev. D}
  {\bfseries 105} no.~5, (2022) 054006},
  \href{http://arxiv.org/abs/2106.10299}{{\ttfamily arXiv:2106.10299
  [hep-ph]}}.

\bibitem{Manohar:2016nzj}
A.~Manohar, P.~Nason, G.~P. Salam, and G.~Zanderighi, ``{How bright is the
  proton? A precise determination of the photon parton distribution
  function},'' \href{http://dx.doi.org/10.1103/PhysRevLett.117.242002}{{\em
  Phys. Rev. Lett.} {\bfseries 117} no.~24, (2016) 242002},
  \href{http://arxiv.org/abs/1607.04266}{{\ttfamily arXiv:1607.04266
  [hep-ph]}}.

\bibitem{Manohar:2017eqh}
A.~V. Manohar, P.~Nason, G.~P. Salam, and G.~Zanderighi, ``{The Photon Content
  of the Proton},'' \href{http://dx.doi.org/10.1007/JHEP12(2017)046}{{\em JHEP}
  {\bfseries 12} (2017) 046},
\href{http://arxiv.org/abs/1708.01256}{{\ttfamily arXiv:1708.01256 [hep-ph]}}.

\bibitem{Ye:2017gyb}
Z.~Ye, J.~Arrington, R.~J. Hill, and G.~Lee, ``{Proton and Neutron
  Electromagnetic Form Factors and Uncertainties},''
  \href{http://dx.doi.org/10.1016/j.physletb.2017.11.023}{{\em Phys. Lett. B}
  {\bfseries 777} (2018) 8--15},
  \href{http://arxiv.org/abs/1707.09063}{{\ttfamily arXiv:1707.09063
  [nucl-ex]}}.

\bibitem{CLAS:2003iiq}
{\bfseries CLAS} Collaboration, M.~Osipenko {\em et~al.}, ``{A Kinematically
  complete measurement of the proton structure function F(2) in the resonance
  region and evaluation of its moments},''
  \href{http://dx.doi.org/10.1103/PhysRevD.67.092001}{{\em Phys. Rev. D}
  {\bfseries 67} (2003) 092001},
  \href{http://arxiv.org/abs/hep-ph/0301204}{{\ttfamily arXiv:hep-ph/0301204}}.

\bibitem{Christy:2007ve}
M.~E. Christy and P.~E. Bosted, ``{Empirical fit to precision inclusive
  electron-proton cross- sections in the resonance region},''
  \href{http://dx.doi.org/10.1103/PhysRevC.81.055213}{{\em Phys. Rev. C}
  {\bfseries 81} (2010) 055213},
  \href{http://arxiv.org/abs/0712.3731}{{\ttfamily arXiv:0712.3731 [hep-ph]}}.

\bibitem{Bosted:2007xd}
P.~E. Bosted and M.~E. Christy, ``{Empirical fit to inelastic electron-deuteron
  and electron-neutron resonance region transverse cross-sections},''
  \href{http://dx.doi.org/10.1103/PhysRevC.77.065206}{{\em Phys. Rev. C}
  {\bfseries 77} (2008) 065206},
  \href{http://arxiv.org/abs/0711.0159}{{\ttfamily arXiv:0711.0159 [hep-ph]}}.

\bibitem{HERMES:2011yno}
{\bfseries HERMES} Collaboration, A.~Airapetian {\em et~al.}, ``{Inclusive
  Measurements of Inelastic Electron and Positron Scattering from Unpolarized
  Hydrogen and Deuterium Targets},''
  \href{http://dx.doi.org/10.1007/JHEP05(2011)126}{{\em JHEP} {\bfseries 05}
  (2011) 126}, \href{http://arxiv.org/abs/1103.5704}{{\ttfamily arXiv:1103.5704
  [hep-ex]}}.

\bibitem{E143:1998nvx}
{\bfseries E143} Collaboration, K.~Abe {\em et~al.}, ``{Measurements of R =
  $\sigma$(L) / $\sigma$(T) for 0.03 \ensuremath{<} x \ensuremath{<} 0.1 and
  fit to world data},''
  \href{http://dx.doi.org/10.1016/S0370-2693(99)00244-0}{{\em Phys. Lett. B}
  {\bfseries 452} (1999) 194--200},
  \href{http://arxiv.org/abs/hep-ex/9808028}{{\ttfamily arXiv:hep-ex/9808028}}.

\bibitem{Accardi:2016qay}
A.~Accardi, L.~T. Brady, W.~Melnitchouk, J.~F. Owens, and N.~Sato,
  ``{Constraints on large-$x$ parton distributions from new weak boson
  production and deep-inelastic scattering data},''
  \href{http://dx.doi.org/10.1103/PhysRevD.93.114017}{{\em Phys. Rev. D}
  {\bfseries 93} no.~11, (2016) 114017},
  \href{http://arxiv.org/abs/1602.03154}{{\ttfamily arXiv:1602.03154
  [hep-ph]}}.

\bibitem{Abt:2016vjh}
I.~Abt, A.~M. Cooper-Sarkar, B.~Foster, V.~Myronenko, K.~Wichmann, and M.~Wing,
  ``{Study of HERA ep data at low Q$^2$ and low $x_{Bj}$ and the need for
  higher-twist corrections to standard perturbative QCD fits},''
  \href{http://dx.doi.org/10.1103/PhysRevD.94.034032}{{\em Phys. Rev. D}
  {\bfseries 94} no.~3, (2016) 034032},
  \href{http://arxiv.org/abs/1604.02299}{{\ttfamily arXiv:1604.02299
  [hep-ph]}}.

\bibitem{Harland-Lang:2019pla}
L.~A. Harland-Lang, A.~D. Martin, R.~Nathvani, and R.~S. Thorne, ``{Ad Lucem:
  QED Parton Distribution Functions in the MMHT Framework},''
  \href{http://dx.doi.org/10.1140/epjc/s10052-019-7296-0}{{\em Eur. Phys. J.}
  {\bfseries C79} no.~10, (2019) 811},
\href{http://arxiv.org/abs/1907.02750}{{\ttfamily arXiv:1907.02750 [hep-ph]}}.

\bibitem{Ball:2013hta}
{\bfseries NNPDF} Collaboration, R.~D. Ball, V.~Bertone, S.~Carrazza,
  L.~Del~Debbio, S.~Forte, A.~Guffanti, N.~P. Hartland, and J.~Rojo, ``{Parton
  distributions with QED corrections},''
  \href{http://dx.doi.org/10.1016/j.nuclphysb.2013.10.010}{{\em Nucl. Phys. B}
  {\bfseries 877} (2013) 290--320},
  \href{http://arxiv.org/abs/1308.0598}{{\ttfamily arXiv:1308.0598 [hep-ph]}}.

\bibitem{Wells:2015uba}
J.~D. Wells and Z.~Zhang, ``{Effective theories of universal theories},''
  \href{http://dx.doi.org/10.1007/JHEP01(2016)123}{{\em JHEP} {\bfseries 01}
  (2016) 123}, \href{http://arxiv.org/abs/1510.08462}{{\ttfamily
  arXiv:1510.08462 [hep-ph]}}.

\bibitem{Grzadkowski:2010es}
B.~Grzadkowski, M.~Iskrzynski, M.~Misiak, and J.~Rosiek, ``{Dimension-Six Terms
  in the Standard Model Lagrangian},''
  \href{http://dx.doi.org/10.1007/JHEP10(2010)085}{{\em JHEP} {\bfseries 10}
  (2010) 085}, \href{http://arxiv.org/abs/1008.4884}{{\ttfamily arXiv:1008.4884
  [hep-ph]}}.

\bibitem{Stump:2003yu}
D.~Stump, J.~Huston, J.~Pumplin, W.-K. Tung, H.~L. Lai, S.~Kuhlmann, and J.~F.
  Owens, ``{Inclusive jet production, parton distributions, and the search for
  new physics},'' \href{http://dx.doi.org/10.1088/1126-6708/2003/10/046}{{\em
  JHEP} {\bfseries 10} (2003) 046},
  \href{http://arxiv.org/abs/hep-ph/0303013}{{\ttfamily arXiv:hep-ph/0303013}}.

\bibitem{Carrazza:2019sec}
S.~Carrazza, C.~Degrande, S.~Iranipour, J.~Rojo, and M.~Ubiali, ``{Can New
  Physics hide inside the proton?},''
  \href{http://dx.doi.org/10.1103/PhysRevLett.123.132001}{{\em Phys. Rev.
  Lett.} {\bfseries 123} no.~13, (2019) 132001},
  \href{http://arxiv.org/abs/1905.05215}{{\ttfamily arXiv:1905.05215
  [hep-ph]}}.

\bibitem{Hammou:2023heg}
E.~Hammou, Z.~Kassabov, M.~Madigan, M.~L. Mangano, L.~Mantani, J.~Moore, M.~M.
  Alvarado, and M.~Ubiali, ``{Hide and seek: how PDFs can conceal new
  physics},'' \href{http://dx.doi.org/10.1007/JHEP11(2023)090}{{\em JHEP}
  {\bfseries 11} (2023) 090}, \href{http://arxiv.org/abs/2307.10370}{{\ttfamily
  arXiv:2307.10370 [hep-ph]}}.

\bibitem{Gao:2022srd}
J.~Gao, M.~Gao, T.~J. Hobbs, D.~Liu, and X.~Shen, ``{Simultaneous CTEQ-TEA
  extraction of PDFs and SMEFT parameters from jet and $ t\overline{t} $
  data},'' \href{http://dx.doi.org/10.1007/JHEP05(2023)003}{{\em JHEP}
  {\bfseries 05} (2023) 003}, \href{http://arxiv.org/abs/2211.01094}{{\ttfamily
  arXiv:2211.01094 [hep-ph]}}.

\bibitem{Liu:2022plj}
D.~Liu, C.~Sun, and J.~Gao, ``{Machine learning of log-likelihood functions in
  global analysis of parton distributions},''
  \href{http://dx.doi.org/10.1007/JHEP08(2022)088}{{\em JHEP} {\bfseries 08}
  (2022) 088}, \href{http://arxiv.org/abs/2201.06586}{{\ttfamily
  arXiv:2201.06586 [hep-ph]}}.

\bibitem{McGowan:2022nag}
J.~McGowan, T.~Cridge, L.~A. Harland-Lang, and R.~S. Thorne, ``{Approximate
  N$^{3}$LO Parton Distribution Functions with Theoretical Uncertainties:
  MSHT20aN$^3$LO PDFs},'' \href{http://arxiv.org/abs/2207.04739}{{\ttfamily
  arXiv:2207.04739 [hep-ph]}}.

\bibitem{NNPDF:2024nan}
{\bfseries NNPDF} Collaboration, R.~D. Ball {\em et~al.}, ``{The path to $\hbox
  {N}^3\hbox {LO}$ parton distributions},''
  \href{http://dx.doi.org/10.1140/epjc/s10052-024-12891-7}{{\em Eur. Phys. J.
  C} {\bfseries 84} no.~7, (2024) 659},
  \href{http://arxiv.org/abs/2402.18635}{{\ttfamily arXiv:2402.18635
  [hep-ph]}}.

\bibitem{Cooper-Sarkar:2024crx}
A.~Cooper-Sarkar, T.~Cridge, F.~Giuli, L.~A. Harland-Lang, F.~Hekhorn,
  J.~Huston, G.~Magni, S.~Moch, and R.~S. Thorne, ``{A Benchmarking of QCD
  Evolution at Approximate $N^3LO$},''
  \href{http://arxiv.org/abs/2406.16188}{{\ttfamily arXiv:2406.16188
  [hep-ph]}}.

\bibitem{Ablinger:2024xtt}
J.~Ablinger, A.~Behring, J.~Bl\"umlein, A.~De~Freitas, A.~von Manteuffel,
  C.~Schneider, and K.~Sch\"onwald, ``{The non-first-order-factorizable
  contributions to the three-loop single-mass operator matrix elements AQg(3)
  and \ensuremath{\Delta}AQg(3)},''
  \href{http://dx.doi.org/10.1016/j.physletb.2024.138713}{{\em Phys. Lett. B}
  {\bfseries 854} (2024) 138713},
  \href{http://arxiv.org/abs/2403.00513}{{\ttfamily arXiv:2403.00513
  [hep-ph]}}.

\end{thebibliography}

\providecommand{\href}[2]{#2}\begingroup\raggedright\endgroup

\end{document}